\documentclass[twocolumn]{aastex62}

\usepackage{color}

\usepackage{pdflscape}

\makeatletter
\AtBeginDocument{\let\LS@rot\@undefined}
\makeatother

\newcommand{\project}[1]{\textsl{#1}} 
\newcommand{\pipeline}[1]{\texttt{#1}}

\newcommand{\npassz}{8086}
\newcommand{\nfailz}{144,779}
\newcommand{\fracpassz}{5.3}
\newcommand{\npasssin}{3057}
\newcommand{\nelimsin}{5029}
\newcommand{\fracelimsin}{62}
\newcommand{\nelimdur}{110}
\newcommand{\npassdur}{2947}

\newcommand{\myoverlap}{72}
\newcommand{\theiroverlap}{54}
\newcommand{\theirmissing}{46}
\newcommand{\nimissed}{192}
\newcommand{\nimissedstars}{182}

\newcommand{\numstars}{152,865}
\newcommand{\nsystemme}{1312}
\newcommand{\nsystems}{1306}
\newcommand{\ericbad}{6}
\newcommand{\nastro}{1702}
\newcommand{\nmulti}{396}
\newcommand{\nfailvet}{1641}
\newcommand{\oddevenebs}{215}
\newcommand{\qatsebs}{259}
\newcommand{\twoqatsebs}{518}
\newcommand{\nfps}{39}
\newcommand{\nfpsys}{30}
\newcommand{\withsingles}{942}
\newcommand{\withsinglessys}{809}
\newcommand{\nsingles}{124}
\newcommand{\nplanet}{818}
\newcommand{\nplanethost}{698}
\newcommand{\nnewplanet}{374}
\newcommand{\ngaia}{648}
\newcommand{\nostellar}{17}
\newcommand{\nostellarczero}{13}
\newcommand{\rserrgaia}{5.7}
\newcommand{\rserrhuber}{17.9}
\newcommand{\smallnep}{69}
\newcommand{\fracsingle}{88}
\newcommand{\singletrantot}{77}
\newcommand{\singletranplan}{21}
\newcommand{\nebs}{579}
\newcommand{\ndoubleebs}{455}
\newcommand{\planettoeb}{1.21}

\newcommand{\newsn}{38}
\newcommand{\oldsn}{62}
\newcommand{\snbelowold}{66}
\newcommand{\newmult}{49}
\newcommand{\nnewinmulti}{93}
\newcommand{\nnew}{374}
\newcommand{\fracnewinmulti}{25}
\newcommand{\ninmulti}{207}
\newcommand{\fracinmulti}{25}
\newcommand{\nplanetpairs}{161}
\newcommand{\nsingletran}{77}
\newcommand{\fracbright}{47}
\newcommand{\nsmallbright}{190}
\newcommand{\rperrgaia}{14}
\newcommand{\rperrhuber}{30}
\newcommand{\lumerrgaia}{2.5}
\newcommand{\lumerrhuber}{37}
\newcommand{\inserrgaia}{48}
\newcommand{\inserrhuber}{86}
\newcommand{\arerr}{23}
\newcommand{\nmultis}{87}
\newcommand{\nothermultis}{57}
\newcommand{\nmdwarfcands}{100}

\newcommand{\nnewusp}{25}

\shorttitle{Planet Candidates in \project{K2} Campaigns 0--8}
\shortauthors{Kruse et al.}

\received{26 March 2019}
\revised{19 June 2019}
\accepted{21 July 2019}
\published{9 September 2019}
\submitjournal{The Astrophysical Journal Supplement Series}

\begin{document}

\title{Detection of Hundreds of New Planet Candidates and Eclipsing
  Binaries in \project{K2} Campaigns 0--8}

\correspondingauthor{Ethan Kruse}
\email{ethan.kruse@nasa.gov}

\author[0000-0002-0493-1342]{Ethan Kruse}
\affiliation{NASA Goddard Space Flight Center, Greenbelt, MD 20771,
  USA}

\author[0000-0002-0802-9145]{Eric Agol}
\altaffiliation{Guggenheim Fellow}
\affiliation{Department of Astronomy, University of Washington, Box
  351580, Seattle, WA 98195, USA}
  
\author[0000-0002-0296-3826]{Rodrigo Luger}
\affiliation{Center for Computational Astrophysics, Flatiron
  Institute, 162 5th Avenue, 6th Floor, New York, NY 10010, USA}

\author[0000-0002-9328-5652]{Daniel Foreman-Mackey}
\affiliation{Center for Computational Astrophysics, Flatiron
  Institute, 162 5th Avenue, 6th Floor, New York, NY 10010, USA}

\begin{abstract}
We implement a search for exoplanets in campaigns zero through eight
(C0--8) of the \project{K2} extension of the \project{Kepler} spacecraft.  We
apply a modified version of the \pipeline{QATS} planet search
algorithm to \project{K2} light curves produced by the \pipeline{EVEREST}
pipeline, carrying out the C0--8 search on $1.5\times 10^5$ target
stars with magnitudes in the range of $Kp = $ 9--15.  We detect
\nplanet{} transiting planet candidates, of which \nnewplanet{} were
undiscovered by prior searches, with $\{64, 15, 5, 2, 1\}$ in
$\{2,3,4,5,6\}$-planet multiplanet candidate systems, respectively.
Of the new planets detected, \nmdwarfcands{} orbit M dwarfs, including
one that is potentially rocky and in the habitable zone.  154 of our
candidates reciprocally transit with our solar system: they are
geometrically aligned to see at least one solar system planet
transit. We find candidates that display transit timing variations
and dozens of candidates on both period extremes with single transits
or ultrashort periods.  We point to evidence that our candidates
display similar patterns in frequency and size--period relation to
confirmed planets, such as tentative evidence for the radius gap.
Confirmation of these planet candidates with follow-up studies will
increase the number of \project{K2} planets by up to 50\%, and characterization
of their host stars will improve statistical studies of planet
properties.  Our sample includes many planets orbiting bright stars
amenable for radial velocity follow-up and future characterization
with JWST. We also list the \nebs{} eclipsing binary systems detected
as part of this search.
\end{abstract}


\section{Introduction}

The \project{Kepler} mission \citep{bor10} produced an abundance of
transiting planet candidates within the 400 deg$^2$ and 4
yr (2009--2013) of its survey of $\approx 2 \times 10^5$ stars
\citep{bor16, cou16, tho18}.  The failure of the spacecraft's second
of four reaction wheels in 2013 halted the prime mission owing to
pointing degradation; three wheels were required to stabilize the
pitch, yaw, and roll of the spacecraft.  Symmetric irradiation by the
Sun on the solar panels was used to balance the telescope along a
third axis, allowing continued operation with only two reaction wheels
\citep{how14}.  However, such a balance was only achieved by pointing
the telescope along the ecliptic --- away from the original
\project{Kepler} mission field of view.

To avoid sunlight in the aperture, the telescope was reoriented to a
new field of view along the ecliptic every $ \sim$80 days (a
``campaign''). These campaigns along the ecliptic constituted a new
mission entitled \project{K2} \citep{how14}; altogether \project{K2} observed 20 campaigns
(C0--19) during its 2014--2018 lifetime. Because the \project{K2} campaigns
covered more varied portions of the galaxy and the target lists were
driven by community proposals, the \project{K2} targets were more diverse than
the mostly solar-like stars of the \project{Kepler} mission. Late-type
M dwarf stars were favored as targets because these gave the best
chance within the limited campaign duration to find small, temperate
planets that may be rocky and lie in the surface liquid water
``habitable zone'' \citep{kas93, kop13}.  Stellar clusters of varying
ages were also targeted, including Praesepe \citep[also called
  Beehive;][]{ lib16, obe16, man17}, M67 \citep{nar16}, Pleiades
\citep{dav16a}, and Hyades \citep{man16}, to uniformly study many stars
at the same age.

While the photon pressure was approximately balanced and kept the
spacecraft relatively stable, the pointing of \project{K2} did drift over a
timescale of 6 hr, followed by a thruster fire that abruptly
brought the pointing back near the initial position.  The pointing
typically drifted by $\approx$3$^{\prime\prime}$--4$^{\prime\prime}$ (corresponding to
$\approx$1 pixel) in early campaigns and by slightly smaller angles in
the later campaigns.  These pointing drifts led to gradual systematic
changes in the measured flux of target stars due to variations in
subpixel sensitivity that cause a different number of photoelectrons
to be detected depending on the location of a star on the detector.
The gradual systematic change in flux was followed by an abrupt
discontinuity caused by the thruster fire changing the pointing more
dramatically between cadences. These effects combined to create light
curves dominated by sawtooth-shaped systematics on timescales of 6--12
hr; the changes in flux due to these systematic effects were about
an order of magnitude larger than the white-noise level of most
targets.

The official \project{Kepler} pipeline used for the prime mission was
not optimized to handle these pointing drift systematics. The team
processed the pixels and produced calibrated pixel files using the
same methods as the original mission, but the sawtooth-shaped effects
were not fully corrected for many of the derived light curves
\citep{how14}. Altogether, the raw photometric precision of an average
12th magnitude star in \project{K2} was $\approx$400 ppm (compared to 100 ppm in
the original \project{Kepler} mission), which is too noisy to detect
small transits \citep{how14}.

Many groups have modeled the additional \project{K2} systematics with custom
pipelines to generate light curves with reduced instrumental noise and
artifacts.  Since the measured flux correlates strongly with the
position of stars on the detector, \citet{van14} developed a \project{K2}
detrending algorithm, \pipeline{K2SFF}, that decorrelates the raw \project{K2}
flux against the centroid position of each star versus time with a
nonlinear function.  \citet{van14} demonstrated a precision within a
factor of two or better of the original \project{Kepler} mission,
depending on the magnitude of the target star.

Other approaches to the detrending utilize common-mode variations
among stars \citep{for15, mon15}, a Gaussian process approach to
decorrelating the flux against time and stellar centroid
\citep[\pipeline{K2SC};][]{aig16,pop16}, and point-spread function fitting to derive
light curves for multiple stars in each postage stamp \citep{lun15}.

Building on these detrended light curves, multiple groups have
searched the \project{K2} data to find planet candidates
\citep[e.g.,][]{bar16a,cro16, pop16,van16b,may18,pet18} in campaigns
0--8.  These different searches contain some overlap in the resulting
planet candidates, but each search also lists planet candidates not
included by the others.  In addition, comparing with the
\project{Kepler} planet population, the \project{K2} dataset has yet to be fully
mined for transiting exoplanets \citep{dot19}.  This leaves open the
opportunity to find planets that have escaped detection to date.

Here we present a new search for planets in the \project{K2} data set utilizing
an approach to detrending that further improves the photometric noise
to within 20\% of the original \project{Kepler} mission for stars in
the magnitude range $11 < Kp < 13$, where $Kp$ is the magnitude
defined in the \project{Kepler} bandpass.  This approach, EPIC
Variability Extraction and Removal for Exoplanet Science Targets
(\pipeline{EVEREST}), solely utilizes information from pixels within
the aperture containing each star and sidesteps estimating the stellar
centroids \citep{lug16}.  \pipeline{EVEREST} provides improved light
curves for \project{K2} target stars, which enables a more sensitive search for
planetary transits.

In \S\ref{methods} we summarize the \pipeline{EVEREST} detrending
pipeline and then discuss our planet search pipeline; we also detail
the vetting of candidates, the search for multiple candidates, and the
Markov chain analysis procedure used for each candidate.  In
\S\ref{results} we summarize the results of our search, the procedure
to separate out eclipsing binaries (EBs), and the properties of the planet
candidates.  In \S\ref{ch:analysis} we discuss the implications of
those results, and we conclude in \S\ref{conclusions} by pointing out
future directions and applications of this planet search.

\section{Methods} \label{methods}

\subsection{Pipeline Overview}

Before diving into details, this section sketches the general outline
of our pipeline --- from raw pixels to planet candidates --- and
points the reader to the relevant sections discussing each piece in
more depth.

To address the systematic variations in the \project{K2} photometry, we have
developed a pipeline for the decorrelation of raw \project{K2} light curves,
\pipeline{EVEREST} \citep{lug16,lug18}, which we briefly review in
\S\ref{everest}.  Our approach makes use of the pixel-level
decorrelation (\pipeline{PLD}) technique, which bypasses centroid
measurements and instead decorrelates the light curve against the
normalized flux in each pixel within the aperture \citep{dem15}. Once
\pipeline{EVEREST} has removed the most significant systematics, we
apply some further filtering to the data to remove spurious outliers
(\S\ref{prep}).

We then use a modified version of the \cite{car13} Quasi-periodic
Automated Transit Search (\pipeline{QATS}) technique to search for
planet candidates. Although designed specifically to find planets
exhibiting transit timing variations (TTVs), \pipeline{QATS} can also
be used as a general purpose planet detection method, which we apply
to the \project{K2} dataset. \pipeline{QATS} relaxes the strictly periodic
constraint assumed in most planet search techniques, such as Box-fitting
Least Squares \citep[\pipeline{BLS};][]{kov02}. Instead, consecutive
transits must only fall within a tunable time window --- the wider the
window, the larger the TTVs that can be captured.  However, wider time
windows come at the cost of raising the noise floor, making the
detection of the smallest planets more difficult. We compromise by
searching over three different windows, from perfectly periodic to a
very large TTV signal.

The original \pipeline{QATS} technique worked directly with the
light-curve fluxes, but we instead calculate the log-likelihood ratio
($\frac{1}{2}\Delta \chi^2$) between a transit of a fixed depth and
duration and a simple polynomial continuum centered at every cadence
in the light curve; in other words, we ask at every cadence whether a
fixed-shape transit centered there is a better fit to the data than a
simple polynomial (\S\ref{likelis}). These likelihood comparisons at
every cadence become the \pipeline{QATS} inputs. \pipeline{QATS} then
searches for (quasi-)periods and phases that maximize the likelihood
improvements of the fixed-shape transit over the polynomial
continuum. We repeat this search for a grid of possible transit depths
and durations. By searching over a range of transit depths, we force
the algorithm to only look for features that have the same depth for
every transit event.

For each depth, duration, period, and \pipeline{QATS} transit window
tuple, we input to \pipeline{QATS} the log-likelihood ratio versus
time to find the maximum likelihood times satisfying these constraints
(\S\ref{qats}).  Over the entire range of duration, depth, period, and
transit window combinations the largest log-likelihood ratio solution
is selected for vetting.

\pipeline{QATS} will return a ``maximum likelihood candidate'' for every
single star regardless of its actual probability of being a planet, so
we perform some automated cuts to select for high-significance
transit-shaped events (\S\ref{autovet}).  Systems that pass the
automated cuts are then manually vetted (\S\ref{vetting}) for
astrophysical significance (i.e., planet transits or EBs). All stars with a candidate planet or EB
passing both the automated and manual vetting are passed back through
the entire process after masking out the known object to search for
further signals; we repeat until nothing else passes both vetting
stages and we have a final list of vetted objects of interest
(\S\ref{multi}).

This final list of astrophysical objects is modeled through a Markov
chain Monte Carlo (MCMC) analysis to determine the systems' parameters
(\S\ref{mcmc}). We also run an MCMC for the odd and even events
separately to help identify EBs (\S\ref{oddeven}) and
to check for period and ephemeris matches between candidates
indicating likely false positives (\S\ref{ephems}). All systems that
are not flagged by these cuts as false positives or EBs
compose our final list of planet candidates
(\S\ref{sec:general_properties}).

\subsection{Data Selection and Light-curve Detrending}
\label{everest}

We used the \texttt{EVEREST 1.0} detrended light curves for \project{K2}
campaigns 0--8 \citep{lug16} as the starting point for our planet
search. These were derived from MAST Data Releases
1-11\footnote{\url{https://keplerscience.arc.nasa.gov/k2-pipeline-release-notes.html}}
of the calibrated pixel-level light curves and a version of the
\texttt{EVEREST} code that is permanently archived on Zenodo
\citep{everest}.  We selected all long-cadence targets with object
type \texttt{star} or \texttt{null}, yielding a sample of \numstars{}
stars in C0--8 (we do not search any short-cadence light curves). The
\pipeline{EVEREST} light curves were generated using aperture number
15 of the \texttt{K2SFF} pipeline \citep{van14,van16b}.

The details of the detrending process are described in \citet{lug16}.
Briefly, the \pipeline{PLD} technique in \pipeline{EVEREST} uses the
normalized flux in each pixel in the aperture as inputs, similar to
how a centroid calculation works. The normalization removes the
astrophysical signal within the aperture (assuming that the flux is
due in total to the target star), leaving only instrumental signals in
the pixel time series. \pipeline{EVEREST} uses normalized pixel fluxes
and products of powers of normalized pixel fluxes as inputs to a
linear model that is subtracted from the data to yield a
systematics-free light curve. Cross-validation is performed to
optimize the number of regressors and prevent overfitting, and a
GP is employed to capture correlated stellar
variability. The \pipeline{EVEREST} light curves recover
\project{Kepler}-like precision for stars brighter than $Kp \sim 13$.

\subsection{Further Light-curve Preparation}
\label{prep}

Prior to searching the publicly available \pipeline{EVEREST} light
curves for planets, we cleaned up any regions that could interfere
with detecting transits. The two main obstacles in the \project{K2} data are (1)
bad data points occurring at the same cadence across multiple light
curves (primarily due to poorly modeled thruster fire events) and (2)
outliers unique to one particular light curve (e.g., stellar flares or
cosmic rays).

\begin{figure}[tbp]
\includegraphics[width=\columnwidth]{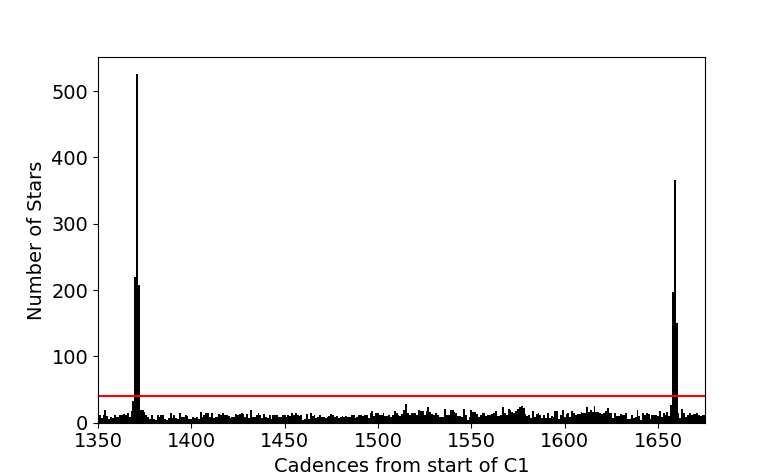}
\caption{Zoom-in on a segment of Campaign 1 (about 1 week of 
long-cadence data).  The histogram presents the number of stars whose top
  signal has a period longer than 9 days and a ``transit'' time at a
  particular cadence. If there were no correlated systematics between
  stars, each cadence should have approximately the same number of
  events. The spikes represent uncorrected systematics, and we remove
  from our search any cadences with bins above the red line (40
  different stars; see Table \ref{badcadstab}), as well as the
  cadences before and after the problematic ones. \label{badcads}}
\end{figure}

A straightforward, albeit time-consuming, way to find bad cadences
impacting many light curves is to run an initial planet search and
make a histogram of the cadences identified as the centers of
transits. As transit times should not correlate between stars, the
distribution of transit times across all stars should be uniform; an
obvious preference for ``transits'' occurring at a particular cadence
across multiple stars is highly indicative of some systematic
effect. In our first search we found such pileups, as shown for a
subset of Campaign 1 in Figure \ref{badcads}.

For each campaign, we removed the most troublesome cadences, as well
as the ones just before and after, in every light curve; we also
manually inspected a handful of the events to see whether a longer-duration
segment needed to be removed to eliminate the problem. To identify the
problem cadences, we used the maximum likelihood signal (MLS) returned by
\pipeline{QATS} for every star, excluding those with periods shorter
than 9 days.  Nine days was chosen as a cutoff because shorter-period
signals are likely less affected by single bad cadences and only add
noise to this measure. A histogram of the number of stars for which a
potential transit event is identified at every cadence is shown in
Figure \ref{badcads}. The threshold for the number of affected stars
to require a cadence's removal (listed in Table \ref{badcadstab})
changes with each campaign and with the total number of stars with
longer-period signals, but in general cadences that affected more
than about 0.25\% of such eligible stars were removed. Altogether, we
removed around 3\% of cadences per campaign at this stage; the full
list of excluded cadences in each campaign can be found in Table
\ref{table:badcads}.

\begin{table}[tbp]
\footnotesize
\caption{Identification of Bad Cadences \label{badcadstab}} 
\begin{center}
\begin{tabular}{c c c c c}
\hline
Camp. & Min Number  & Removal \%  & Number  & Percent of \\
& Affected for & of Possible & of Cads & Campaign \\
 & Removal &  Stars & Removed & Removed \\
\hline
0 & 20 & 0.43\% & 34 & 2.10\% \\
1 & 40 & 0.29\% & 115 & 2.93\% \\
2 & 30 & 0.26\% & 108 & 2.84\% \\
3 & 30 & 0.27\% & 123 & 3.63\% \\
4 & 30 & 0.26\% & 86 & 2.48\% \\
5 & 50 & 0.28\% & 116 & 3.17\% \\
6 & 50 & 0.22\% & 221 & 5.72\% \\
7 & 30 & 0.27\% & 89 & 2.24\% \\
8 & 50 & 0.31\% & 100 & 2.60\% \\
\hline
\end{tabular}
\end{center}

{\bf{Note.}} For each campaign, we list
  the minimum number of stars showing a ``transit'' at a cadence
  required for us to remove that cadence (and what percentage of
  possible stars that limit is). We also list how many total cadences
  are removed per campaign. See Table \ref{table:badcads} for the
  complete list of removed cadences.
\end{table}

\begin{table*}[tbp]
\scriptsize
\caption{\project{Kepler} Cadence Numbers Removed from All Light
 Curves before Searching for Planets.
\label{table:badcads}}
\centering
\begin{tabular}{| c | c | c | c | c |}
\hline
C0 & C1 & C2 & C3 & C4 \\
\hline
89616--89618&91495--91499&95552--95558&99621--99624&103773--103779\\
89630--89632&91809--91811&95971--95973&99636--99670&103783--103790\\
89659--89661&91940--91942&96021--96023&99762--99765&103804--103806\\
89665--89669&92000--92004&96332--96334&99858--99862&103808--103814\\
89725--89728&92071--92075&96378--96471&100050--100053&104004--104006\\
89794--89797&92083--92086&96546--96550&100243--100245&104016--104018\\
90288--90290&92336--92338&97208--97212&100445--100449&104052--104054\\
90365--90393&92371--92375&97353--97355&100459--100461&104088--104090\\
&92504--92507&97625--97627&100902--100904&104316--104318\\
&92528--92531&98265--98267&101105--101108&104965--104967\\
&92803--92807&98313--98315&101131--101134&105071--105074\\
&93091--93095&98373--98376&101411--101413&105179--105181\\
&93464--93466&98960--98964&101489--101494&105359--105363\\
&93798--93801&&101947--101949&105443--105445\\
&94075--94079&&102066--102069&106032--106034\\
&94460--94462&&102078--102081&106127--106131\\
&94471--94473&&102343--102345&106617--106621\\
&94490--94492&&102404--102407&106635--106637\\
&94640--94642&&102476--102478&106702--106707\\
&94762--94765&&102559--102561&106800--106802\\
&94771--94775&&102769--102771&\\
&94844--94847&&102857--102860&\\
&94868--94871&&102864--102866&\\
&95036--95038&&102875--102877&\\
&95072--95074&&&\\
&95120--95123&&&\\
&95144--95148&&&\\
&95155--95159&&&\\
\hline
\end{tabular}

\vspace{0.7em}

\begin{tabular}{| c | c | c | c |}
\hline
C5 & C6 & C7 & C8 \\
\hline
107579--107590&111362--111554&115890--115893&119938--119947\\
107592--107629&112003--112007&116753--116755&119950--119953\\
107823--107825&112126--112129&117401--117404&119956--119958\\
107918--107922&112161--112165&117415--117417&119985--119989\\
108124--108126&112197--112200&117437--117440&120046--120048\\
108182--108185&112303--112306&117737--117739&120149--120160\\
108291--108294&112473--112477&117978--117980&120274--120276\\
108687--108689&112508--112512&117984--117986&120671--120674\\
109071--109073&112653--112657&118000--118003&120763--120771\\
109551--109553&112702--112705&118017--118020&120945--120948\\
109815--109819&112736--112740&118025--118029&121030--121034\\
109924--109926&112779--112781&118073--118076&121426--121430\\
110020--110022&112795--112799&118083--118087&121470--121489\\
110045--110047&112813--112816&118096--118099&121632--121634\\
110211--110214&112880--112884&118113--118116&122278--122282\\
110234--110237&112941--112945&118169--118172&122374--122378\\
110415--110418&113180--113182&118288--118291&122566--122568\\
110932--110934&113266--113268&118306--118308&122586--122595\\
111041--111044&113686--113689&118361--118364&122673--122678\\
&113878--113882&118672--118675&122686--122688\\
&113925--113929&118864--118867&122758--122761\\
&113937--113940&119022--119024&123010--123014\\
&114225--114228&119238--119240&123526--123528\\
&114415--114421&&123622--123626\\
&114428--114432&&\\
&114434--114436&&\\
&114669--114672&&\\
&114789--114792&&\\
&115089--115095&&\\
\hline
\end{tabular}

\end{table*}

Next, we removed outliers within each individual light curve. 
Single-point outliers are especially common in \project{K2} light curves, frequently
cannot be removed by \pipeline{EVEREST} detrending, and need to be
mitigated.  To identify the outliers, we ran a median filter through
the light curve (kernel size of five cadences) and compared the flux at
any given cadence to its median filter value. The residuals between
the observed flux and the median filter were treated as a normal
distribution centered on the median whose width we fit with the median
absolute deviation (MAD) --- scaling appropriately with
\[ \sigma = \frac{1}{\Phi^{-1}\left(\frac{3}{4}\right)} \text{MAD} 
\approx 1.4826 \cdot \text{MAD} \] where $\Phi^{-1}$ is the quantile
function of the normal distribution.

Any cadence whose flux deviated more than 10 standard deviations
(positive or negative) from the median filter prediction was labeled
as an outlier candidate, as were any adjacent cadences that were more
than three standard deviations from the median filter value (this
captures, e.g., the exponential decay of flares). Any single cadence
or set of two consecutive cadences identified in this manner was
immediately removed as an outlier. However, short-duration transits
also get classified as large flux deviations from the median baseline
value. We made use of a transit's symmetry to retain any potential
transit signals through the following method.  In a series of
consecutive cadences all labeled as possible outliers as describe
above, we identified the central cadence(s): the median cadence (two
median cadences for the even case) or central 20\% of cadences in the
rare cases of 10 or more consecutive outliers. If the minimum flux in
that central group was within five times the light-curve noise level of
the overall minimum flux for the entire group, we retained the group
and did not remove the cadences as outliers. This worked to keep
transits with durations of at least three cadences; however, removing
all one- and two-point outliers will bias the search against deep,
short-duration (1 hr or less) transits, and we discuss this
limitation in \S\ref{usps}.

Quiet stars usually had fewer than 10 cadences removed as outliers at
this stage; active stars with flares had a higher fraction of cadences
removed. Stars where \pipeline{EVEREST} performed poorly (due to,
e.g., saturation or crowding from nearby bright stars) also had an
increased number of outliers removed in this process.

\subsection{Calculating Likelihoods}
\label{likelis}

With potential outliers removed, we began the transit search using
\pipeline{QATS}.  The original \pipeline{QATS} algorithm makes the
assumption that the input data have a zero-mean ``continuum'' and white
noise, which requires perfect instrumental and astrophysical
detrending at every cadence in the light curve simultaneously and in
practice is not achievable for all \project{K2} stars.  We instead only locally
detrend with a polynomial continuum and have modified the input to the
\pipeline{QATS} algorithm by replacing the observed flux at each
cadence in the light curve with the log-likelihood ratio
$\left(\frac{1}{2}\Delta \chi^2\right)$ between a local polynomial fit
to the data with and without an added transit (of fixed depth and
duration) centered there. Positive log likelihoods indicate that the
local light curve is better fit by adding the transit, centered on
that cadence, to the modeled polynomial continuum. In this section we
describe how we chose our grid of transit depths and durations and
calculated the likelihood ratio between a transit of that fixed scale
and a polynomial continuum. In \S\ref{qats} we describe how each of
those series of likelihoods was input to \pipeline{QATS} to search
for planets.

To calculate the likelihood ratio between a transit and polynomial
continuum, we first need to choose an appropriate polynomial order and
continuum width (nearby out-of-transit cadences on either side of a
putative transit at the cadence of interest). These values need to be
chosen independently for each light curve to account for the different
timescales of variability present in each star.

For a given light curve, we tested several possible combinations of
continuum widths, from 0.2 to 0.6 days with a spacing of 0.1 days, and
polynomial order, from 2 to 5. For each combination, we calculated the $
\frac{1}{2}\Delta \chi^2$ as described below for 500 randomly chosen
points in the light curve. We fit these 500 $\frac{1}{2}\Delta \chi^2$
values for their median $\mu_w$ and standard deviation $\sigma_w$
(calculated robustly using $\sigma_w = 1.4826 \cdot \text{MAD}$) and
chose as the optimal continuum width $w$ and polynomial order $M$ the
combination that minimized the coefficient of variation
$\frac{\sigma_w}{\mu_w}$.  This criterion is intended to give the most
uniform likelihood ratio throughout the light curve so that the
transit events are easier to identify.

In all of our $\chi^2$ calculations, the uncertainties on each data
point were taken to be uniform throughout the light curve and set to
our estimate of the white-noise level in the light curve. We
calculated the white-noise level by determining the MAD of the
residuals of a linear fit through 1000 random 4 hr segments of the
light curve.  The median value of these 1000 residual MADs was taken
to be the white-noise level $\sigma$, after scaling by the normal
factor of 1.4826.

\begin{figure}[tbp]
\includegraphics[width=\columnwidth]{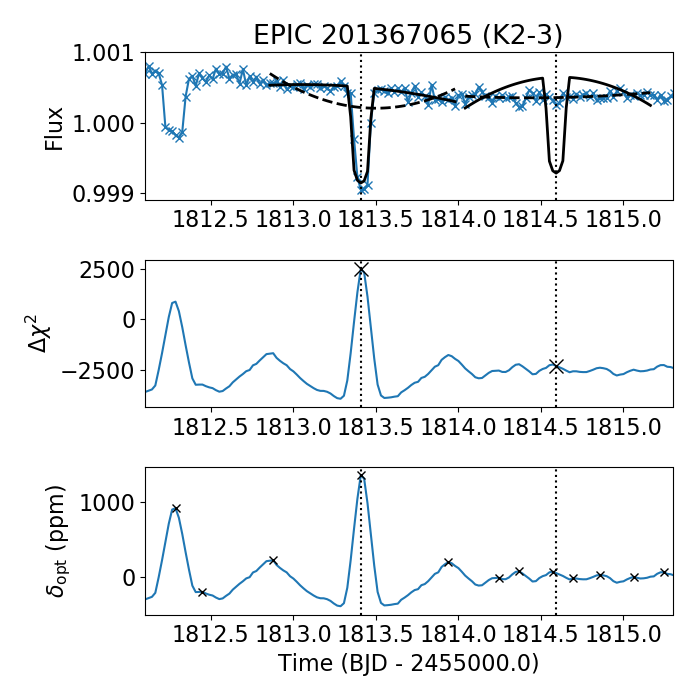}
\caption{Example $\Delta\chi^2$ calculation at a transit of the planet
  K2-3 b and an arbitrary continuum region. The top panel depicts a
  region of the light curve around a specific transit time and a time
  with no transits (dotted lines). The dashed line is the polynomial
  continuum fit to the region (i.e., transit depth $\delta = 0$), while
  the solid line is the polynomial continuum fit plus an assumed
  transit at the particular cadence (of fixed duration $T = 3.04$ hr
  and depth $\delta = 1365$ ppm). The difference in $\chi^2$ between
  these models is plotted as a cross in the middle panel, which
  also shows how this $\Delta\chi^2$ varies with the transit model's
  $t_0$ at every cadence.  The bottom panel depicts the depth at which
  the $\Delta\chi^2$ is maximized when the transit model (still at
  fixed duration $T=3.04$ hr) is centered there. Local maxima used
  for determining which depths to search for planets are marked with
  a cross. Note that one of the transits of K2-3 c is visible on the
  left. \label{dchis}}
\end{figure}

Having chosen the polynomial continuum parameters for a given star, we
can proceed to calculate the likelihood of a transit at every cadence
in the light curve: we model a potential transit as a polynomial
continuum plus a transit shape centered on that cadence ($t_0$). A
worked example of this process is demonstrated in Figure \ref{dchis}
and we recommend referring to it throughout this section.

The transit plus polynomial modeled flux $m$ at cadence $t_i$ (using
all cadences within the transit duration $T$ and continuum region $t_i
\in \Big[ t_0 - w - \frac{T}{2}, t_0 + w + \frac{T}{2} \Big]$) is
determined by the sum of the transit model and the polynomial
continuum
\begin{equation} \label{chi2model}
m(t_i) = \frac{\delta}{\delta_0} \Big(\mathcal{F}\left(t_i,t_0, T,
b,u_1,u_2,\delta_0\right)-1\Big) + \sum_{j=0}^M a_j (t_i-t_0)^j
\end{equation}
where the continuum polynomial has order $M$ with coefficients $(a_0,
..., a_M)$, and $\mathcal{F}$ is a Mandel--Agol transit model
integrated over the cadence duration with mid-transit time $t_0$,
duration $T$ (first to fourth contact), impact parameter $b$, linear
and quadratic limb-darkening parameters $(u_1,u_2)$, and maximum depth
of transit $\delta_0$.

In all cases we used a transit shape with fixed parameters
$(b,u_1,u_2,\delta_0) = (0.3,0.4,0.25,0.02)$.  The depth of a transit
can be well approximated by the following function of limb-darkening
parameters, impact parameter, and planet-to-star radius ratio
$\frac{R_p}{R_*}$
\begin{equation}
\delta_0 =
\left(\frac{R_p}{R_*}\right)^2\frac{1-u_1(1-\sqrt{1-b^2})-u_2(1-\sqrt{1-b^2})^2}{1-u_1/3-u_2/6}.
\end{equation}
Inverting this equation gives $\frac{R_p}{R_*} \approx 0.13$ for our
chosen transit parameters, appropriate for a Jupiter-sized planet
transiting the Sun or an Earth-sized planet transiting an M dwarf ---
roughly what one would expect to find in the 80-day-duration \project{K2}
campaigns.  This transit model was subsampled by a factor of 7
  and binned to the 30-minute-long cadence. To maintain linearity
(and therefore greatly reduce computation time), this fixed transit
shape is scaled by a multiplicative factor to the appropriate depth
$\delta$, rather than adjusting, e.g., the $\frac{R_p}{R_*}$, which
would slightly change the shape of the transit as the depth changes
and break the linearity with depth.  Our method is robust to the
precise shape of the transit model, however, and even a simple box
model would work nearly as well.

While the transit shape is not critical to the search and thus held at
one set of fixed values, we do need to search over a variety of
transit durations. Because the model $m(t_i)$ is a nonlinear function
of the transit duration, $T$, we hold the transit duration fixed and
instead repeat the search over a grid of 16 fixed transit durations
from 2 to 17 hr, with each duration 15\% wider than the previous
one.  With the transit time, shape, and duration fixed, the model is
linear in the remaining free parameters: the transit depth and
polynomial coefficients $(\delta, a_0, ..., a_M)$.

The solution to this linear equation produces the transit depth (and
polynomial continuum parameters) that best fits the data (minimizes
the $\chi^2$) in the local region near the $t_0$. In many cases this
optimal transit depth, $\delta_{\text{opt}}$, will be near 0,
indicating that the region is best fit without a transit at all since even
in systems with planets the transits occur in a small fraction of the
total cadences in the light curve.  Because of the nature of linear
equations, the $\chi^2$ at any other depth $\delta$ is a quadratic
function and can be computed later after solving the linear equation
just once for that $t_0, T$ combination.  We solved for the linear
coefficients $(\delta,a_0,..,a_M)$ at every duration in our grid and
with the transit model midpoint $t_0$ at every cadence for which there
was at least one valid cadence both in transit and on each side
of the continuum; we then saved the coefficients at each such grid
point to later calculate the $\chi^2$ at any depth of interest.

Finally, to calculate our desired log-likelihood ratio
$\left(\frac{1}{2}\Delta \chi^2\right)$ between a specific transit
model and pure polynomial continuum (equivalent to setting the transit
depth $\delta = 0$), we need to select a fixed depth for the transit
model. As with the transit durations, we use a grid of different
depths and repeat the search for each depth and duration combination
of transits. Unlike the durations, however, we adaptively choose our
grid of transit depths based on the optimal depths
$\delta_{\text{opt}}$ we calculated at every cadence in the light
curve.

\begin{figure}[tbp]
\includegraphics[width=\columnwidth]{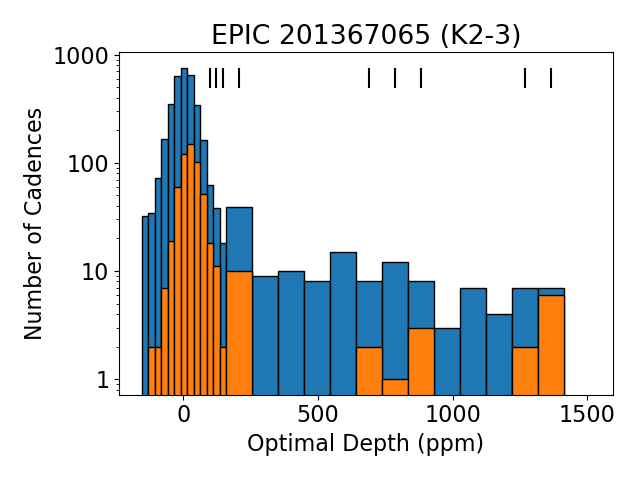}
\caption{Optimal depth distribution of all cadences (blue) and
  cadences at local maxima (orange; see bottom panel of Figure
  \ref{dchis}) for the search of K2-3 with fixed transit duration
  $T=3.04$ hr. Bin width is $0.5\sigma$ between
  $[-3\sigma,3\sigma]$ and 2$\sigma$ beyond that.  Tick marks indicate
  those depths chosen for the \pipeline{QATS} planet search, as
  described in the text. Note that the three planets in this system
  have depths of 1416, 822, and 782 ppm, falling nicely within our
  chosen search depths.  \label{optdepths}}
\end{figure}

We used the distribution of optimal depths throughout the light curve
to decide at which transit depths to search for planets: we should
only search for planets at depths where at least some cadences suggest that
there might be such a transit. As an example, the distribution of the
optimal depths at all cadences for one particular duration of the
Campaign 1 star EPIC 201367065 (the confirmed three-planet system K2-3) is
shown in Figure \ref{optdepths}. Because in-transit cadences are rare,
even for stars with planets, the majority of the light curve has
optimal transit depths near 0, with a roughly Gaussian spread
(different for each duration). We searched every light curve at every
duration for small transits at depths corresponding to 2, 2.5, and 3
standard deviations. Above 3$\sigma$, the optimal depths were binned
with a width of 2$\sigma$, and any bins containing a local maximum (a
cadence where the $\delta_{\text{opt}}$ is larger than at the cadences
immediately preceding and following; identified in the bottom panel of
Figure \ref{dchis}) were also searched.  This procedure allowed us to
adapt the grid of depths to the individual noise properties of each
star and to avoid spending time searching for transit depths that
would be undetectably small or transit depths larger than the data
suggest exist. On average, this method produced around eight depths to
search for each duration, or about 100--150 combinations of depth and
duration per star.

Having chosen the depths to search for planets at each transit
duration, we finally need to calculate at every cadence the
log-likelihood ratio between pure continuum ($\delta = 0$) and a transit
at that chosen depth. Because of the simplicity of the linear model,
these $\chi^2$ calculations are straightforward and we generated the
$\Delta\chi^2$ for every cadence, duration, and depth combination from
the previously saved coefficients.

\begin{figure}[tbp]
\includegraphics[width=\columnwidth]{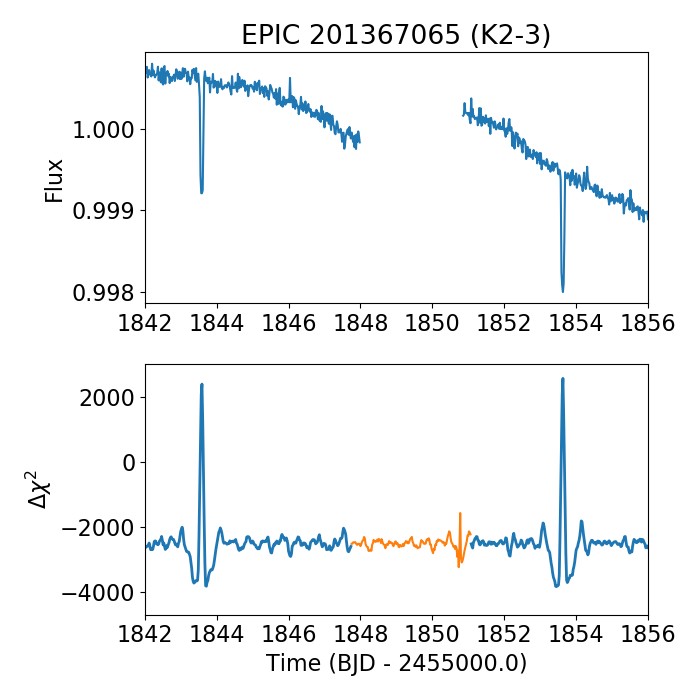}
\caption{Filling data gaps with simulated $\Delta\chi^2$ values using
  an autoregressive model. Top: flux of the K2-3 system near the data
  gap in the middle of C1. Transits of K2-3 b can be seen on either
  side.  Bottom: calculated $\Delta\chi^2$ at every cadence (blue)
  between the continuum model and a continuum model plus transit of
  fixed duration $T=3.04$ hr and depth $\delta = 1365$ ppm. The
  orange line filling in the data gap is the simulated $\Delta\chi^2$
  values.  Overall it does a reasonable job of reproducing the correlated
  structures seen in the real values. A slight artifact from the
  transit can be seen near the end of the gap, but a large spike is
  prevented owing to the clipping of inputs to the
  model. \label{datagap}}
\end{figure}

An example of this process is demonstrated in Figure \ref{dchis} with
K2-3 for a fixed duration and depth combination.  The top panel shows
the polynomial fit compared to a polynomial plus transit fit (of the
fixed depth and duration) for a $t_0$ centered on one of the transits
of the 10-day period K2-3 b, as well as for an arbitrary continuum
region.  The middle panel shows how the $\Delta\chi^2$ between the two
models changes depending on which cadence the fixed-scale transit is
centered.  Finally, the bottom panel depicts the $\delta_{\text{opt}}$
--- showing at every cadence what transit depth (still with the fixed
transit duration) maximizes the $\Delta\chi^2$. Cadences centered on
real transit events will have optimal depths comparable to the true
transit depth, while continuum regions will have optimal depths near
0, indicating that the region is best fit without a transit at all.

One final step is required since \pipeline{QATS} requires evenly
spaced data: we need to handle data gaps and locations in the light
curve where the $\Delta\chi^2$ were not computed.  Because of the
correlated structure of the log likelihoods, simply filling in gaps
with randomly chosen likelihoods biases the planet search. To do a
better job at capturing this correlation, we used an autoregressive
technique to interpolate from surrounding regions similar to the 
gap-filling technique of \project{Kepler}'s Transiting Planet Search
\citep[\pipeline{TPS};][]{jen10b}.

We first clipped from the $\Delta\chi^2$ series any points more than
$\pm5\sigma$ from the median value to avoid predicting anomalous
values (such as transits) in gaps without any data. On each side of a
data gap, we fit an autoregressive model using the
\pipeline{statsmodels} Python package's
\pipeline{statsmodels.tsa.ar\_model.AR} and a maximum lag length of 30
cadences to predict $\Delta\chi^2$ values within the data gap
\citep{sea10a}. The two models were then combined using inverse linear
weights: the model based on data before the gap had weights linearly
declining from 1 to 0, while the model from data after the gap
contributed weights increasing from 0 at the beginning of the gap to 1
at the end. Cadences within the gap were assigned simulated
$\Delta\chi^2$ values based on the appropriate weighted average of the
two autoregressive models. The results for the data gap in the K2-3
system are shown in Figure \ref{datagap}.

To summarize our procedure, we calculated the log-likelihood ratio
($\frac{1}{2}\Delta \chi^2$) between a local polynomial and a
polynomial plus a transit of fixed duration and depth at every cadence
in the light curve. This likelihood ratio is what was fed into the
\pipeline{QATS} algorithm instead of raw fluxes.  The likelihood ratio
calculation was repeated over a grid of transit durations (always the
same between 2 and 17 hr) and transit depths (varying values with each
duration depending on the $\delta_{\text{opt}}$ distribution). On
average, this method produced 100--150 different duration and depth
combinations per star, each of which was fed to \pipeline{QATS} to
search for a planet with a transit of that scale.

Our likelihood ratios are akin to the \project{Kepler}
\pipeline{TPS}'s single-event statistics, although they use
wavelet-based matched filters instead of polynomials to remove stellar
variability \citep{jen10b, sea13}. Our total likelihoods after running
the likelihood ratio time series through \pipeline {QATS} similarly
mirror \pipeline{TPS}'s multiple-event statistics.

There are two notable limitations in our chosen implementation. First,
by setting our minimum depth at two standard deviations above 0, we will
have reduced sensitivity to planets smaller than this level. This
mostly affects ultra-short-period planets, which could transit over 200
times in a single \project{K2} campaign; even with single-transit depths of one
standard deviation, combined together they would add up to a
detectable signal.

Second, in setting the cadence of our likelihood evaluations equal to
the \project{Kepler} cadence, we will have reduced sensitivity to
shorter-duration transits whose center might fall in between
cadences. The maximum $\frac{1}{2}\Delta \chi^2$ naturally occurs when
the center of our transit model aligns perfectly with the center of
the actual transit. The difference between this true maximum and the
measured value at our discrete cadences depends on the transit
duration, most severely affecting short-duration transits whose shape
changes significantly over the course of a cadence. Our model
evaluation cadence was chosen because the original \pipeline{QATS}
algorithm operated directly on fluxes and thus required its
evaluations to occur at the same cadence as the data. Because our new
technique can evaluate the $\frac{1}{2}\Delta \chi^2$ in between
cadences, we could revisit our model evaluation cadence in future
searches to increase sensitivity to the shortest-duration transits.

\subsection{\pipeline{QATS}}
\label{qats}

With a complete, evenly spaced measure of the log-likelihood of a
transit of a certain depth and duration at every cadence in the light
curve, we can run \pipeline{QATS}. As a brief summary of how it works,
given a chosen minimum and maximum allowable time span between
transits, $T_{min}$ and $T_{max}$, \pipeline{QATS} returns the maximum
total likelihood, as well as the set of cadences that maximize it, from the
input series of likelihoods. The parameters $T_{min}$ and $T_{max}$
are integer multiples of the \project{Kepler}/\project{K2} long cadence.  We
refer to the ``period'' as $P=(T_{min}+T_{max})/2$, and the number of
cadences between them, $W = T_{max} - T_{min}$, as the transit
window --- indicating the allowable range of TTVs. For a periodic
search we set $W = T_{max} - T_{min} = 1$; we cannot set $T_{max} =
T_{min}$ as this would require the period of the planet to be an exact
integer multiple of the number of cadences. The original version of
\pipeline{QATS} contains a transit duration parameter in units of the
number of cadences; we set this to unity, as the log-likelihood already
accounts for an assumed transit duration.

Because we do not know the period of any potential planets in advance,
we searched over possible periods ranging from a minimum of 0.3 days
($T_{min}$ = 14 cadences) up to the entire duration of each campaign.
Any vetted candidate with a period between 0.3 and 0.6 days was rerun with
a minimum period of 0.1 days (2.4 hr) to ensure that we found the
correct period and were not biased by the lower limit.  We also
searched for periodic planets ($T_{max} = T_{min} + 1$), as well as two
quasi-periodic allowances $\big( T_{max} = \text{ceil}\big[1.0033
  T_{min} \big]$ \text{and} $T_{max} = \text{ceil} \big[1.0066 T_{min}
  \big] \big)$.  The more we relax the periodic constraint by widening
$W$, the larger TTVs \pipeline{QATS} can find, but at the cost of
extra noise because \pipeline{QATS} is also more easily able to string
together random positive deviations. Under the widest 0.66\%
quasi-periodic window, the first case in which $T_{max} = T_{min} + 2$
occurs at $154 = 152 + 2$ cadences, or a period of 3.1 days.

\begin{figure}[tbp]
\includegraphics[width=\columnwidth]{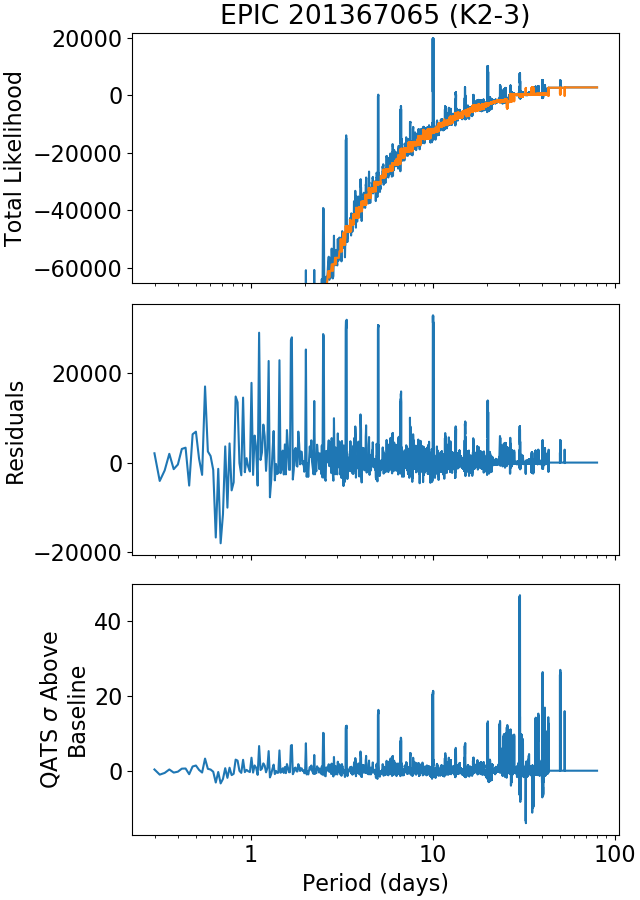}
\caption{Top: total \pipeline{QATS} likelihoods (blue) in the search
  of K2-3 with a transit of fixed duration $T=3.04$ hr and depth
  $\delta = 1365$ ppm. The highest likelihood period is at 10.03 days
  --- the period of the largest and shortest-period planet in the
  system. Other spikes of high likelihood occur at aliases of the true
  period (e.g., 1:N). The orange line is our estimate of the baseline
  level, ignoring the spikes at significant periods (see text for
  details). Note that the search continued to a minimum period of 0.3 days,
  but we trim for clarity of the longer periods.  Middle: 
  residuals after subtracting robust baseline estimate from the
  \pipeline{QATS} spectrum. The baseline residuals decrease in
  magnitude at longer periods.  Bottom: \pipeline{QATS} total
  likelihood above the baseline, in standard deviations. Periods
  beyond about 20 days have three or fewer events contributing to the
  total likelihood, and this measure loses its
  effectiveness. \label{qatsspec}}
\end{figure}

For every unique combination of $T_{min}$ and $T_{max}$ ($\sim$7000
total), \pipeline{QATS} returned the maximum total likelihood
improvement possible for a transiting planet at the specified depth
and duration.  We do not need to discuss the phase, because for a given
period \pipeline{QATS} implicitly searches over all phases.  For a
given depth and duration combination, we plot the \pipeline{QATS}
total likelihoods over all periods (and thus also phases) and transit
windows to create a \pipeline{QATS} ``spectrum,'' as in the top panel of
Figure \ref{qatsspec}. We have a similar spectrum for each of the
$\sim$150 depth and duration combinations per star. For most cases,
the maximum log-likelihood improvement is negative (a transiting
planet is less likely than a pure stellar continuum), but the
combinations where the maximum likelihood is positive could be a
planet. For every star we saved the depth, duration, period, and
transit window combination that produced the maximum likelihood
improvement (greater than 0) across all possible combinations as the
best result in the system.

\subsection{Automated Cuts}
\label{autovet}

 For clarity, we refer to every star's top \pipeline{QATS} result as a
 ``maximum likelihood signal'' (MLS) and withhold the ``candidate''
 label until it has been fully vetted.
    
The maximum likelihood and period for every star's MLS in C1 are shown
in the top panel of Figure \ref{likelihoods}. Each \project{K2} target with a
maximum likelihood above 0 is shown at its period, with the color
corresponding to the best-fit transit duration. The blue cloud (short
transit durations) at lower total likelihood improvements and
intermediate periods shows the baseline level for stars without
planets. It is almost always possible to stitch together a handful of
regions where the stellar continuum is randomly lower than average for
a few cadences, so a shallow-depth, short-duration transit can
marginally improve the fit. The sudden onset of signals at 3.1 days is
the result of the widened \pipeline{QATS} window, making it easier for
random noise to add up.

The larger likelihoods at long periods and longer transit durations
tend to be caused by ``transits'' being fit to local stellar
activity. In some cases a ``transit'' is fit to a local minimum owing to
stellar rotation, while in other cases the fits are skewed by flares
or cosmic rays that were missed in our outlier rejection step.

To narrow the results from the top signal in every star to only the
ones most likely to have a real transiting planet, we made several
automated cuts, which we describe next.

\subsubsection{\pipeline{QATS} Peaks}
\label{qatspeaks}

An indicator of a transiting planet candidate is that the
\pipeline{QATS} spectrum should show much-improved likelihoods at the
planet's period and its aliases compared to other periods. As seen in
the top panel of Figure \ref{qatsspec} for the 10-day planet in K2-3,
the total likelihood smoothly increases with increasing period, but
overlaid on that smooth background is a series of sharp likelihood
increases at the 10-day period and its aliases (1:N, 3:2, etc.).

\begin{figure}[tbp]
\includegraphics[width=\columnwidth]{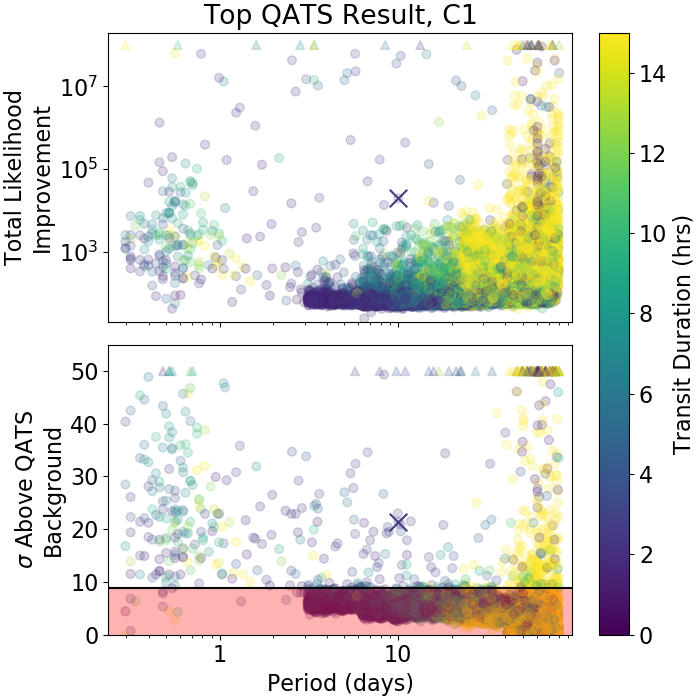}
\caption{Top: period and total likelihood improvement for every
  star's MLS in C1, color-coded by the transit duration. The sudden
  density of results at 3.1 days is due to the first allowed increase
  of the \pipeline{QATS} window. Any values above $10^8$ are clipped
  down to $10^8$ for clarity and marked with a triangle. The cross
  marks the first planet in the K2-3 system that we have used as an
  example.  Bottom: results from the top panel after normalizing
  the peak event to standard deviations above the \pipeline{QATS}
  spectrum background level. The line and shaded region at 8.86
  indicate our cutoff: any points below that are rejected from manual
  examination.  Any values above $50$ are clipped down to $50$ and
  marked with a triangle. The long-duration, high-significance cloud
  at periods longer than 40 days is due to the limited effectiveness
  of this measure for one- and two-transit events (see
  text). \label{likelihoods}}
\end{figure}

Since nearly every star will produce a total likelihood above zero at
some small depth, short transit duration, and long period, we need a
way to distinguish planets from noise.  Our first automated cut
focused on the fact that planets will have these spikes in their
\pipeline{QATS} spectrum at the planet's period and aliases thereof,
while other stars show only the smoother spectrum. To distinguish
between the two, we measured how high above the surrounding ``baseline''
level the maximum likelihood in a \pipeline{QATS} spectrum is.

\begin{figure}[tbp]
\includegraphics[width=\columnwidth]{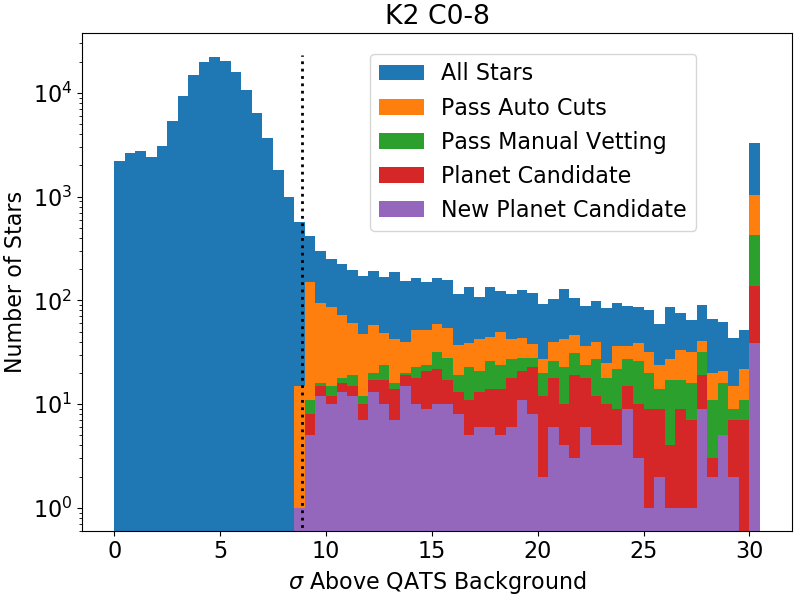}
\includegraphics[width=\columnwidth]{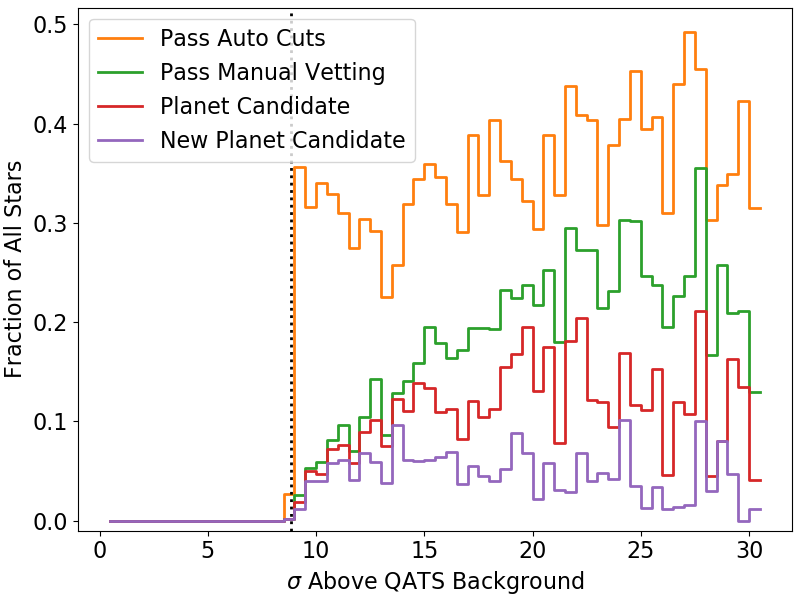}
\caption{Top: for all stars (C0--8, blue), the distribution of heights
  above the \pipeline{QATS} spectrum baseline value for the MLS, in
  standard deviations (the z-score; see Figure \ref{qatsspec}). The
  vertical dotted line indicates our cutoff in C1 (see Table
  \ref{zscorecuts} for other campaigns), below which results are
  removed from our search; this cutoff corresponds to 3.5 standard
  deviations away from the Gaussian core of the distribution. The
  orange bars show stars that pass all further automated cuts; green
  are those that pass the manual vetting as well; red are the planet
  candidates (after removing EBs); purple shows only our new planet
  candidates (those not found by any previous groups). All results
  above the maximum value (30) are grouped together in the last bin.
  Bottom: same as top panel, but showing the fraction of stars in the
  bin that pass each respective level of cuts.
\label{passvet}}
\end{figure}

Unfortunately, there is no theoretical means to predict the baseline
for a \pipeline{QATS} spectrum, so we must make an empirical
estimate. To do so, we robustly fit a second-order polynomial to a
local region of the \pipeline{QATS} spectrum, while including a linear
step parameter for changes in the \pipeline{QATS} window size, as well
as the number of transits used to create the maximum likelihood. The
robust estimate (an iterative fit where points 3 standard deviations
away from the fit are down-weighted) was crucial to avoid biasing the
baseline estimate owing to the spikes in likelihood from planets.

The robust baseline estimate is shown in the top panel of Figure
\ref{qatsspec} as the orange line, and the residuals after subtracting
this baseline are shown in the middle panel. Aside from the spikes from the
planet's period and its aliases, the residuals decrease with period
with a power law of $\alpha \approx -0.5$, which we accounted for by
measuring the MAD of the residuals as a function of period. Using
those MAD estimates, we normalized the \pipeline{QATS} spectrum
against the baseline level and calculated the z-score (standard
deviations above the baseline) for each total likelihood, as shown in
the bottom panel of Figure \ref{qatsspec}. Because real planets have
spikes rising high above the surrounding regions, we expect this
z-score measure to separate planets with high values from the
remainder of the stars without significant transit signals.

The distribution of z-scores for the MLS of every star in C1 is shown
in the bottom panel of Figure \ref{likelihoods}; indeed, most of the
stars show relatively low values, while the high z-scores often
contain astrophysical events --- either EBs or transiting planets.

The distribution of each star's MLS in standard deviations above the
\pipeline{QATS} spectrum baseline level (this time for all C0--8) is
shown in the top panel of Figure \ref{passvet}. Most results form a
Gaussian distribution at the core with z-scores of 3-8, with long
tails in both directions.  The upper tail contains the planets and
EBs, which exhibit the sharp spikes in the
\pipeline{QATS} spectrum and should be further followed up.

We pause to note, however, as seen in the bottom panels of Figures
\ref{qatsspec} and \ref{likelihoods}, that the z-score measure begins to
break down at the longest periods ($\gtrsim40$ days).  When the period
becomes large enough for only one or two cadences to contribute to the
total likelihood, many \pipeline{QATS} quasi-periodic period windows
(with a choice of, e.g., two cadences near the ends of a campaign or a
single cadence near the middle) will end up with the same total
likelihood --- choosing only the single cadence in the light curve
with the highest likelihood. This clumping of many periods with the
same total likelihood affects our \pipeline{QATS} spectrum baseline
estimate and the standard deviation estimate of the residuals, causing
many periods to have either a z-score near 0 or an artificially high
value (and is thus the cause for the excess of results at the low end
of the Gaussian tail in Figure \ref{passvet}).

More careful handling of one- and two-transit events will be addressed
in updated versions of our search pipeline; for now we accept these
z-score estimates and the increased likelihood of false positives and
false negatives at the longest periods in this stage of automated
vetting. The main effects are (a) an increased likelihood that one- and
two-transit planets are missed in this search (having z-scores of
$\sim0$) and (b) the occasionally inflated z-scores at long periods,
which also mean that an increased number of false positives pass our automated
threshold --- however, these are subsequently culled in the manual
vetting and ultimately have minimal impact on the validity of the
final candidate list.

\begin{figure*}[tbp]
\includegraphics[width=\textwidth]{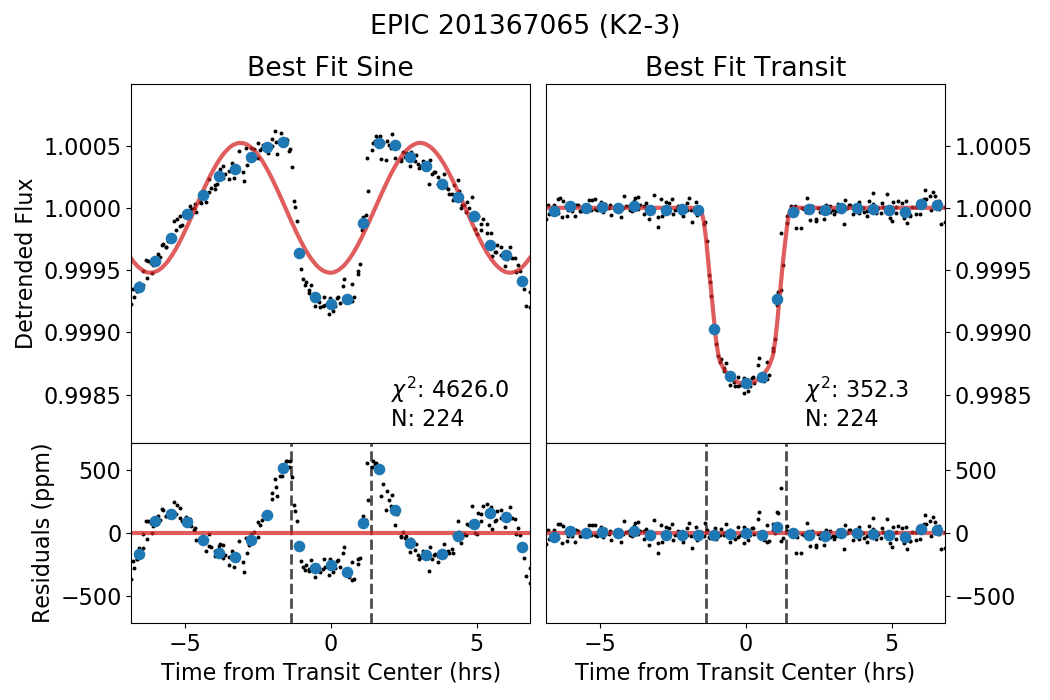}
\caption{Comparison between a local polynomial plus sine curve fit
  (left) and a local polynomial plus transit fit (right) to the real
  planet K2-3 b.  In both cases the local polynomial has been divided
  out of each transit before plotting the folded fit of all transits
  together. The individual cadences are shown in the background in
  black points, while the binned median is displayed by larger blue
  points. The best-fit model is the red line. The $\chi^2$ of the fit
  and number of cadences contributing are displayed in the lower right
  corner. Bottom: residuals of each fit on the same scale, emphasizing
  how much better the transit fit is in this case. The vertical dashed
  line shows the beginning and end of the best-fit
  transit. \label{realtran}}
\end{figure*}

To select a minimum z-score threshold for an MLS to pass this stage,
we looked at the distribution (top panel of Figure \ref{passvet}) for
each campaign individually. We fit for the median and standard
deviation of the z-score distribution (using the MAD) for each
campaign; however, to avoid biasing the fit with the long tails, we
only used those with periods less than 10 days and z-scores between
1.5 and 10.  We then selected as our minimum z-score threshold a value
of 3.5 standard deviations above this fit --- the exact value is
listed for each campaign in Table \ref{zscorecuts} but is near a
z-score of 9.

\begin{table}[tbp]
\caption{Minimum z-score Cutoff for Each Campaign
\label{zscorecuts}}

\begin{center}
\begin{tabular}{c c}
\hline
Campaign & Min z-score \\
\hline
0 & 8.71  \\
1 & 8.86  \\
2 & 9.61  \\
3 & 8.93  \\
4 & 8.93  \\
5 & 8.87  \\
6 & 8.99  \\
7 & 9.54  \\
8 & 8.99  \\
\hline
\end{tabular}
\end{center}

{\bf{Note.}} Any star with
  an MLS below this threshold (Figure \ref{passvet}) is
  removed from our search.
\end{table}

We chose 3.5 standard deviations to balance eliminating nearly all
stars from the core distribution indicative of no planets with not
eliminating potentially real planet candidates from the long tail too
early. As this is just the first of several rounds of automated and
manual cuts, we wanted to be lenient at this first stage. In Figure
\ref{passvet} we also show how many stars pass all our automated cuts
(described below), then those that also pass the manual vetting, then those
that become planet candidates (not EBs), and finally the new planet
candidates that have not been found by any other groups. The bottom
panel shows the same, except as a fraction of stars in the bin.

Above z-scores of $\sim20$, our results stay consistent: about 35\% of
all stars will end up passing all the automated cuts, and about half
of those will go on to pass the manual vetting as well. As the z-score
drops from 20 to our lower limit near 9, the rate of systems passing
the automated cuts remains consistent, but fewer systems pass our
manual vetting (\S\ref{vetting}), indicating more contamination by
false positives. We discuss the lowest signal-to-noise (S/N) planet
candidates further in \S\ref{previous}.

In total, we identified \npassz{} MLSs across all campaigns above our
z-score cutoffs (\fracpassz{}\% of all stars searched). However, the
majority of these are spurious owing to the \pipeline{QATS} algorithm
picking up on stellar variability (which is also quasi-periodic and
will produce a similar \pipeline{QATS} spectrum to periodic transits)
or other sources of false positives. To separate planets from
quasi-periodic stellar variability, we made a cut based on the
candidate transit morphology.

\begin{figure*}[tbp]
\includegraphics[width=\textwidth]{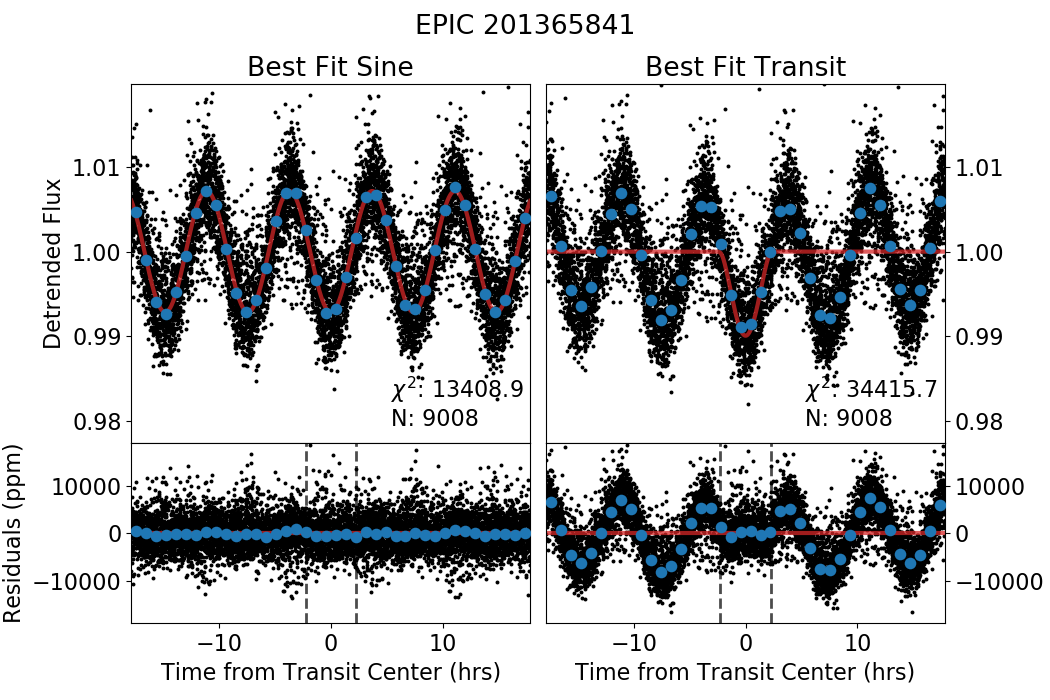}
\caption{Same as Figure \ref{realtran}, but for a star in which the
  sinusoidal fit was better than the transit model fit. This MLS was
  discarded as being caused by stellar variability. \label{vartran}}
\end{figure*}

\subsubsection{Morphology Cut: Sine Fit versus Transit Model}
\label{sinecut}

As seen in Figure \ref{likelihoods} the fraction of systems passing
the z-score cut is exceptionally high at periods of less than 2 days.
The majority of these systems are variable stars where a ``transit'' can
be well fit at the local minimum of stellar oscillations or
variability. Although stars display a wide diversity of variability,
in general stellar variability is roughly sinusoidal while transits
are more localized in time and phase. To automatically eliminate the
most obvious cases of stellar variability, we compared a folded
transit fit to a folded sinusoidal fit and eliminated systems where
the sinusoidal fit is better.

We selected each transit and a continuum region twice the transit
duration on either side (or 0.2 days, whichever is larger for
especially short transits). We fit a transit model plus local
polynomial to each transit and found the optimal global transit
parameters. We then repeated the procedure but fit a local polynomial
continuum plus a sine wave, optimizing for the amplitude, phase, and
period of the sine curve.  The sine curve fit was initialized with a
semi-amplitude of negative the transit depth and a phase such that the
sine curve's minimum is at the center of transit. We tested three
different initial sine curve periods to start the optimization (0.5,
1, and 2 times the transit duration) and used the best result. This
process is similar to the Sine Wave Event Evaluation Test performed by
the \project{Kepler} Robovetter \citep{tho18}.

A comparison between the transit and sine fit for a real transiting
planet (K2-3 b) is shown in Figure \ref{realtran}, while a false
positive due to stellar variability is demonstrated in Figure
\ref{vartran}.  Any MLS in which the sine fit had a smaller $\chi^2$,
and thus higher likelihood, than the transit fit was rejected as being
caused by stellar variability. This cut eliminated \nelimsin{} of the
\npassz{} systems passing the z-score test (\fracelimsin{}\%), leaving
\npasssin{} MLSs to continue to the next stage.

\subsubsection{Duration Limits}
\label{durcut}

After the z-score and sine-versus-transit fit comparisons, the final
automated cut we made is on the transit duration. The motivation
behind this cut was a subset of obvious false positives that were
difficult to differentiate via our other cuts but had poor fits and
exceptionally long durations.

We chose an empirical cut to eliminate events of long durations by
comparing the period--duration relation of the \project{Kepler} Objects of Interest 
\citep[KOIs;][]{tho18} and ensuring that the cut was well above them.  Our chosen
relation for the maximum allowed duration $T_{max}$ to pass the
automated cut was (in hours) $T_{max} = 0.017 \cdot P + 7$.  Thus,
anything under 7 hr will pass no matter its period, while at a 
10-day period, the MLS must have a duration of less than 11 hr.

This ad hoc decision will be revisited in future searches, but in this
case the cut removed \nelimdur{} stars that would otherwise have
passed on to manual vetting. The most likely real astrophysical
victims of this cut are short-period, near-contact binaries rather
than a large population of genuine planet candidates.

\begin{figure}[tbp]
\includegraphics[width=\columnwidth]{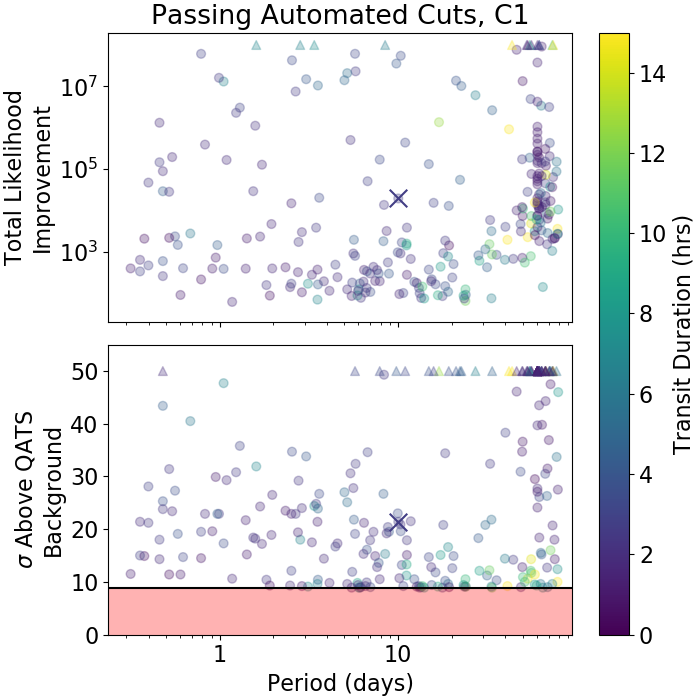}
\caption{Same as Figure \ref{likelihoods} but only for those systems
  that passed the automated cuts and were passed onto the manual
  vetting stage. The cross marks the first planet in the K2-3
  system that we have used as an example. One remaining artifact is the
  vertical line of MLSs around 60 days. This is a result of a bad
  cadence affecting a small number of stars that was not identified
  and left uncorrected by our outlier removal in \S\ref{prep}.  These
  systems were all removed in the manual vetting stage, however (see
  Figure \ref{allcuts}). \label{autocuts}}
\end{figure}

\begin{figure}[tbp]
\includegraphics[width=\columnwidth]{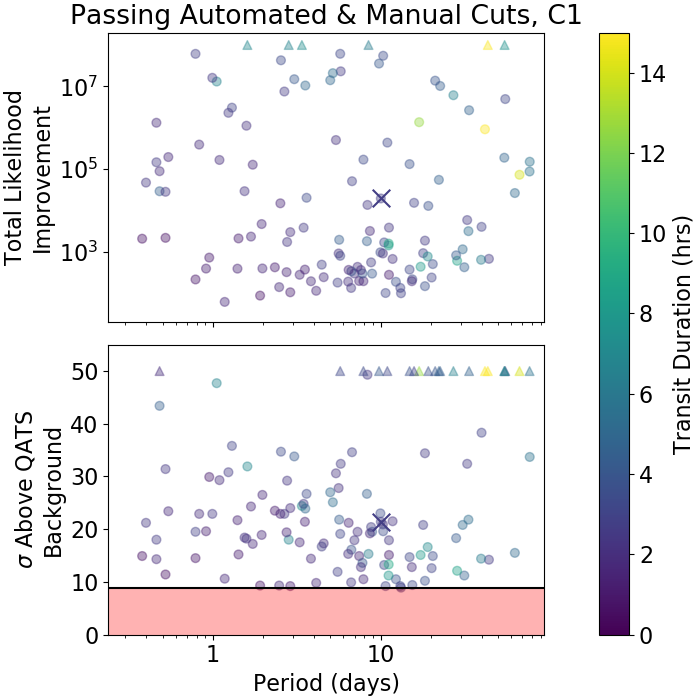}
\caption{Same as Figures \ref{likelihoods} and \ref{autocuts} but only
  for those systems that passed both the automated and manual vetting
  stages.  The cross marks the first planet in the K2-3 system that we
  have used as an example. \label{allcuts}}
\end{figure}

After all three automatic cuts (z-score, sine-versus-transit fit
comparisons, and transit duration), we were left with \npassdur{}
MLSs.  These were then subjected to diagnostic plots and manual
inspection. The distributions of these MLSs after the automated cuts
and after both the automated and
manual vetting are shown in Figures \ref{autocuts} and \ref{allcuts}, respectively.

\subsection{Manual Vetting}
\label{vetting}

To date, only three transit searches have been fully automated: (1)
the \cite{for16} search of quiet, bright \project{Kepler} stars for
single-transit events, and (2) the final two official \project{Kepler}
planet lists that used a Robovetter to automatically test for
completeness and reliability \citep{cou16, tho18}. Those searches were
only able to accomplish full automation by building on years of
efforts to understand and quantify the instrumental systematics of
\project{Kepler} and using the results of manual vetting and sorting
of candidates by previous groups to help train and verify the
automated pipelines.

\project{K2}, while using the same telescope as \project{Kepler}, has
dramatically different and more extreme instrumental artifacts owing to
the thruster fires and pointing drift. Further, each \project{K2} reduction
pipeline introduces its own biases. Thus, as with nearly every other
planet search to date, we visually inspected all \npassdur{} remaining
MLSs that passed our automated cuts and removed systems manually
judged to be unlikely to be due to a transiting planet or EB.
This process is inevitably subjective, but in this
section we describe our manual vetting tools and diagnostic plots used
to discriminate false positives.

The \pipeline{QATS} algorithm returns an array of transit time
estimates for an MLS, but only to the nearest cadence; it also returns
the transit depth and duration grid points that generated that MLS.  We first refined the transit times and
solved for the best transit parameters, including refining the impact
and limb-darkening parameters, which were fixed to the ``default
transit shape'' in the \pipeline{QATS} search.

We used an iterative procedure to find the best times and transit
parameter values. We started with the input \pipeline{QATS} transit
shape and let each transit time float individually to allow for
subcadence precision. We made no assumptions about periodicity and
thus can account for TTVs. We then fixed those transit times and
optimized the five transit parameters ($R_p/R_*, T, b, u_1, u_2$). We
iteratively continued this approach until we reached ``convergence,''
which we defined as neither the depth nor duration changing by more
than 0.1\% between iterations. We found this to be sufficiently
accurate for visual inspection and use in further diagnostics;
however, we note that we use a more rigorous (but much more
computationally intensive) MCMC approach for all reported parameter
measurements (see \S\ref{mcmc}). In most cases we achieved convergence
within a few iterations, although we capped it at 40 to prevent
infinite loops (most common in false positives not actually well fit
by a transit shape or planets with very shallow transits whose
individual times cannot be well determined).

\begin{figure}[tbp]
\includegraphics[width=\columnwidth]{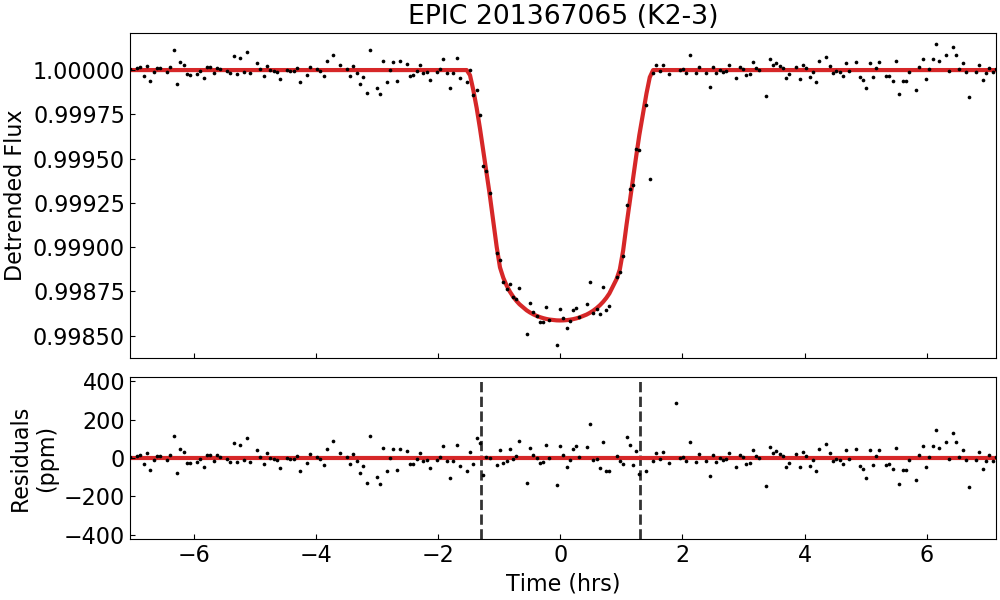}
\caption{Folded transit of the first planet in K2-3 used for our
  manual vetting. Black points are the individual fluxes, and the red
  line is the best-fit model. The bottom panel shows the residuals from
  this best-fit model. Each transit is folded and aligned based on its
  best-fit individual transit time, allowing for correct folding of
  all transits even with TTVs. \label{k2-3fold}}
\end{figure}

With better transit parameters and transit time estimates, we created
diagnostic plots for vetting planet candidates. The first and most
important diagnostic figure is the folded transit plot, demonstrated
in Figure \ref{k2-3fold}. This figure allows us to quickly evaluate
how well all the transits combine to fit a transit shape and reject
false positives that are visually poor fits. The folded transit figure
is especially critical for the low-S/N candidates where
individual transits are not visible by eye, and thus many of the other
diagnostic plots that rely on individual transits are not helpful.
  
\begin{figure}[tbp]
\includegraphics[width=\columnwidth]{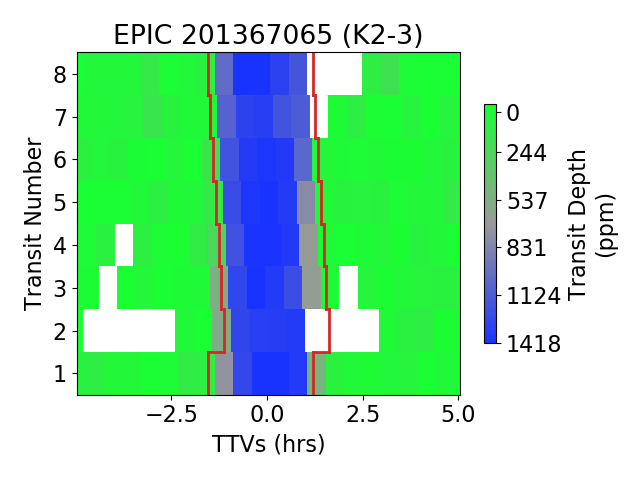}
\caption{ ``River plot'' diagnostic for the first planet in the K2-3
  system. Each detrended transit is plotted along a row with time 0 at
  the mean period. Green points are cadences near the continuum level,
  while blue points are those with fluxes near the best-fit transit depth.
  White cadences are missing or outlier cadences that were removed
  before the search. The red line shows ingress and egress for a
  transit centered on the cadence \pipeline{QATS}
  identified. \label{river}}
\end{figure}

Next, we generated a ``river plot'' for each MLS (Figure
\ref{river}). We fit our measured transit times with a periodic
ephemeris $t_N = t_0 + N P$. We then plot each detrended transit (the
best-fit polynomial continuum divided out) as a row in the figure,
with time 0 the best-fit periodic ephemeris.  Perfectly periodic
planets will show a vertical river of blue transits centered at 0
offset in the river plot. Planets with TTVs cannot be fit by a linear
ephemeris, so some transits will arrive early or late, creating a more
meandering river. Regardless, real planets will have correlated
transit timing, with subsequent transits not occurring too much
earlier or later than the one immediately previous. In contrast,
essentially random individual transit timing relative to the periodic
ephemeris indicates that \pipeline{QATS} picked up on noise and the MLS is
unlikely to be real; we rejected such systems even if the folded
transits looked promising.

\begin{figure}[tbp]
\includegraphics[width=\columnwidth]{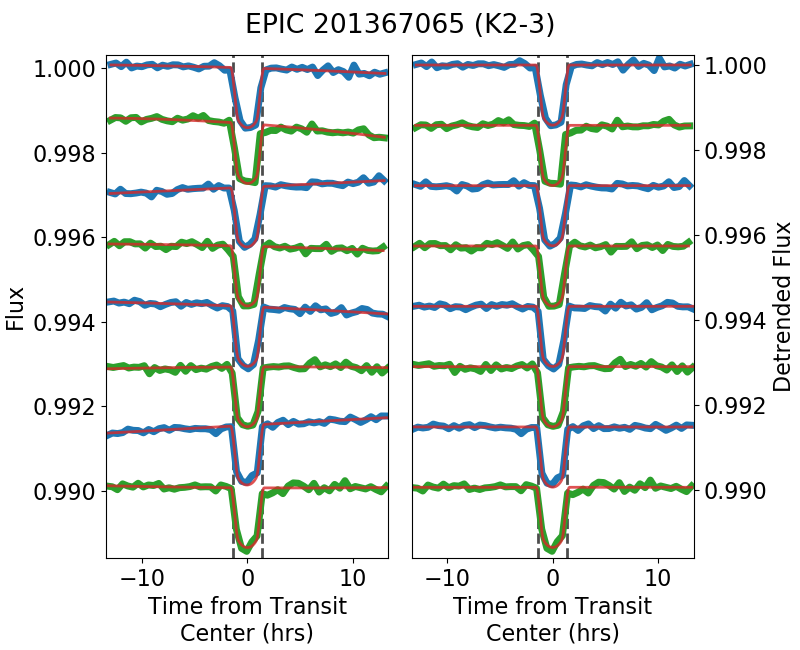}
\caption{Transit stack for the first planet in K2-3. The left panel is
  the raw fluxes with the transit plus polynomial model (red line),
  and the right panel is only the transit model (red line; polynomial
  stellar continuum model divided out). Odd transits are in blue, while
  even transits are in green.  The dashed lines behind depict ingress
  and egress of the transit model. Each transit is vertically offset
  for clarity. \label{stack}}
\end{figure}

Our next diagnostic plot is a transit stack of all transits (or the
first 20 for extremely short period planets) as in Figure
\ref{stack}. The left column shows the raw fluxes of each transit
(vertically offset for clarity), while the right column shows the
detrended fluxes (polynomial continuum removed). This figure allows us
to individually analyze every transit, and the alternating colors
allow us to compare the even and odd transits to the overall best-fit
model.  From this information we can visually determine whether all the
transits appear consistent or whether perhaps one or two false-positive
events are dominating the detection; we can also check for likely
binary-star false positives if the even and odd transits have
different depths or durations. This plot, in conjunction with the
folded transit figure, allows us to discard false positives where the
individual ``transits'' have wildly different and non-transit-shaped
events, often due to outliers or systematic or stellar variability.

\begin{figure}[tbp]
\includegraphics[width=\columnwidth]{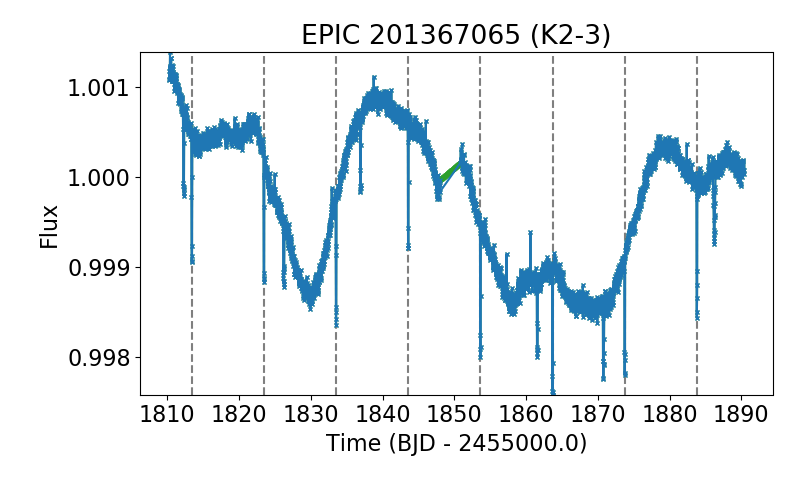}
\caption{Full \pipeline{EVEREST} light curve of the K2-3 system,
  with the first planet's transits marked by the dashed
  lines. \label{lightcurve}}
\end{figure}

We also plot a snapshot of the entire light curve (Figure
\ref{lightcurve}) with the times of transit identified by vertical
dashed lines. This allows us to easily identify broader trends in the
light curve and to quickly see whether the identified ``transits'' happen to
occur in a bad region of data or whether the entire light curve may have
problems that put the validity of the MLS in doubt (e.g.,  due to
\pipeline{EVEREST} having problems with blended sources and saturated
stars).

\begin{figure}[tbp]
\includegraphics[width=\columnwidth]{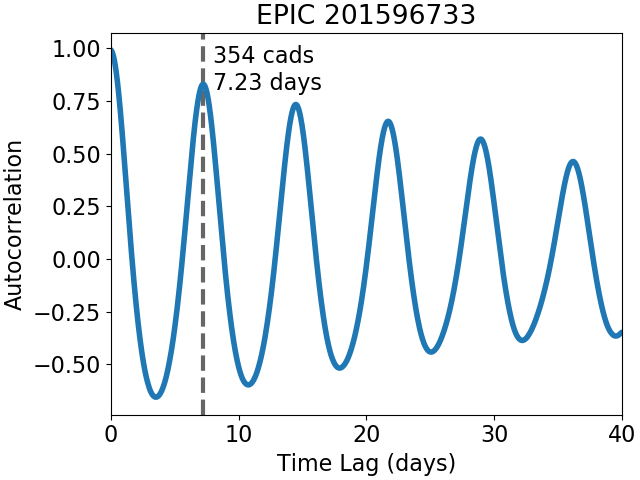}
\caption{Autocorrelation function of the light curve of the planet
  candidate host EPIC 201596733. Our identified local maximum
  indicating the star's rotation period is marked by the dashed line
  and annotated in both days and \project{Kepler} cadences. (Note that this
  is not the K2-3 system, as that one does not have an identifiable
  rotation period.) \label{autocorrelation}}
\end{figure}

Finally, we calculated a simple autocorrelation of the entire light
curve and identified the period of the first local maximum as in
Figure \ref{autocorrelation}. The autocorrelation function helps to
identify likely periods of stellar variability and provokes extra
skepticism if an MLS is at the same period as the autocorrelation
period.

Taking all of these diagnostic plots into consideration along with the
\pipeline{QATS} spectrum, we vetted each MLS that passed the automated
cuts. In this manual vetting stage, only one decision was made: is the
MLS astrophysically real, or is it a false positive. In other words,
planet candidates and EBs are grouped together and
separated from stellar variability, systematic errors, and other false
alarms. The EB versus planet candidate decision is made
later through automated cuts (\S\ref{ebsort}).

While no quantitative limitations can be placed on what passes and
what does not (or else manual vetting would not be necessary), some
examples of systems that did not pass the manual cut are as follows.
On the short-period end, the dominant false-positive cases were a
result of stellar rotation and/or oscillations that are asymmetric or
otherwise able to have the transit model fit marginally better than a
simple sine curve; these often have the same autocorrelation period as
the MLS event period.

At long periods, the biggest problem was outliers failing to be
completely removed in our light-curve cleaning. Many of these appear
to be systems where \pipeline{EVEREST} was unable to properly model a
large number of cadences. A quick examination of the full light-curve
diagnostic plot showed a large number of outliers throughout the light
curve, not just at the ``periodic'' events picked up by \pipeline{QATS},
and the individual transits are only one or two cadences and have
varied depths. These false alarms are often due to crowding or
saturation, which version 1 of \pipeline{EVEREST} often handles
poorly; version 2 of \pipeline{EVEREST} does better with these stars,
and a future planet search of the \pipeline{EVEREST 2.0} light curves
may turn up additional planet candidates.

Most of these long-period (one to three transits) and short-duration
(one to three cadences per transit) MLS decisions were easy to reject,
but some were less clearly false positives. Still, we erred on the
side of caution and rejected them because of the preponderance of
outliers in \project{K2} light curves. In the end, none of the MLSs having both
durations shorter than 2 hr and periods longer than 30 days passed
through our manual vetting (31 of \project{Kepler}'s 4500 candidates
meet those criteria), although this was not an intentional rule. Those
longer-period systems that did pass were deep enough and long enough
to show distinct shaping due to limb darkening, as opposed to a 
three-cadence dip with no nuanced shape.

While most vetting decisions were relatively easy, the most common
borderline cases were low-S/N systems: MLSs with a shallow
transit depth near the edge of our sensitivity.  These are the most
subjective, and we again erred on the side of caution, as evidenced by
the declining fraction of systems passing our manual vetting in Figure
\ref{passvet}. If anything seemed dubious, the system was rejected.
The most common causes for rejection were stellar variability at a
period near that of the MLS, small flares near individual events that
skewed the fits and likely caused the spikes in $\Delta\chi^2$ that
\pipeline{QATS} picked up on, and transits showing significant and
uncorrelated TTVs.

We only accepted systems with TTVs if the individual transits were
easily identifiable. Over long baselines with many transits, small
transits below the noise with TTVs are exactly what \pipeline{QATS} is
designed for; however, over the short baseline of \project{K2} (which itself
does not allow enough time for most real TTVs to develop), and with
only a dozen or fewer transits, \pipeline{QATS}'s flexibility more
often allows it to string together noise than find actual planets with
TTVs. In fact, that is mostly what the MLS looks like in the thousands
of systems that did not pass our \pipeline{QATS} z-score cut.

Manually reviewing \npassdur{} MLSs is a time-intensive process (24
hr of work if given just 30 s per star), and E.K was the only
one to vet every single one. Of these, he passed \nsystemme{} as being
astrophysically real (planets or EBs). E.A. reviewed those
\nsystemme{} and agreed with the assessment for all but \ericbad{} ---
a sign of the conservative approach taken for the vetting. Those
\ericbad{} systems were removed, leaving \nsystems{} stars with at
least one candidate.

\subsection{Multiplanet Searching}
\label{multi}

For all systems that passed the first stage of manual vetting
(\nsystems{} of the \npassdur{}), we iteratively searched the light
curve again for further planets until the newest MLS no longer passed
the automatic and manual vetting stages. The only change we made to
the process was lowering the z-score threshold for subsequent events
to 8; this change was made because of the significantly smaller number
of systems required to analyze and a desire to pass as many
potentially interesting MLSs onto the manual vetting process as
feasible.

To avoid rediscovering candidates, we masked all of them out before
searching the light curve again. This was accomplished by fitting the
continuum region around the transit as above, except this time
ignoring all the cadences within the duration of the transit. We
filled those masked in-transit fluxes with the polynomial continuum
interpolation and added white noise to match the rest of the light
curve.

We note that because \pipeline{QATS} is able to identify the specific
times of transit, regardless of the presence of TTVs, we have no need
to mask out wider windows around the transit than just the transit
itself.  This is in contrast to, e.g., the \project{Kepler} pipeline,
which removes a total of 3$\times$ the transit duration around every
transit, which has been shown to lead to a reduced sensitivity to
multiplanet systems \citep{zin19}.  The \project{K2} searches of \cite{cro16},
\cite{pet18}, \cite{van16b}, and \cite{may18} all follow the
\project{Kepler} example and remove 3$\times$ the transit duration of
data at every transit. This could partially explain why we find more
multiplanet systems than all of these groups (see \S\ref{previous}).

\subsection{Modeling Candidates}
\label{mcmc}

In all, \nastro{} candidates around \nsystems{} stars passed both the
automated and manual vetting stages. Many of these are EBs,
however, and we used MCMC results to help separate the true
planet candidates from likely EBs and false positives
(due to e.g., ephemeris matching indicating that the same source is
contaminating multiple targets).

We modeled each astrophysical candidate and obtained the optimal
physical parameters and associated uncertainties with an
affine-invariant Markov chain
\citep[\pipeline{emcee};][]{for13}. Before running the MCMC, we
recalculated the light curve with \pipeline{EVEREST}, this time
masking out all transiting signals to prevent their partial removal by
the detrending process, and thus to obtain accurate transit depths.

In the case of multiplanet systems, any transits by other planets in
the continuum regions were masked and removed before we started the
following detrending and fitting. However, for cadences where other
planets' transits overlap with the one being fit, no attempt was made
to disentangle them: that transit will be deeper. Any candidates with
a significant fraction of transits overlapping those of other planets
may have biased parameters.

Before starting the MCMC, we detrended each individual transit and then
fed the detrended transits into the MCMC. We modeled the local
variability around each transit (centered on the original cadence
\pipeline{QATS} predicted) with a third-order polynomial and a continuum
region on each side with a width given by the larger of 0.2 days or
two transit durations. In some cases third order was not enough to
capture the stellar variability, so we continued to increment the
polynomial up to a maximum of sixth order while the model's reduced
chi square was above 1.5.  We then fed each of these detrended
transits into MCMC to find the optimal transit model parameters.

Although our search allowed for TTVs, in this stage we modeled all
candidates as periodic and used only a period and initial time of
transit as parameters owing to computational constraints.  The initial
transit time and period have priors such that the center of every
transit must fall within the detrended transit window $W$. The other
parameters in our MCMC model were the transit duration $T$ (uniform
prior between 0 and $W$), planet-to-star radius ratio $R_p/R_*$ (log
uniform prior with a maximum of 10), two quadratic limb-darkening
parameters using the transformed \cite{kip13} $q_1, q_2$ formalism
(uniform prior between 0 and 1), and impact parameter $b$.

As with many prior analyses, we found that low-S/N
transits produced almost no constraints on the transit shape. This
simply means that the limb-darkening parameters default to the distribution
of the prior, but the impact parameter can cause extra
complications. Without a S/N large enough to distinguish
between ``U"-shaped planet transits and ``V"-shaped grazing EBs,
finely tuned planet-to-star radius ratios and impact
parameters can match the transit depth for any impact parameter. As
this study is focused on planet candidates, and to help the MCMC
converge, we limited the impact parameter to a uniform prior between
$[0, 1 + 2 \sqrt{\delta}]$, where $\delta$ is maximum depth of transit
measured from the numerical light-curve model.

We added five additional parameters to account for further uncertainty
in the light curves.  A systematic error parameter (log uniform prior
between $10^{-7}$ and $10^{-1}$) allows for inflating the white noise
on each data point. A flux dilution parameter accounts for the
fraction of flux in the light curve due to additional light from
another star (a binary companion, background star, or extra light
spilling into the aperture) diluting the transit depth (log uniform
prior between $10^{-6}$ and 1).  By adopting a maximum flux dilution
value of 1, we implicitly assume that the planet candidate transits the
brightest star in the aperture; if additional observations suggest that
this is not true, the MCMC should be rerun with a more informative and
accurate prior. Further, this prior, by definition, will not hold when
modeling an EB when the brighter star eclipses the
dimmer star; in such cases the radius ratio can also exceed our limit
of 10.  We therefore urge caution in using the fit parameters for the
eclipsing binary systems. Complete and accurate modeling of all \project{K2} EBs
is beyond the scope of this study.

\renewcommand{\arraystretch}{1.2}
\begin{table}[tbp]
\caption{The Parameters and Prior Ranges for the Transit Model
  Fitting.
\label{tab:params}}
\begin{center}
\begin{tabular}{|llc|}
\hline
Parameter & Description & Range \\
\hline
$t_0$ & Initial time of transit & $\cdots$ \\ 
$P$   & Orbital period & $\cdots$ \\ 
$T$   & Transit duration & ${\cal U}(0,W)$\\
$\log(R_p/R_*)$ & Log radius ratio & ${\cal U}(-8,1)$\\ 
$q_1,q_2$ & Quadratic limb darkening & ${\cal U}(0,1)$\\
$b$ & Impact parameter & ${\cal U}[0,1+2\sqrt{\delta}]$\\
$\log(\sigma_{\text{sys}})$ & Log systematic error & ${\cal U}(-7,-1)$\\
$f_{\text{dilution}}$ & Log flux dilution & ${\cal U}(-6,0)$\\
$Q$ & Fraction of normal data & ${\cal U}(0,1)$\\
$\mu_{\text{outlier}}$ & Median of outliers & ${\cal U}[1-\delta,1+\delta]$\\
$\log(\sigma_{\text{outlier}})$ & Log std. dev. of outliers &$ {\cal U}(-7,0)$\\
\hline
\end{tabular}
\end{center}

{\bf Note.} The ephemeris parameters, $t_0$ and $P$, are allowed to
  vary such that every transit time falls within the detrended window
  $W$ centered on the original \pipeline{QATS} predicted time.
\end{table}

\renewcommand{\arraystretch}{1.}

The final three parameters in our MCMC fits are for an outlier mixture
model \citep{for14b}. Because thruster fires, flares, etc., cause
numerous outlier events in \project{K2} data, we needed a way to prevent the
outliers from skewing the transit model. We accomplished this with a
mixture model, for which the three parameters are the fraction of
data points in the normal transit model (uniform prior between 0 and
1) and the outlier points' median (uniform prior between $[1 - \delta,
  1 + \delta]$) and standard deviation (log uniform between $10^{-7}$
and 1).

All of the parameters describing the model used for transit fitting,
along with their priors, are summarized in Table \ref{tab:params}.  A
total of 12 parameters are allowed to vary within the range of
priors.  Other parameters, such as the \pipeline{EVEREST} detrending
parameters, the continuum windows, and the coefficients of the
polynomials used in detrending each transit, are held fixed during the
fitting.

We ran the MCMC (using 60 walkers) initially for a burn-in phase of
10,000 steps. Every 10,000 steps thereafter, if any chains got stuck
in regions of lower probability (the chain's maximum likelihood was
smaller than a factor of half of the overall maximum likelihood), we
stopped and reset the sampler near the previous maximum likelihood
solution to bring all chains into similar parameter space. All
iterations up to this point were also discarded as part of the burn-in.
We terminated the MCMC chains once we achieved 2000 independent
samples of each parameter (measured using the number of steps times
the number of walkers divided by the longest autocorrelation length)
or 1 million iterations if convergence seemed unlikely. Lack of
convergence can occur when the \pipeline{EVEREST} light curves are not
well detrended and each transit or eclipse has varying and inaccurate
depths. This is most common for deep EBs and is noted below.

Finally, in addition to calculating full chains for the parameters of
the model, we calculated three derived parameters at every step for
each walker.  We measured the transit depth (\cite{man02} value at the
step's impact parameter, adjusted for the quadratic limb-darkening
parameters and extra light flux dilution), ratio of semimajor axis to stellar
radius ($a/R_*$, calculated assuming a circular orbit using
Eq.\ 7 and 14 of \cite{win10}), and the subsequent stellar density
(Eq. 30 of \cite{win10}), neglecting the contribution of the planet to
the density.

The ratio of semi-major axis to stellar radius is given by
\begin{equation}
\frac{a}{R_*} = \frac{\left[\left(1 + \frac{R_p}{R_*}\right)^2 - b^2 \cos^2\beta\right]^{1/2}}
{\sin \beta}
\end{equation}
where $\beta = \pi T/P$.  The incident flux on the planet can be
computed from this and the stellar luminosity via $L_* (a/R_*)^{-2}$,
assuming a circular orbit for the planet.

\section{Results} \label{results}

From the initial sample of \numstars{} stars in \project{K2} C0--8, \nastro{}
candidates around \nsystems{} stars passed our automated and manual
vetting for astrophysical significance. Table \ref{tab:cuts}
summarizes the results of our cuts for astrophysical significance,
showing how many stars were cut and remained in the pipeline after
each stage of the automatic and then manual vetting. Also, the
iterative search for additional candidates around the \nsystems{} that
passed the first stage of manual vetting turned up a further \nmulti{}
astrophysical candidates, as indicated in the table. However, many of
these objects are actually EBs or other false
positives. With our full list of candidates set, we turn toward
discriminating between the likely EBs and planet
candidates.

\subsection{Identifying EBs and False Positives}
\label{ebsort}

In doubly-eclipsing binaries where the primary and secondary eclipses
have significantly different depths or where the secondary occurs too
far from phase 0.5 to be picked up as TTVs by the \pipeline{QATS}
transit window, \pipeline{QATS} will find the deeper eclipse first and
then the shallower secondary in our follow-up search for additional
signals. In this case, we have two distinct candidates at the same
period with significantly different depths, and we can confidently
label the system as an EB (discounting the rare
circumstances of detectable occultations of hot Jupiters).

\renewcommand{\arraystretch}{1.2}
\begin{table}[tbp]
\caption{Summary of the Search and How Many Candidates Pass Each Stage
  in the First Search of Every Star
\label{tab:cuts}}
\begin{center} 
\begin{tabular}{
  |p{\dimexpr.57\linewidth-2\tabcolsep-1.3333\arrayrulewidth}
  |p{\dimexpr.2\linewidth-2\tabcolsep-1.3333\arrayrulewidth}
  |p{\dimexpr.23\linewidth-2\tabcolsep-1.3333\arrayrulewidth}|
  }
\hline
Stage  & Change at This Step & Number after This Step  \\
\hline
All C0--8 stars (\S\ref{everest}) & --- & \numstars{} \\
\hline
\pipeline{QATS} signal $> 9\sigma$ above baseline (\S\ref{qatspeaks}) & -\nfailz{} & \npassz{} \\
\hline
Transit fit better than sine fit  (\S\ref{sinecut}) & -\nelimsin{} &  \npasssin{} \\
\hline
Duration not too long for period (\S\ref{durcut})  &  -\nelimdur{} & \npassdur{} \\
\hline
Manually vetted as astrophysical (\S\ref{vetting})& -\nfailvet{}  & \nsystems{}  \\
\hline\hline
Additional candidates passing all vetting found in multiplanet search (\S\ref{multi}) & +\nmulti{} & \nastro{} \\
\hline
\end{tabular}
\end{center}

{\bf Note.} The final line gives the number
  of candidates found after masking the \nsystems{} candidates, and
  subsequently searching for additional transit signals.
  
\end{table}
\renewcommand{\arraystretch}{1.}

When the primary and secondary eclipses approach the same depth,
however, \pipeline{QATS} will often combine them together into a
single candidate at half the true binary orbital period. Careful
modeling of the odd and even events independently can determine that
the two sets do indeed have different depths or ephemerides (i.e., the
secondary eclipse is not precisely at phase 0.5), which then allows us
to once again label the system as an EB instead of a
planet candidate.

Another confounding factor in the search for planet candidates is the
source of false positives due to the same binary signal appearing in
many different targets at varying depths. This contamination can be
caused by internal reflections within the telescope, by electronic
cross talk between readout channels, and/or by sources affecting the
signal of other targets on the same CCD column; these false positives
were identified in the original \project{Kepler} mission and found to
be the cause of $\sim$10\% of the KOIs \citep{cou14}. These false
positives thus need to be seriously considered and can be found by
looking for multiple candidates that have the same period and
ephemeris --- indicating they are ultimately caused by the same
source.

Taking all of these contaminants into consideration, we ran three
different tests to discriminate and remove the obvious EB cases and
false positives from our planet candidate sample:
\begin{enumerate}
\item odd--even tests to check for consistent parameters between odd
  and even candidate transit events;
\item a period and ephemeris collision match to test for the same
  signal appearing in multiple targets; and
\item a maximum depth cut to remove singly-eclipsing EBs.
\end{enumerate}
We describe each of these in the following sections.

\subsubsection{Odd--Even Tests}
\label{oddeven}

For every candidate, in addition to a full fit with all transits, we
carried out the MCMC fit (\S\ref{mcmc}) to the odd and even transits
separately and then compared the posterior distributions.  Any depth
variations or ephemeris offsets are indicative of the candidate
actually consisting of both the primary and secondary eclipses of an
EB.

\begin{figure}[tbp]
\includegraphics[width=\columnwidth]{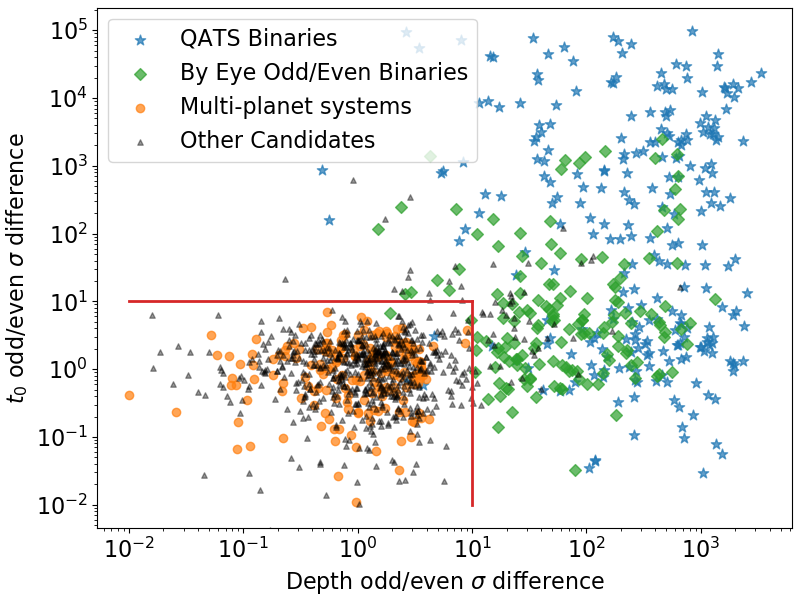}
\caption{Odd--even test for EBs. All of our candidates
  are shown, sorted into categories (see text for details). The red
  lines indicate our cut between EB and planet candidates --- only
  systems in the lower left pass and are labeled planet candidates at
  this stage.  \label{binary_match}}
\end{figure}

Figure \ref{binary_match} compares every candidate's odd and even
transits, showing how many standard deviations apart the periods and
depths are. We display all candidates separated into four classes: (1)
systems \pipeline{QATS} identified as two different signals at the
same period (we treat the first as the ``odd'' and second as the ``even''
here), (2) systems manually noted to have possible depth or ephemeris
variations during the manual vetting (before the MCMC was run,
\S\ref{vetting}), (3) candidates in multiplanet systems (multiple
candidates at different periods, each showing more than one transit),
and (4) all other candidates.

The \pipeline{QATS} and manually identified EBs should show
significant depth and/or ephemeris differences between their odd and
even fits, indicating they are caused by two different eclipse
events. The multiplanet systems --- which in the original
\project{Kepler} sample were found to be much more likely to be real
planets \citep{row14} --- should show no differences between the odd
and even depths and ephemerides since they will predominantly be
complementary subsets of the same planetary transit. Indeed, that is
what we find in Figure \ref{binary_match}, further motivating this
cut.

Because \pipeline{EVEREST 1.0} can sometimes get transit depths a
little wrong, even after masking them out before running the pipeline,
we were conservative in our cuts; we set our limits at 10$\sigma$ for
both depth and transit ephemeris. If the odd and even depths or
ephemerides deviated from one another above this value, we
automatically categorized the system as an EB.  This cut
eliminated \oddevenebs{} candidates and thus created \oddevenebs{}
doubly-eclipsing EBs.

\renewcommand{\arraystretch}{1.2}
\begin{table}[tbp]
\caption{EPICs Identified as Hosting False Positives (the Same Signal
  Appearing in Multiple Targets)
\label{tab:fps}}
\begin{center}
\begin{tabular}{ | c c c c |}
\hline
\multicolumn{4}{| c |}{False-positives EPICs} \\
\hline
201467358 & 201467521 & 202073124 & 202073125 \\
202627452 & 202627721 & 203728604 & 203730072 \\
204676803 & 204676841 & 205916737 & 205916793 \\
210386880 & 210386883 & 210512752 & 210512842 \\
211078690 & 211079188 & 212409658 & 212497267 \\
212516916 & 212516935 & 212563850 & 212563954 \\
212844216 & 212844260 & 220222029 & 220222060 \\
220391408 & 220391520 & & \\
\hline
\end{tabular}
\end{center}
{\bf Note.} These systems are removed from both
  our candidates and EB lists.
\end{table}
\renewcommand{\arraystretch}{1.}

The only exception to this occurred in EPIC 203771098, which was
originally a two-planet system with an outer 42.4-day planet and an
inner 20.9-day companion. However, the outer planet had just two
transits, and those transits differed in depth by 36$\sigma$ (3150 and
4370 ppm), making it a supposed odd/even binary. This is the only case
of our odd/even systems being composed of two single-transit events,
both with depths below 4\%, so we instead call them single transits of
planets 1 and 3, with planet 2 remaining the inner 20.9-day
candidate. We note, however, that it is likely that these two transits are
a single planet, since these depth variations are not present in other
reductions, and the planet has been confirmed at 42 days as a single
radial velocity planet, K2-24 c \citep{pet16, pet18b}.

We also made the same tests when multiple candidates around a star
have nearly the same period.  This accounts for instances where
\pipeline{QATS} found primary and secondary eclipses individually
(usually because the depths are wildly different or the secondary
eclipse phase is far from 0.5).  When two or more candidates were
found around the same star, we checked to see whether the periods were
the same within 5\% tolerance.  We identified \qatsebs{} target stars
that showed two transiting candidates each (\twoqatsebs{} candidates
in total) for which the periods were consistent within this tolerance
(the \pipeline{QATS} binaries in Figure \ref{binary_match}). In only
one pair of the \twoqatsebs{} objects (one of the \qatsebs{} targets)
were the depths and ephemerides consistent within our cuts. That pair
(EPIC 214984368) was combined and reclassified as a single-planet
candidate; however, we note that it appears to be an EB with depths
slightly too shallow to discriminate the difference, due to our high
threshold, with a single campaign of data. All others were removed
from the planet candidates list and labeled EBs.

\subsubsection{Period and Ephemeris Matching}
\label{ephems}

Nearby targets whose point-spread functions overlap can cause one EB
to be detected at varying depths in nearby sources. Similarly, bright
stars can have their light reflect across to widely separated areas of
the detector, again causing the same signal to appear at different
depths in many stars \citep{cou14}.

\begin{figure}[tbp]
\includegraphics[width=\columnwidth]{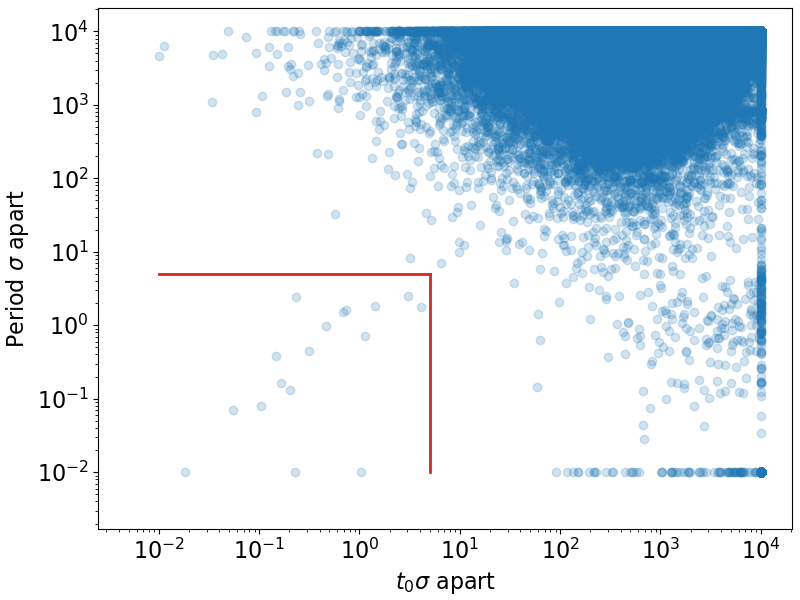}
\caption{Difference in period and $t_0$ between all pairs of
  candidates in the same campaigns (clipped between $[10^{-2}, 10^4]$
  for clarity). Pairs where at least one candidate is a single transit
  and thus does not have a defined period are given a period
  difference of 0 and clipped to the $10^{-2}$ line along the
  bottom. Any pairs with both parameters within 5$\sigma$ (lower left
  of red lines) are labeled false positives. \label{collisions}}
\end{figure}

Within each campaign, we compared the period and time of first transit
for every pair of candidates (Figure \ref{collisions}). Any pair where
both the periods and first transit times were consistent within 5
standard deviations were flagged as false positives. It is beyond the
scope of this work to track down which (if any) of the targets is the
true source --- sometimes the actual EB is a bright star whose light
is not downloaded in the postage stamps. Through this method we
identified \nfps{} candidates or EBs (around \nfpsys{} distinct EPICs,
the nine overlaps were primary and secondary eclipses of a single EB)
with matching periods and transit times and labeled them as false
positives. They are listed in Table \ref{tab:fps}.

\subsubsection{Depth Cut}
\label{depthcut}

After eliminating the obvious false positives and doubly-eclipsing
EBs, we were left with \withsingles{} candidates around
\withsinglessys{} stars. However, none of our previous cuts account
for singly-eclipsing EBs. Rather than make any judgments about transit
shape, we imposed a simple depth cut of 4\%, which comfortably
encompasses all confirmed \project{Kepler} planets to date\footnote{As
  listed on the NASA Exoplanet Archive}. Only 11 of the 2297 confirmed
\project{Kepler} planets have depths above 2\%, and just two are above
3\%.  Any candidate with a transit deeper than 4\% was labeled as a
likely EB for our purposes. This cut eliminated
\nsingles{} candidates.

After all our cuts, the only star with objects labeled both planet
candidates and EBs is EPIC 220598367. This star has
both a 7.65-day planet candidate and primary and secondary eclipses of
a 5.27-day binary with significant phase offsets between the two. This
can be explained by examining the postage stamp, which shows light
from two 14th magnitude stars contributing to the light curve. One of
them likely hosts the planet candidate, while the other is the
EB source. All other stars are hosts to exclusively
planet candidates or EBs.

\begin{figure*}[tbp]
\includegraphics[width=\textwidth]{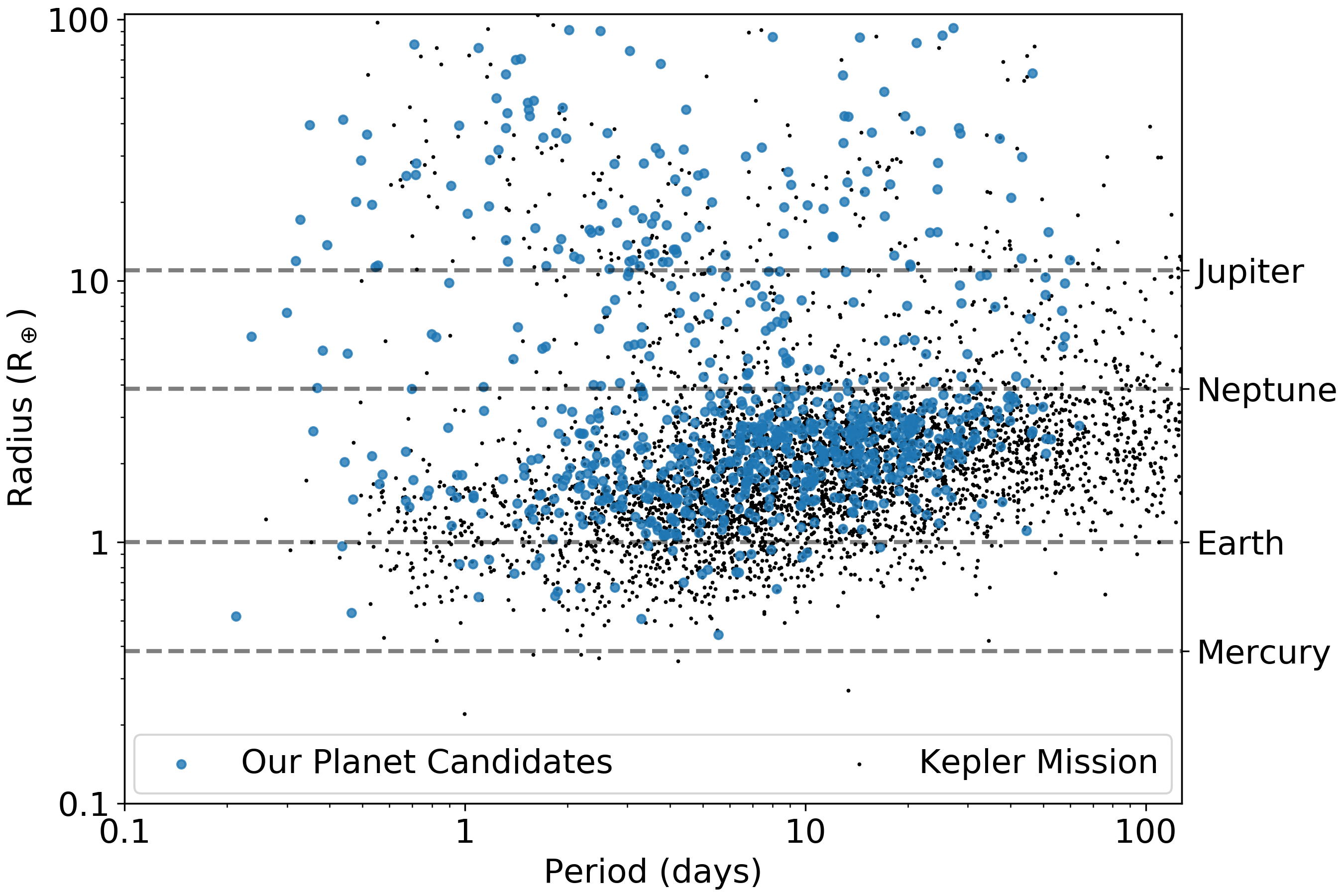}
\caption{Our planet candidates (larger blue points) as a function of
  period and radius compared to the \project{Kepler} planet candidates
  (smaller black points). The right-hand scale shows the size of some
  solar system planets for references. Our planet candidates with a
  single transit, and thus no well-defined period, are left off this
  figure. \label{candidateperiod}}
\end{figure*}

\subsection{Stellar Parameters}
\label{stellarpars}

To convert the observationally measured planet-to-star radius ratio to
a physical planetary radius, we require parameters of the host
star. While the target stars in the \project{K2} fields have not been studied as
extensively as the original \project{Kepler} field, some work has been
done to characterize all of the \project{K2} stars.

Most recently, the \project{Gaia} second data release (DR2) has
provided stellar parameters (radius, temperature, and luminosity) for
over 70 million sources, though notably excluding all stars with radii
less than 0.5 R$_\odot$ \citep{gai16, gai18, and18}. For stars they do
constrain, they have stellar radius errors of about 10\%.

For our stellar parameters, we adopt the \project{Gaia} DR2 radii,
luminosities, and effective temperatures when available. As
recommended by \cite{and18}, we only trust the \project{Gaia}
parameters when the parallax uncertainty is less than 20\% and the
Priam flag values suggest the effective temperature can be trusted. We
also only use the \project{Gaia} values if the \project{Gaia} source
is within 3$^{\prime \prime}$ of the EPIC listed
coordinates. Altogether, \project{Gaia} provides parameters for
\ngaia{} of our \nplanet{} planet candidates.

For stars not in the clean \project{Gaia} sample, our secondary source
of stellar parameters comes from \cite{hub16}, who classified 88\% of
the stars targeted in C1--8. They provide stellar radii to an
uncertainty of about 40\%. They do not directly give a luminosity
estimate (needed for incident flux calculations), so we derive one
from the effective temperature and radius. The \cite{hub16} values
fill in stellar parameters for all but \nostellar{} candidates, of
which \nostellarczero{} are in C0 and not covered by their work.  For
those \nostellar{} candidates, we simply leave blank any field relying
on stellar parameters and report only those deduced directly from the
light curve.

The median error in stellar radius for our \project{Gaia} stars (most
of those above 0.5 R$_\odot$) is \rserrgaia{}\%, although see
\cite{and18} for reasons this may be an underestimate. For the
\cite{hub16} stars, the stellar radius uncertainty has a median of
\rserrhuber{}\%.

Our planet radius and incident flux ranges are calculated using the
$-\sigma$, median, and $+\sigma$ values from our MCMC analysis,
combined with the respective stellar uncertainties. In other words,
the lower and upper errors are each treated as one-sided normal
distributions and propagated independently.

\subsection{General Properties of the Planet Candidates}
\label{sec:general_properties}

After applying our selection cuts to eliminate EBs and false
positives, we are left with \nplanet{} planet candidates around
\nplanethost{} stars, listed in Table \ref{table:candidates}.  Our
recovered multiplicity rate is tabulated in Table \ref{tab:multis};
most notably we have five four-planet systems (EPIC 205071984,
EPIC 206135682, EPIC 211939692, EPIC 212157262, and EPIC 220221272), two five-planet systems
(EPIC 211413752 and EPIC 211428897), and a six-planet system (EPIC
210965800), discussed below. We compare our ratio of multiplanet to
single-planet systems to those from \project{Kepler} and other \project{K2} searches in
\S\ref{previous}.

All of our planet candidates are shown as a function of period and
radius in Figure \ref{candidateperiod} and compared to the
\project{Kepler} candidates. As with \project{Kepler}, the majority
(\smallnep{}\%) of our candidates are smaller than Neptune, and the
period and radius distribution is qualitatively similar between the
two samples. The notable difference is that our candidates cut off at
periods of about 50 days owing to \project{K2}'s limited campaign duration;
similarly, we are not as efficient at finding small, long-period
planets without the benefit of several years to build up enough low-S/N
transits for detection.

\begin{figure}[tbp]
\includegraphics[width=\columnwidth]{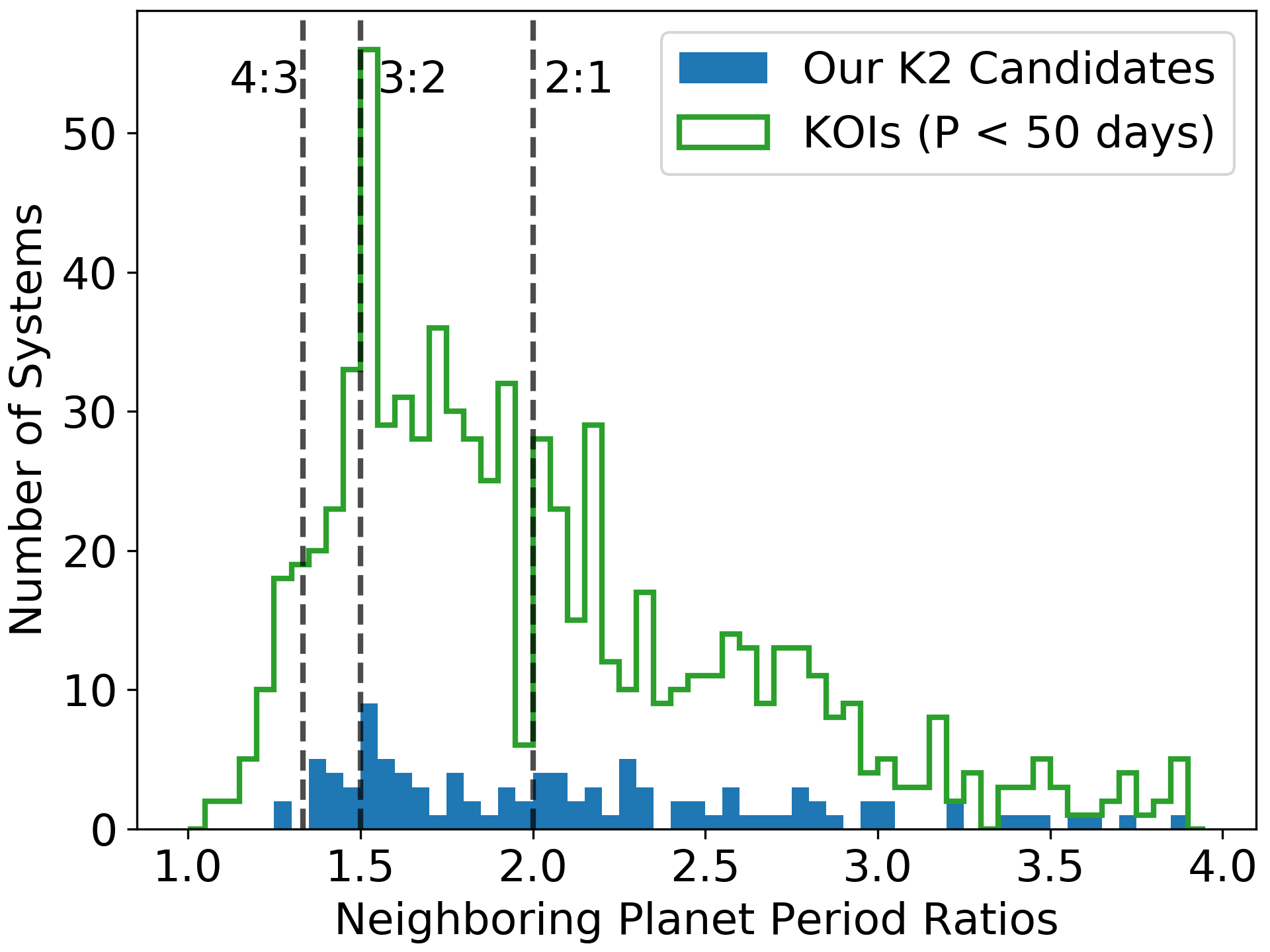}
\caption{Period ratios of neighboring planets in our multiplanet
  systems compared to those of the \project{Kepler} planet candidates
  with periods less than 50 days. \label{periodratios}}
\end{figure}

We also show how the period ratios of neighboring planets in our
multiplanet systems compare to the \project{Kepler} sample (Figure
\ref{periodratios}). While we have many fewer systems, we find the
same pileup of planet pairs just outside of a 3:2 period ratio that
was uncovered from \project{Kepler} \citep{lis11b, fab14}.

\subsection{Comments on Individual Systems}

We pause briefly to discuss and call attention to some individual
systems uncovered in our search that show unusual or especially
interesting properties.

\subsubsection{Transit Timing Variations}
\label{sec:ttvs}

The main reason \pipeline{QATS} was developed was to detect planets
with TTVs. However, TTVs will not be nearly as evident in the brief 80
days of a \project{K2} campaign as they were in the 4 yr of
\project{Kepler}. In \project{Kepler}, a TTV planet's transit times
would vary on timescales of many hundreds of days
\citep{maz13}. Trimmed to an 80-day window, those same variations
would not show their full amplitude, and the planet would be difficult
to distinguish from one without TTVs. Still, we use the full power of
\pipeline{QATS} and allow for it to pick up large TTV systems
(\S\ref{qats}).

% planet candidates table is table 7, but listed at end.
\setcounter{table}{7}
\begin{table}[tbp]
\caption{Summary of Our Planetary System Multiplicity --- the Number
  of Systems We Found with a Given Number of Planet
  Candidates. \label{tab:multis}} \centering
\begin{tabular}{| c | c |}
\hline
Multiplicity & No. of Systems \\
\hline
1 & 611 \\ 
2 & 64 \\ 
3 & 15 \\ 
4 & 5 \\ 
5 & 2 \\ 
6 & 1 \\ 
\hline 
\end{tabular}
\end{table}

A quantitative search separating small-amplitude TTVs from periodic
systems is beyond the scope of this work, but we did make note of any
candidates with TTV amplitudes visible by eye in the manual vetting
stage. Because our MCMC model does not account for TTVs, the
parameters of these candidates will not be completely accurate. In
total, four single-planet systems were noted as showing signs of TTVs:
EPIC 201561956, EPIC 212639319, EPIC 220639177, and EPIC 211924657, the last of
which is the most notable and we discuss further.

\begin{figure}[tbp]
\includegraphics[width=\columnwidth]{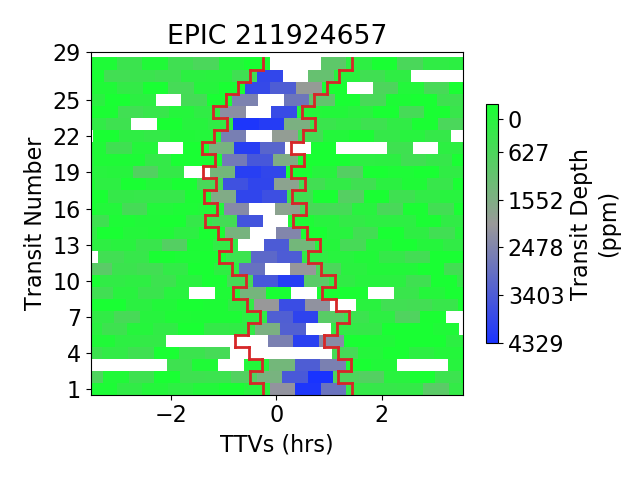}
\caption{Same as Figure \ref{river}, a ``river plot'' showing the large
  transit timing variations of our planet candidate EPIC 211924657.1
  (period 2.64 days). This planet has been previously validated as
  K2-146 b. \label{ttvriver}}
\end{figure}

The C5 star EPIC 211924657 hosts a single 2.64-day candidate that
shows considerable TTVs (TTV amplitude on par with the transit
duration, shown in Figure \ref{ttvriver}); it was previously detected
by several other groups and validated as K2-146 b by
\cite{hir18}. This star is considered an M dwarf and thus does not
have \project{Gaia} parameters, but with the \cite{hub16} stellar
radius, the planet is 1.5 R$_\oplus$.  No additional transit signals
have been detected from C5 data alone, but TTV modeling may be able to
constrain the properties of the perturbing planet, especially because
this target was also observed in C16 and C18 (and in short cadence)
--- providing a 3 yr baseline for the TTVs.

\subsubsection{Single-transit Candidates}
\label{sec:singletran}

As discussed at the end of \S\ref{qatspeaks}, our chosen metric to
separate planet candidates from noise based on the \pipeline{QATS}
spectrum peaks above the baseline breaks down at the longest periods,
resulting in a reduced sensitivity to one- and two-transit
candidates. Furthermore, we only search for events with durations up
to 17 hr, and systems with periods of 50 up to hundreds of days
that would only produce one or two transits in a campaign might have
durations longer than that limit, causing us to miss them (the most
notable example of which is the 54 hr transit found by
\cite{gil18}).

Nevertheless, we detected \singletrantot{} astrophysical single
``transit'' events in total. Of those, the majority are exceptionally
deep and labeled as EBs via our depth cut (\S\ref{depthcut}) or likely
secondary eclipses for a deep EB that eclipsed twice. Just
\singletranplan{} are labeled planet candidates, and most of those are
V-shaped and likely to be shallow secondary eclipses of long-period
EBs.

However, one of our single-transit events (211087003.1) also hosts an
inner candidate with three transits (period 28 days), which provides
strong evidence that the single transit is from an outer planet in the
system. In addition, one of our four-planet systems, EPIC 211939692,
consists of two planets showing two transits each at periods of 27 and
39 days, and two further outer single transits with significantly
different depths and durations. All of these candidates are newly
discovered by our search.

\subsubsection{High-multiplicity Systems}
\label{sec:multis}

Perhaps the most exciting aspect of \pipeline{EVEREST}'s detrending is
the capability to find systems with several planets. To date, one
five-planet system has been announced from \project{K2}'s C0--8, but it contained three
single-transit events, which makes the periods hard to constrain
\citep[and also caused our search to miss them;][]{van16a}. Recently,
another five-planet system has been announced from C12, and all five
planets have well-determined periods with two potential transits of a
sixth planet \citep{chr18}.  Here we add two more five-planet systems, as
well as a new six-planet system, in which all planets have at least
three transits.

EPIC 211413752 hosts five planet candidates, of which only the
shortest period (2.15 days) and deepest (9.33-day period) had been
previously reported (and validated as K2-268 b \& c by
\cite{liv18}). Our five planets have periods of 2.15, 4.53, 6.13,
9.33, and 26.27 days; the middle three are very near to 3:4:6 period
ratios, a common trait of short-period multiplanet systems
\citep{fab14}.  It is possible that these three planets form a Laplace
resonance as $1/4.53 - 2/6.13 + 1/9.33 \approx 0.0018$ day$^{-1}$.
The host star is uncertainly labeled a giant by \cite{hub16}, but
\project{Gaia} classifies it as a K dwarf, which is more consistent
with the stellar densities of $\approx1$ g/cm$^3$ derived from each of
the planet candidates. With the \project{Gaia} value, the planets
range between 1.6 and 3.6 R$_\oplus$.

Similarly, we report that EPIC 211428897 hosts five planet candidates, of
which the first, second, and fourth had been previously found
\citep{dre17}. This star has been characterized by both \cite{dre17}
and \cite{hub16} as an M dwarf, and it hosts five planets, all with
periods less than 7 days: 1.61, 2.18, 3.29, 4.97, and 6.27 days. These
planets also appear to be near a resonant chain of periods, with
corresponding ratios near 3:4, 2:3, 2:3, and 4:5, respectively, and are
all smaller than Earth (between 0.5 and 0.8 R$_\oplus$).  Both of
these five-planet systems are in C5 and have been observed again in C16
or C18.

Finally, the C4 star EPIC 210965800 is host to a six-planet candidate
system. Only the deepest planet (8.75 days) had been previously found
(and validated as K2-178 b by \cite{may18}), but we add five
additional planets, bracketing it at periods of 1.83, 4.28, 13.16,
21.09, and 30.29 days. The second through fourth planets form a chain
near 1:2:3 period ratios, but the innermost and outer two planets are
relatively far from first-order resonances.  It may be that the outer
planets form a chain of Laplace resonances, with $1/4.28-4/8.75+3/13.16
\approx 0.0045$ day$^{-1}$, $1/8.75-2/13.16+1/21.09 \approx 0.01$
day$^{-1}$, and $1/13.16-3/21.09+1/30.29 \approx -0.00027$ day$^{-1}$,
reminiscent of Kepler-80 \citep{mac16}, Kepler-223 \citep{mil16}, and
Trappist-1 \citep{lug17}.  The host star is slightly subsolar, and
the planets range from 1.5 to 3.8 R$_\oplus$.

\subsection{Eclipsing Binaries}
\label{sec:ebs}

In our search for planets, we inevitably uncovered EBs as well, which
we have done our best to separate out (\S\ref{ebsort}). Compared to
our \nplanet{} planet candidates around \nplanethost{} stars, we found
\nebs{} EB systems (\ndoubleebs{} of which we found both primary and
secondary eclipses). Our ratio of \planettoeb{} transiting planet host
stars per EB compares very favorably to the original \project{Kepler}
mission, which has a ratio of 1.19 (3456 planet host
stars\footnote{\url{https://exoplanetarchive.ipac.caltech.edu/}} to
2909 EBs\footnote{\url{http://keplerebs.villanova.edu/}} [\citealp{kir16}]
listed in the online catalogs).

\begin{figure}[tbp]
\includegraphics[width=\columnwidth]{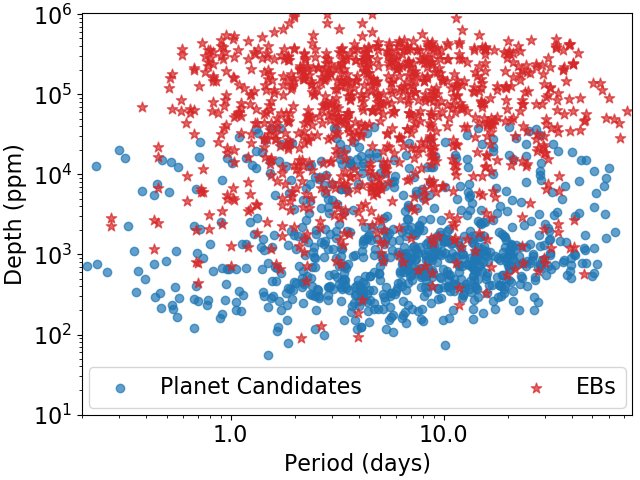}
\caption{Distribution of our planet candidates and EBs
  (with primary and secondary eclipses as separate points) as a
  function of period and transit or eclipse depth. By our definition,
  anything above 4\% transit depth ($4\times10^4$~ppm) is labeled as an
  EB, forcing that upper limit for planets
  (\S\ref{depthcut}). \label{ebdist}}
\end{figure}

While not the focus of this paper, we list all of our EBs in Table
\ref{table:ebs} and briefly discuss some of the highlights and
caveats. We compare the EB sample to the planet candidates in Figure
\ref{ebdist}. The two have similar period distributions, but of course
the EBs have considerably deeper eclipses than planet transits.

Despite EBs typically having deeper eclipses, we do not expect our
list to be complete. First, our search only extends out to a maximum
duration of 17 hr. Second, we impose an upper duration limit as a
function of period for any astrophysical candidates (\S\ref{durcut}),
established to reduce false positives. Those restrictions are both
loose enough not to eliminate the vast majority of planets but may
more substantially reduce EB yields. Any evolved or very large stars
will have much longer transit and eclipse durations, but corresponding
planetary transit depths would be so small around such large stars
that we would not expect to find them anyway. However, EB depths can
still be substantial around large stars, yet we would either not find
or prematurely eliminate any eclipse signals because of their long
durations. Finally, \pipeline{EVEREST} can have trouble with deep
eclipses; while it tries to mask them out before detrending, some
eclipses can slip by and cause the pipeline to overfit them, changing
their depths or even inverting them. \pipeline{QATS} may then have
trouble recovering the system when the depths are not consistent.

We also emphasize that the parameters listed in Table \ref{table:ebs}
may not be fully accurate for the EBs. In our MCMC analysis that
calculates the values in the table (\S\ref{mcmc}), we include a
``second light'' parameter that allows for contaminating flux in the
aperture, but we limit that ratio of additional to host starlight to
1 --- the majority of flux is assumed to be from the star being
transited/eclipsed. By definition, however, one star in an EB is
brighter than the other (except in the limiting case of perfectly
equal stars), and when the brighter star eclipses the other, our
assumption is broken. Similarly, our prior on $R_p/R_*$ only allows
for values up to 10; some EBs may have one star more than $10\times$
bigger than the other, which would prevent it from being modeled
accurately. Our limits on impact parameter may also prevent EBs from
being modeled accurately, and in some cases quadratic limb darkening
is not precise enough to fully model the eclipses.

All of these poor assumptions for EB modeling combine to cause the EB
fits to not be fully accurate. The periods and ephemerides can usually
be trusted, but the depths and durations should only be used as guides
to select various systems for follow-up and analysis.

\subsubsection{Eclipsing Binaries of Note}
\label{ebnotes}

\begin{figure}[tbp]
\includegraphics[width=\columnwidth]{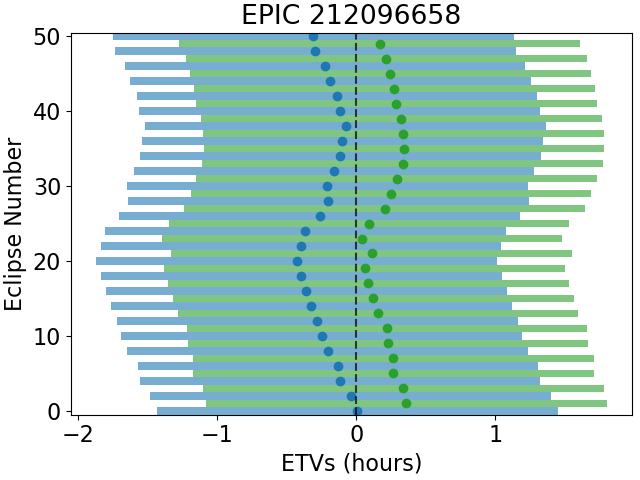}
\caption{Eclipse timing variations evident in the 2.93-day binary EPIC
  212096658. The best-fit times of the individual primary (blue, left)
  and secondary (green, right) eclipses are shown as points, with the
  shading indicating the duration of the eclipses. The ETVs are
  calculated based on the combined average eclipse period of 1.47
  days, with the approximately 20-minute horizontal timing offset
  between primary and secondary caused by the secondary eclipse's
  slight phase offset from 0.5. \label{etvs}}
\end{figure}

In this section, we briefly mention some EBs that stood out during the
manual vetting stage and may warrant individual follow-up:

\begin{enumerate}

\item EPIC 212096658 is a 2.93-day binary that shows eclipse timing
  variations in both the primary and secondary eclipses (Figure
  \ref{etvs}).  By the end of C5, both show a full sinusoidal cycle of
  ETVs from a perturbing third body in the system. This star was
  observed again in long cadence in C16 and short cadence in C18.

\item EPIC 203878683 is a 13.75-day binary that also shows signs of
  ETVs (amplitude $\approx10$ minutes). In this case, the ETV signal
  is only parabolic, and the ETVs of the primary and secondary
  eclipses are anticorrelated.

\item EPIC 201740472 is a short-period EB candidate with a period of
  just 0.96 days. Despite the short orbital period, the system's orbit
  is far from circular, and the secondary eclipse occurs at phase
  0.72. The durations also significantly differ, with the primary
  eclipse lasting $82\pm6$ minutes and the secondary just $23\pm5$
  minutes.

\item EPIC 216814711 is a 4.5-day binary and a classic example of a
  ``heartbeat'' star --- eccentric, short-period binaries where
  periastron passages induce tidal pulsations in the stars and cause
  them to brighten \citep{tho12}.

\item EPIC 212432524 is a 4.95-day binary whose secondary eclipse is
  noticeably asymmetric. The secondary eclipse is 10\% deep at its
  midpoint, but the eclipse depths one-quarter and three-quarters of
  the way through eclipse differ by about 1\%, and this asymmetry is
  consistent throughout the campaign.

\item EPIC 210766835 is a 24.78-day EB that shows large eclipse depth
  variations. The three primary eclipses in the campaign have depths
  of around 7\%, 10\%, and 11\%, while the secondary depths change by
  about 0.2\%. The star appears isolated in the aperture, and the
  depth variations are evident in other reductions of the light curve,
  so they appear to be astrophysical and not an artifact of
  \pipeline{EVEREST}.

\item EPIC 220374480 presents a single 60\% deep eclipse, but it is
  obviously asymmetric. The eclipse appears as if it were an inner
  binary with both stars nearly simultaneously eclipsing a third
  companion.

\end{enumerate}

\section{Discussion}\label{ch:analysis}

In this section we place the candidate planets we have found in the
context of prior searches of \project{K2} data (\S \ref{previous}), including a
discussion of the nature of the planet candidates that were missed by
our pipeline but found by other surveys (\S \ref{missed}).  Although
the planet candidates we have identified have yet to be confirmed, the
properties of the multiplanet candidate systems suggest that most of
these are real planet systems (\S \ref{sec:validity}).

Among the planet candidates, there are several categories that
stand out.  We have discovered new ultra-short-period planet
candidates, several of which are potentially smaller than $4 R_\oplus$
(\S \ref{usps}).  We have found dozens of single-transit candidates
(\S \ref{singletransits}) and planet candidates orbiting stars that
are brighter and smaller than the typical stars in the original \project{Kepler}
survey (\S \ref{sec:brighter_smaller}).  Although we have not refined
the stellar parameters of the host stars in our sample, we see a
tantalizing hint of the radius gap found by \citet{ful17} in our data
(\S \ref{sec:radius_gap}).  Finally, a handful of planet candidates
orbiting late-type stars are candidates for habitable zone exoplanets
(\S \ref{sec:HZ}), and those planets and stellar hosts need further
vetting to validate this categorization.

\subsection{Comparison to Previous Searches}
\label{previous}

The \project{K2} mission began with a trial campaign (C0) from 2014 March to May.
However, the spacecraft did not enter fine guidance pointing
until the latter half, so only the final 35 days of the campaign
produced higher-quality data. The data were released publicly in
2014 September, and several groups began developing ways to correct
for the pointing drift due to the loss of the two reaction wheels
\citep{van14, lun15}. Individual planets began to be published
\citep[e.g.,][]{pet15}, and \cite{for15} published the first candidates
from a systematic search of C1.

% account for EB and PC tables mentioned before but listed at end.
\setcounter{table}{9}
\begin{table*}[tbp]
\caption{Comparison of Our Results to Previous Groups with at Least
  Two Campaigns of Overlap and More Than 100 Candidates  \label{tab:compare}}
\begin{center}
\begin{tabular}{| l c c c |}
\hline
Candidate List  & Campaigns & Number They Found & Number We Found  \\
 & Searched  & (We Also Found) & (They Also Found) \\
\hline
\cite{bar16a} & 1--6 & 175 (134) & 586 (116) \\ 
\hline
\cite{cro16} & 0--4 & 167 (133) & 380 (132) \\
\hline
\cite{van16b} & 0--3 & 231 (182) & 301 (157) \\
\hline
\cite{pop16} & 5--6 & 168 (126) & 236 (119) \\ 
\hline
\cite{pet18} & 5--8 & 151 (128) & 438 (110) \\ 
\hline
\cite{may18} & 0--8 & 231 (191) & 818 (189) \\ 
\hline
All previous searches\tablenotemark{a} & 0--8 & 692 (500) & 818 (444) \\ 
\hline
\end{tabular}
\end{center}
\tablenotetext{a}{As listed on
  \href{https://exoplanetarchive.ipac.caltech.edu/}{NASA Exoplanet
    Archive's} \project{K2} Candidates Table, including those found by others
  not individually listed above.}
  
{\bf Note.} The columns
  indicate how many planet candidates are on one list and in
  parentheses how many overlap between their search and ours. The
  final row indicates our search compared to every published \project{K2}
  candidate in C0--8, including smaller searches not included
  individually in this table.

\end{table*}

\begin{figure}[tbp]
\includegraphics[width=\columnwidth]{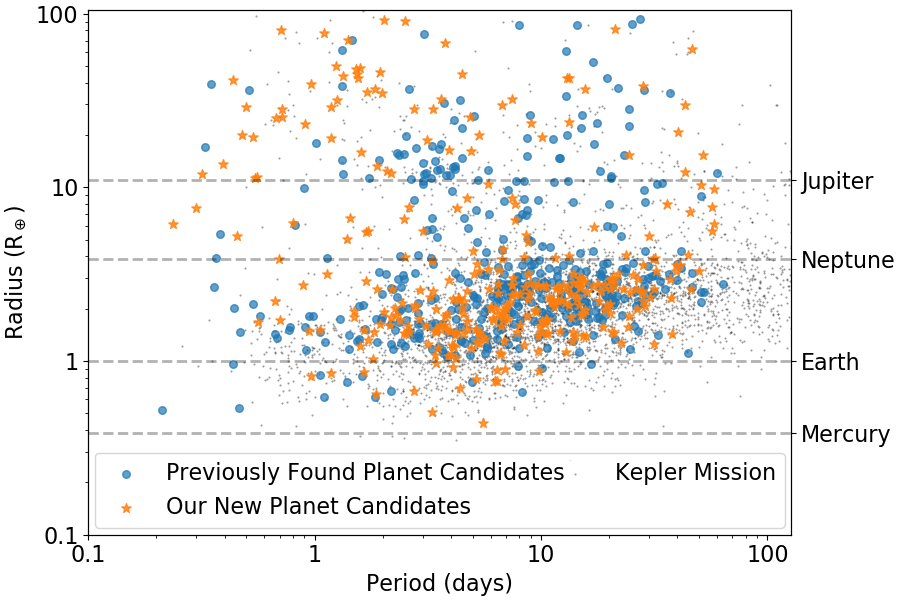}
\caption{Same as Figure \ref{candidateperiod}, but dividing our planet
  candidates between those that had been previously found by at least
  one group and our new candidates. \label{candidateperiodnew}}
\end{figure}

Since then, there have been hundreds of planet candidates announced,
but we choose to compare our \pipeline{QATS} pipeline to those with
more than one campaign of overlap and at least 100 candidates on their
list. The list of other searches and the detailed comparison of the
overlap between our lists of planet candidates can be found in Table
\ref{tab:compare}. In general, we find approximately 75-85\% of the
candidates found by other groups; however, they tend to find just
20--50\% of our candidates. Compared to every published \project{K2} C0--8 planet
candidate to date from all searches (even those not individually
listed in the table), we find \myoverlap{}\% of them, while
\theirmissing{}\% of our candidates are new and unlisted in every
other search. We split our planet candidates into those that are new
and those that had been previously reported by at least one other
group in Figure \ref{candidateperiodnew}.

\begin{figure}[tbp]
\centering
\includegraphics[width=0.8\columnwidth]{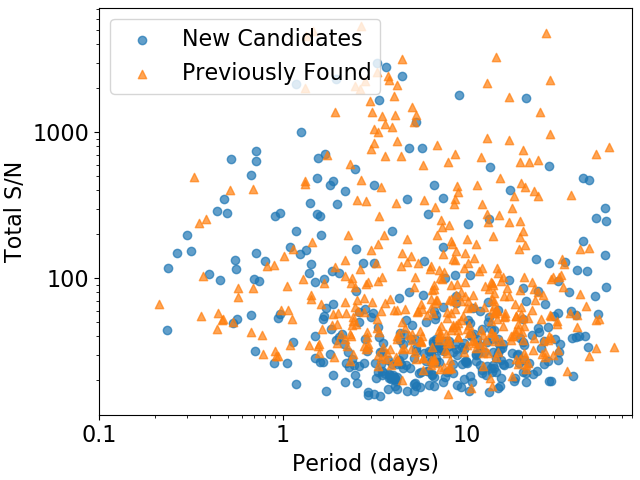}
\includegraphics[width=0.8\columnwidth]{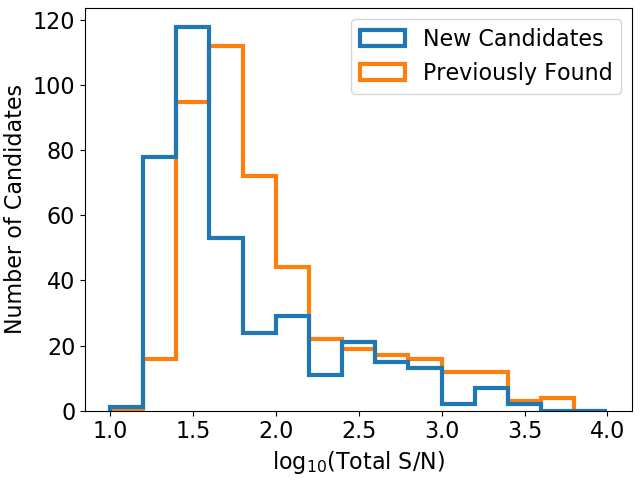}
\includegraphics[width=0.8\columnwidth]{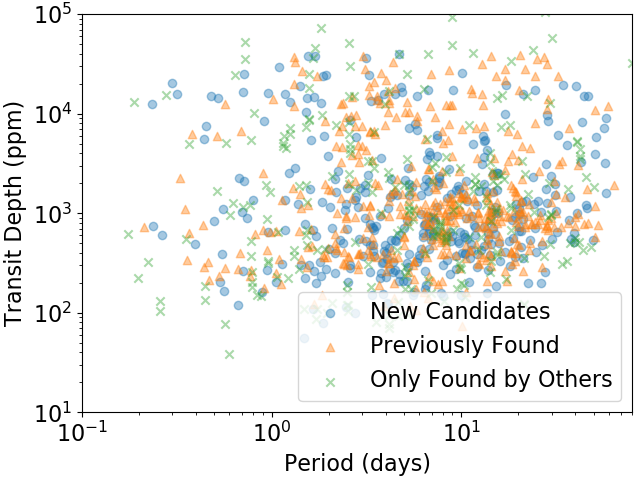}
\caption{Our new candidate sample S/N compared to those
  found by other groups. Top: our new planet candidates (circles)
  compared to those any other group has found as well (triangles) as a
  function of period and total transit S/N. Middle: same as the top panel,
  but marginalized over period to highlight the lower-S/N
  transits of our new planet candidates. Bottom: transit depth vs.
  orbital period parsed by candidates that are new (blue circles), found
  by our search and other searches (orange triangles), and only found
  by other searches (green crosses). \label{other_searches}}
\end{figure}

Our new candidates (those missed by other groups) are unlikely to be
dominated by high-significance false positives or EBs that we include
in our lists but other groups found and discarded. Figure
\ref{other_searches} highlights how our new planet candidates differ
from the candidates shared between our search and previous
efforts. Our planet candidates tend to be found at lower total transit
S/N values, indicative of our better sensitivity to small
planets. We estimate the total S/N as
$\frac{\delta}{\sigma} \sqrt{NT}$ with $\delta$ the transit depth,
$\sigma$ the \pipeline{EVEREST} light-curve noise level, $N$ the
number of transits, and $T$ the transit duration (in cadences). Our new
planet candidates have a median total transit S/N of
\newsn{}, and \snbelowold{}\% of our new planet candidates have a
total transit S/N below the median value of \oldsn{} for
the planet candidates found by at least one other group.

On the low-significance end, our new candidates are also unlikely to
be false positives found by digging into noise owing to the similarity
between our new candidates and previously found ones. For one, we find
\newmult{} new candidates in systems that already had at least one
planet previously found (e.g., five new planets in the previously
single EPIC 210965800 and two new planets in the previously
three-planet system EPIC 211428897). In total, \fracnewinmulti{}\%
(\nnewinmulti{}/\nnew{}) of our new planet candidates are found in
multiplanet systems, consistent with the overall \fracinmulti{}\%
rate (\ninmulti{}/\nplanet{}) for all our planet candidates.

\begin{figure}[tbp]
\includegraphics[width=\columnwidth]{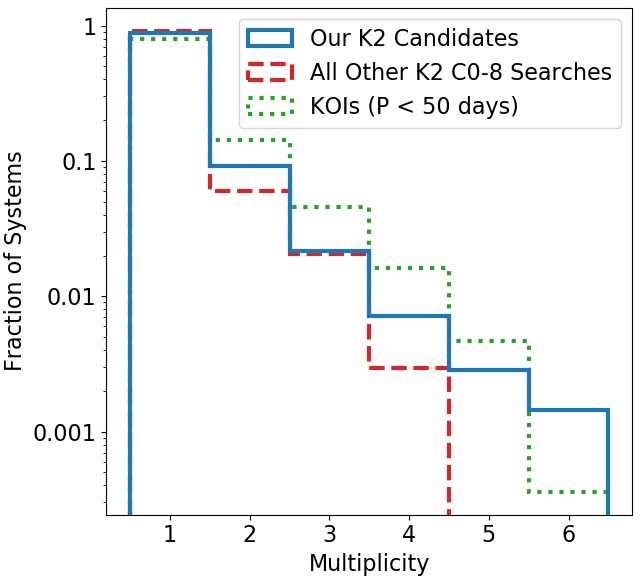}
\caption{Multiplicity of our planet candidate systems compared to
  the \project{Kepler} planet candidates (restricted to periods less
  than 50 days) and the results from all previous \project{K2} searches
  combined. \label{multiplicity}}
\end{figure}

We compare our multiplicity to that from the \project{Kepler} planet
candidate sample and the combined results of previous \project{K2} searches in
Figure \ref{multiplicity}. To make a more uniform comparison, we only
consider KOI multiplicity at periods less than 50 days --- KOIs with
periods beyond 50 days are not counted in the systems. Still, the KOIs
have relatively fewer singles and more multiplanet systems:
\fracsingle{}\% of our systems are single, compared to 79\% of the KOI
sample with periods less than 50 days.  This could be a result of
\project{Kepler}'s longer baseline, allowing the detection of smaller
planets inside periods of 50 days. It could also indicate that our sample
contains slightly more false positives in the single-candidate
systems, and/or that the stellar target selection results in different
planet populations being compared. On the other hand, all previous \project{K2}
searches combined have 92\% of their systems with just a single
candidate.

We believe that the vast majority of our newly discovered planet candidates
are in fact real planets that went undetected by other groups for any
of several reasons.  First, \pipeline{EVEREST} produces higher-precision
light curves than other groups. A full comparison between
\pipeline{EVEREST} and other groups can be found in \cite{lug16}, but
in general \pipeline{EVEREST 1.0} (as used in this work) produced
light curves of 20--50\% higher precision than \pipeline{K2SFF}
\citep{van14} and \pipeline{K2SC} \citep{aig16}.

Secondly, our better results could also be due to differences in the
planet search methods themselves. One possible difference is the
approach used to reject outliers before searching -- an especially
critical step in \project{K2} with its frequent thruster fires causing
additional outliers. Other searches did not first run a full search to
identify problematic cadences and remove them, likely leading to many
real signals being overwhelmed by residual thruster fires and
instrumental noise. Our version of \pipeline{QATS} also includes a
grid search in depth, helping enforce similar transit depths between
events; this differs from most other searches that only use a
3D grid of period, phase, and transit duration. Our
inclusion of depth may help boost low-S/N candidates over
the threshold of detection. Our method of selecting significant peaks
in our \pipeline{QATS} spectrum based on their height above the
background level (\S\ref{qatspeaks}) differs from other groups and
could help us better distinguish candidates.  Finally, when searching
for multiple planets in a system (\S\ref{multi}), we only mask exactly
the transit duration rather than a window three times the transit
duration, potentially allowing us to detect additional planets that
others missed as a result of lower duty cycles of real data in their
multiplanet searches \citep{zin19}.

As a brief test, we searched the publicly available \pipeline{K2SFF}
light curves for 20 of our planet candidates that have not been found
by any other groups. In about half the cases, the resulting signal is
weak enough that it would not have passed our tests, and we also would
have missed these candidates if we had used the \pipeline{K2SFF} light
curves. There are still hints of the individual transits at the times
we found them, but the signal gets lost in the additional noise
present in the light curve. In the other half of cases, our search of
the \pipeline{K2SFF} light curves clearly rediscovers our planet
candidates exactly where we found them in the \pipeline{EVEREST} light
curves. In many cases, the individual transits are identifiable by eye
in both the \pipeline{EVEREST} and \pipeline{K2SFF} light curves. We
therefore attribute an estimated half of our new planet candidates to
the increased precision of the \pipeline{EVEREST} light curves and the
other half to the increased sensitivity of our \pipeline{QATS} planet
search. We leave a detailed and more quantitative analysis of the
differences between all planet search techniques to future
investigation.

\subsubsection{Planets We Missed}
\label{missed}

If indeed our planet search is more sensitive to the smallest planets
and we are finding almost twice as many planets as had been found by
all previous searches combined, then it merits pausing to discuss why
there were any planets found by other groups that we missed. In this
section we explore those cases and discuss the different ways they
fell out of our search.

The \nimissed{} planets found by other groups that were missed in our
search (bottom row of Table \ref{tab:compare}) orbit \nimissedstars{}
unique stars. Of those \nimissedstars{} stars where others find
planets that we missed, the simplest explanation is one of differing
target lists: we did not search eight of those stars, leaving 174 systems
to explain. Another 10 systems are not major problems; we found the
same candidates, but we disagree on the period (almost always by a
factor of 2). For example, EPIC 201577112 has a candidate with two
transits separated by 44 days, which we call the planet's
period. However, if the planet's period had been 22 days, a potential third
transit would have fallen in the C1 data gap, and \cite{van16b} used
22 days as the period (they were the only other group to find this
candidate).

\paragraph{Missed Planets as Top \pipeline{QATS} Result.}

Of the 164 remaining systems with planets we ``missed,'' one-third of
them (58) were in fact found by \pipeline{QATS} as the top MLS, but
they were dropped somewhere along the way. Thirty-four of those 58 were dropped
at the very first of our automated cuts: the \pipeline{QATS} spectrum
peak above the baseline level (\S\ref{qatspeaks}). The most common
reason (20) was that the candidate only had one or two transits in the
campaign, which we have a documented difficulty detecting. Six of the
candidates were dropped because the transit duration is shorter than
1 hr, and many of the transits were removed as outliers, making the
overall signal much weaker. These systems may be recovered in future
versions of \pipeline{QATS} as we improve the pipeline, especially for
the one- and two-transit systems.

Six more systems were discarded in the other two stages of automated
vetting. Two were discarded for having the sine curve fit better than
a transit. EPIC 203533312 is a 0.18-day likely EB, and
EPIC 206152015 is a 0.81-day EB with out-of-eclipse
variability at a larger amplitude than the eclipses (\cite{bar16a}
only found the primary eclipse and called it a planet candidate). We
might revisit the sine curve test in future versions, but recovering
short-period binaries with large-amplitude variability is not a
present priority.

The other four systems (EPIC 206247743, EPIC 206439513, EPIC 210365511,
EPIC 210609658) were removed by our duration limit: they all have
exceptionally long transit durations for their orbital periods, most
likely indicating that they are orbiting evolved stars and could be EBs
(and two of the four have V-shaped transits, also indicative of
EBs). Removing the duration cut would allow these systems to be
detected, but it will place extra burden on the manual vetting stage
(an extra 110 systems to vet to find these four candidates) unless we are
able to better remove results due to stellar variability without the
duration limit.

Finally, 17 planet candidates found by other groups were removed by us
at the manual vetting stage (without any prior knowledge of the
candidacy status from other groups). One reason they were removed was
because the planet candidate has a very similar period to the obvious
stellar rotation period, and we were not confident enough that the
transits were distinct from stellar variability (EPIC 204346718,
EPIC 210843708, EPIC 210857328, EPIC 211834065). Most of the candidates were removed
in manual vetting, though, because they were on the low-S/N
end, and we were not confident enough to pass them.

Our manual rejection of 17 systems approved by other groups highlights
the subjectivity of manual vetting at the lowest-S/N
limit. On the other hand, it bolsters the idea that our new candidates
are not due to laxer vetting standards on our part by allowing through
lower-S/N false positives than others.

\paragraph{Missed Planets Overlooked by \pipeline{QATS}.}

Of the remaining 106 systems where our search does not find a planet
claimed by other groups, in 2 of them (EPIC 203823381 and EPIC 212737443)
the top result from \pipeline{QATS} was instead a long-period
candidate (one or two transits) that appears to be an outer companion to
the inner planet others found; because that outer companion did not
pass our automated cuts (as with other one- and two-transit systems),
we never masked it out and searched for the inner planet.

Four more systems (EPIC 201488365, EPIC 202072965, EPIC 204489514, EPIC 212579424)
are clear, deep EBs, but \pipeline{EVEREST} had trouble detrending
them. \pipeline{EVEREST 2.0} handles deep EBs better, and these would
be found in a search of the newer light curves. On the other hand,
these systems are definitively not planet candidates, despite their
label as such by other groups.

Finally, we have the 100 systems in which others claim there to be
planet candidates but our search comes up with nothing. In roughly
one-third of these cases, the \pipeline{EVEREST} light curve shows
excessive amounts of outliers indicative of poor detrending; most of
these stars show evidence of crowded apertures, saturated pixels, or
stellar light drifting out of the aperture during the campaign --- all
of which are known to cause degraded quality in our v1.0 light
curves. Inspection of v2.0 light curves shows better quality and often
the transits visible by eye, indicating that we will be able to
recover a number of these ``missing'' candidates using the newest
detrending.

Even in cases where there is nothing obviously wrong with the light
curve, we find that using the updated \pipeline{EVEREST} 2.0 light
curves can help improve detectability. For example, EPIC 211784767 is
an isolated, 12th magnitude star whose light curve looks normal in
v1.0, but \pipeline{QATS} finds no evidence of a 3.58-day planet as
claimed by \cite{bar16a}. However, running an identical version of
\pipeline{QATS} on the \pipeline{EVEREST} 2.0 light curve turns up the
planet exactly as expected, at the same depth and duration claimed.

We cannot find every ``missing'' planet simply by upgrading
\pipeline{EVEREST}, however. Around 50 candidates found by any one of
the other groups elude detection in our search, even with v2.0. We
note, however, that each such candidate was only found by one of the
various groups, even though every campaign (C0--8) has been searched by
at least three different teams, often more.

Consistency of planet yields between groups has been a recurring
problem for \project{K2} searches (see, e.g., \cite{pet18}'s comparison to
\cite{pop16} and \cite{bar16a} where catalogs overlap on about 60\% of
each other's candidates). Our recovery rate of 75-85\% of planets
found by other groups is better than these previous overlaps, and
improving our pipeline and updating \pipeline{EVEREST} will only
bolster that result.

\subsubsection{Evidence Supporting the Validity of Our Multiplanet Systems}
\label{sec:validity}

In this section we will provide two lines of evidence that the planets
in our multiplanet systems in particular are indeed real and orbit
the same star, even if they were missed by other searches.  Both rely
on the Keplerian orbits of planets and the fact that real planets
orbiting the same star are expected to follow those laws.  On the
other hand, if the candidates were not real, but false positives
resulting from digging into noise, the resulting planetary
architectures would have no reason to obey the Keplerian
relationships.

Kepler's laws relate a planet's transit duration ($T$, first to fourth
contact) and orbital period to the stellar and planetary properties
via the relation
\begin{equation}
T P^{-\frac{1}{3}} = \left( \frac{4}{\pi G M_*} \right)^\frac{1}{3}
\left(R_* + R_p \right),
\end{equation}
\citep{sea03}.  This equality assumes a circular orbit ($e=0$), zero
impact parameter ($b=0$), and negligible planet masses ($M_p \ll
M_*$). In the small planet approximation ($R_p \ll R_*$), the
right-hand side depends only on the stellar density
\begin{equation}T P^{-\frac{1}{3}} = \left( \frac{3}{\pi^2 G} \right)^\frac{1}{3} \rho_*^{-\frac{1}{3}} \label{eq:density} \end{equation}
We call this value $T P^{-\frac{1}{3}}$ the (period) normalized
transit duration.

\paragraph{Normalized Duration Ratios.}

We can define the normalized duration ratio $\xi$ as the ratio of the
normalized transit durations between two planets:
\[ \xi \equiv \frac{T_1  P_1^{-\frac{1}{3}}}{T_2 P_2^{-\frac{1}{3}}} \]
where we will always consider planet 1 the innermost of the two
planets.  For planets orbiting the same star, hence having the same
$\rho_*$, the normalized transit duration must be the same constant
value for each planet and $\xi = 1$ --- but only under the
simplifications made in this derivation, namely, that $e=0$, $b=0$,
$M_p \ll M_*$, and $R_p \ll R_*$.  These assumptions are of course not
always true, but $\xi$ should be close to 1 for all planet pairs in
multiplanet systems. Significant deviations from $\xi=1$ hint that
planets are not orbiting the same star, or that at least one of the
candidates may not be real but a false positive due to instrumental
systematics or digging too deep into noise. This normalized duration
ratio has been used in the past to provide evidence that multiplanet
systems are real planets orbiting the same star \citep{ste10, fab14}.

\begin{figure}[tbp]
\includegraphics[width=\columnwidth]{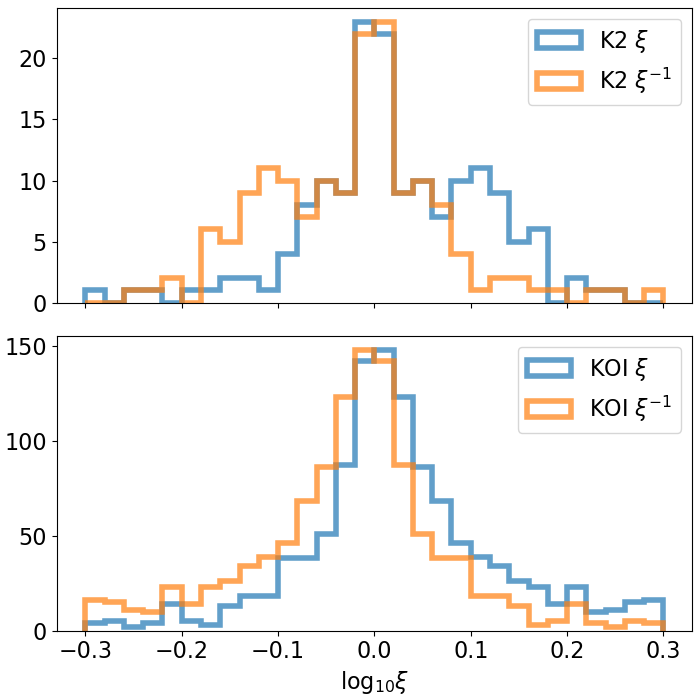}
\caption{Top: our candidate normalized duration ratios ($\xi$) and its
  inverse for all pairs of planets in multiplanet systems. The ratio
  is always the inner planet divided by the outer planet. Bottom:
  same as the top panel, but for the KOIs with periods less than 50 days. The asymmetry
  in the ratio and its inverse indicates a small inclination
  dispersion between planets (see text for discussion).
\label{normed_duration}}
\end{figure}

We compare our normalized duration ratio distribution for all our \project{K2}
multiplanet pairs to those of the \project{Kepler} KOIs with periods
less than 50 days in Figure \ref{normed_duration}. Because our sample
size is much smaller than the KOIs (\nplanetpairs{}--1186 planet
pairs), our metric is noisier. However, both distributions show a
similar pattern: a peak in the normalized duration ratio at 1,
indicating that most multiplanet pairs are real candidates that orbit
the same star.

Another trend appears in both our \project{K2} and the \project{Kepler} sample:
an asymmetry favoring normalized duration ratios slightly above 1. In
Figure \ref{normed_duration} we plot both $\xi$ and $\xi^{-1}$; if the
true distribution were values of $\xi=1$ with random scatter, the two
would be symmetric, but instead we see $\xi$ slightly favoring values
above 1.

This asymmetric distribution was used by \cite{fab14} as evidence of
multiplanet systems having small mutual inclinations --- orbiting in
nearly, but not exactly, the same planes.  If planets are perfectly
coplanar, then any inclination of that mutual plane as viewed from
Earth will cause the outer planets in a system to have higher impact
parameters than the inner planets; therefore, the outer planets will
have slightly lower normalized durations, causing $\xi$ to be slightly
above 1. Allowing for a small eccentricity and mutual inclination
distribution can create the smaller tail of $\xi$ ratios below 1, while
the majority are still distributed just above 1.

\paragraph{Stellar Density Comparisons in Multiplanet Systems.}

As demonstrated in Equation \ref{eq:density}, a measure of a transit's
duration and period can provide insight about the average density of
the host star.  Once again, these measures should provide identical
stellar densities under the stated assumptions, but eccentricity,
inclination, and non-negligible planet masses or radii will cause the
stellar density estimates to deviate slightly from one planet to
another. Nonetheless, when comparing multiple planets in a system, we
should get similar stellar densities; vast deviations would suggest that
our candidates were not real or not orbiting the same star.

As described in \S\ref{mcmc}, we generate the derived stellar density
for every step in our MCMC chains for every planet candidate, using
equations from \cite{win10} and \cite{sea03}. This allows us to get
full posterior distributions for the stellar density estimates and
compare the results for pairs of planets in our multiplanet systems.

\begin{figure}[tbp]
\includegraphics[width=\columnwidth]{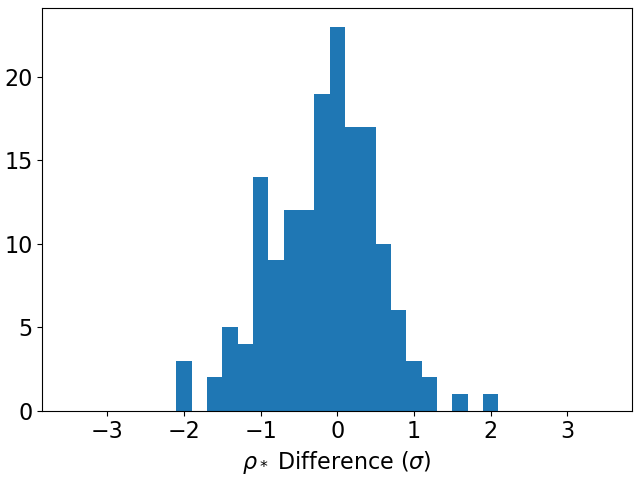}
\caption{Difference in stellar density estimates between pairs of
  planets in our multiplanet systems (in standard deviations apart).
\label{normed_density}}
\end{figure}

The difference in density between each pair of planets is shown in
Figure \ref{normed_density}.  Like the normalized duration ratios, the
derived stellar densities for most pairs of planets are consistent,
indicating that the planets in the multiplanet systems are real and
orbiting the same star.  The broad consistency between the period
ratios discussed in \S \ref{sec:general_properties} is yet another
indication that the multiplanet candidates are likely real planets.

Finally, the similarity in the numbers of single-planet and
multiplanet candidate systems discussed in \S
\ref{sec:general_properties} indicates that by extension the
single-planet systems are likely mostly real planets, although with a
possible small false-positive contamination.  Although the candidates
we have found are still candidates, we are optimistic that further
vetting and follow-up will reveal the majority of these to be
confirmed transiting planet systems.

\subsubsection{Ultra-short-period Planets}
\label{usps}

One goal of our new \pipeline{QATS} pipeline is to detect all planets,
even those edge cases that cause the most trouble for most methods. In
this section and the following we look at \pipeline{QATS}'s
performance in the two period limits of a transit search:
ultra-short-period (USP) planets and single-transit events.

USP planets, usually defined as those with periods less than a day,
are a unique class of planet predominantly found with \project{Kepler}
\citep{san14}. Many pipelines set the minimum period to start their
planet searches above 1 day, making USP detections
challenging. Furthermore, USP transit durations are also very short,
less than an hour in many cases, which means that they span just one or two
\project{Kepler} cadences and can be difficult to find even if the
pipeline searches for such short-period events. The official
\project{Kepler} search \citep[TPS;][]{jen10b}, for example, sets a
minimum transit duration of 1 hr and a minimum period of 0.5 days,
so many USP planets were missed before a dedicated search for them
\citep{san14}.

In our search of C0--8, we used a minimum transit duration of 2 hr
and a minimum period of 0.3 days, so the especially short-duration
events will be harder to detect. Because we also removed one and two
cadence outliers, deep USP candidates may also have had many transits
removed and thus been overlooked (our median star's outlier removal
level was $\sim 1500$ ppm or 4.2 R$_\oplus$ assuming a solar-sized
star). Nevertheless, we compare our search to the dedicated USP search
of \cite{ada16}, who found 19 USP candidates in C0--5.

Of their 19 candidates, we found 13. However, we also found eight new
planet candidates in C0--5 not in their catalog (or in their list of
likely EBs). Two of these candidates (EPIC 202092559 and EPIC 210801536)
have depths less than 200 ppm. We also find 17 previously unreported
USP candidates in C6--8.  Of these \nnewusp{} combined new USP
candidates, after applying our stellar parameters, 8 have radii less
than 4 R$_\oplus$, indicative of small planets worth following up.

\subsubsection{Single-transit Events}
\label{singletransits}

One of the benefits to our version of \pipeline{QATS} is that it can
be used to identify both periodic and single-transit events
simultaneously. As discussed in \S\ref{qatspeaks}, this first search
has a reduced sensitivity to single-transit events, often labeling
them as having no signal above the \pipeline{QATS} spectrum baseline
and prematurely cutting them from the pipeline. This will be adjusted
in future versions, but even so, we have detected \nsingletran{}
single-transit events in our C0--8 search. The other reason our
\pipeline{QATS} search will have difficulty with some single-transit
events is our duration limit. We only search for events up to a
maximum of 17 hr, and planets or EBs that only eclipse once in a \project{K2}
campaign may be on orbital periods of hundreds of days with event
durations much longer than 17 hr. For example, the longest-duration
transiting planet so far was found in C14 and lasts for 54 hr
\citep{gil18} --- longer than our longest complete search window that
includes a 17 hr transit and 0.6 days of continuum on either side.

Single-transit searches have historically been done independently of
periodic searches, and the most common approach is not systematic but
rather scanning all light curves by eye for potential transit
events. \cite{wan15} found 41 long-period systems in the original
\project{Kepler} by eye with citizen scientists from the Planet
Hunters initiative; \cite{ueh16} similarly found 28 single transits
manually.  \cite{for16} performed the most comprehensive systematic
single-transit search in \project{Kepler}, creating an automated
search and validation procedure and finding 16 candidates.

The only systematic search for single-transit events in \project{K2} has been
\cite{osb16}, who searched C1--3 and found seven candidates, of which we
found two (EPIC 203311200 and EPIC 201635132). Two of the \cite{osb16}
candidates have durations significantly longer than a day, explaining
our lack of detection, one is very shallow and does not appear in the
\pipeline{EVEREST} light curve, and the other two were found but
rejected by \pipeline{QATS} in its automated vetting stage as
described above.  We also find 20 additional single-transit candidates
in C1--3 not mentioned by \cite{osb16}; however, these are much deeper,
and it is not clear whether they filtered likely EBs out from their sample.

The most comprehensive single-transit search of \project{K2} to date is another
manual one. \cite{lac18} searched C0--14 by eye and found 164
single-transit events, 101 of which are in C0--8 around 100 different
stars. Of those 101 events, we find just 34 --- a sign that we can
improve our detection of single-transit events significantly. Most are
identified by \pipeline{QATS} but eliminated in the automatic cuts
stage, while about a dozen have durations longer than a day and are
missed altogether. On the other hand, we find 34 systems in C0--8 with
a single-transit event that is not listed in the \cite{lac18} catalog.

Altogether, our \nsingletran{} single-transit events represent the
largest such catalog to date in either \project{Kepler} or \project{K2} found
via systematic search (i.e., not by eye). \pipeline{QATS} is also the
only transit search that is able to find both single transits and
periodic planets simultaneously --- not to mention those with
TTVs. This capability has proven useful in the short 70- to 80-day \project{K2}
campaigns, but it will take on even more significance when applied to
the \project{TESS} mission data, where the majority of stars are only
observed for just 27 days.

\subsection{\project{K2}'s Brighter and Smaller Stellar Sample}
\label{sec:brighter_smaller}

The \project{Kepler} mission was designed for a very specific task:
finding Earth-sized planets around stars like the Sun. To accomplish
its goal, \project{Kepler}'s stellar sample was carefully crafted to
include as many FGK-type main-sequence stars as possible that were
bright enough to detect small transits. By stellar type, only about
4000 of \project{Kepler}'s 200,000 targets were M dwarfs, and they
yielded just 150 of the mission's over 4000 planet candidates
\citep{dre13, dre15}.  On the magnitude side, 60\% of
\project{Kepler}'s observed targets and 65\% of its planet candidates'
host stars were fainter than 14th magnitude, making follow-up
difficult. Radial velocity follow-up of small planets is typically only
possible for stars brighter than 12--13th magnitude \citep{mar14}.

On the other hand, the \project{K2} targets were chosen via community proposal,
and a much more diverse set of stars were selected around typically
brighter stars. A total of 50\% of \project{K2}'s targets are brighter than 14th magnitude,
and 30\% are listed in the EPIC as M dwarfs (radii below 0.6
R$_\odot$) by \cite{hub16}.

\begin{figure}[tbp]
\includegraphics[width=\columnwidth]{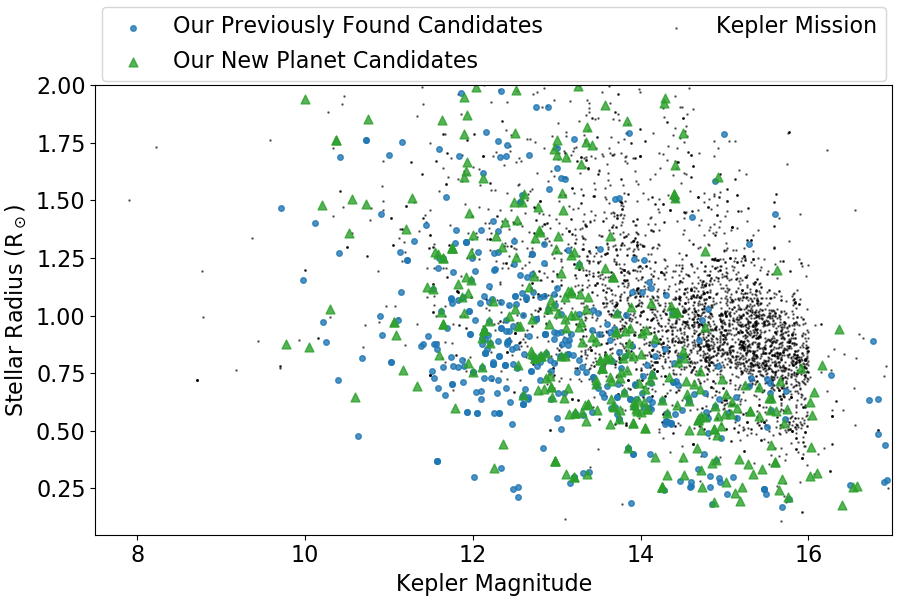}
\includegraphics[width=\columnwidth]{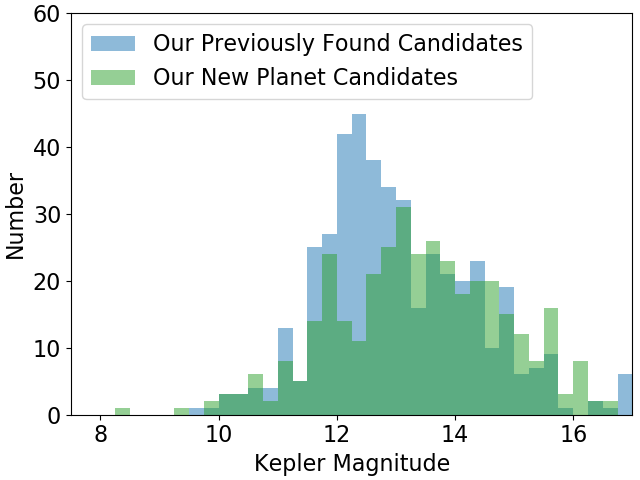}
\caption{Top: \project{Kepler} magnitude vs. host star radius for
  our \project{K2} planet candidates (circles, previously found by other groups;
  triangles, newly found in our search) and the \project{Kepler}
  KOIs. \project{K2} is dramatically expanding the number of planets around
  bright M dwarfs. The sharp cutoff at 16th magnitude is an artifact
  of \project{Kepler}'s target selection process. \label{mag_rstar}
  Bottom: top panel marginalized over host star radius showing
  that our search is finding new planets at all magnitudes but
  becomes increasingly fruitful at the fainter magnitudes, likely due
  to \pipeline{EVEREST}'s better performance.}
\end{figure}

Figure \ref{mag_rstar} shows how our \project{K2} planet candidates skew
brighter: while 85\% of \project{Kepler} candidates are fainter than
13th magnitude, \fracbright{}\% of our \project{K2} planet candidates orbit
stars brighter than 13th magnitude. At all magnitudes brighter than
13, our \project{K2} candidate count from C0--8 is within a factor of 2 of the
original \project{Kepler}; after a complete search through all 20 \project{K2}
campaigns, \project{K2} will have found more planets around stars brighter than
13th magnitude and amenable to follow-up than \project{Kepler} did. In
addition, \project{K2}'s planets are on average shorter period than
\project{Kepler} because of the limited \project{K2} campaign duration, so the
RV signal will be larger for the average \project{K2} planet.

As a function of host star radius, \project{K2} has focused on M dwarfs
considerably more than \project{Kepler} did. Figure \ref{mag_rstar}
also shows the \textit{stellar} radius distribution for the planet
candidates' host stars. Above 0.7 R$_\odot$, \project{Kepler} has
found 5-10$\times$ as many candidates as our \project{K2} C0--8 sample. Yet
around small M dwarfs, \project{K2}'s planet candidate count already outnumbers
that from \project{Kepler}. \project{K2} has already matched \project{Kepler}'s
total planet candidate count for stars less than 0.5 R$_\odot$ and
will more than double it after all 20 campaigns are thoroughly
searched.  Combining magnitude and host star size, \project{K2} has already
produced more planets transiting stars brighter than 13th magnitude
and smaller than the Sun (\nsmallbright{} in our \project{K2} candidate list
vs. 130 from \project{Kepler}).

\subsection{Exploring the Radius Gap} \label{sec:radius_gap}

The possible boundary between rocky and gaseous exoplanets is one of
the most intriguing results to come from the \project{Kepler}
population. \cite{ful17} found a gap in the radius distribution of
small exoplanets at periods less than 100 days, indicating that planets
with radii of 1.5--2.0 R$_\oplus$ are relatively rare. \cite{lop18} and
\cite{van18b} provide tentative evidence that this gap is due to
photoevaporation of the atmospheres because the gap appears to move
to smaller planet radii at lower stellar fluxes incident on the
planets (although see \citealp{gup19} for an alternative
explanation). These short-period rocky planets therefore may have
started off with gaseous atmospheres and only later come to have their
rocky cores exposed after intense stellar radiation blew off the
entire gaseous envelope.

Introduced in \S\ref{stellarpars}, the nominal errors for our
\project{Gaia} stellar radii (most of the stars above 0.5 R$_\odot$)
are just \rserrgaia{}\% \citep[better than the 11\% of ][]{ful17},
while the M dwarf sample with parameters from \cite{hub16} has larger
uncertainties of \rserrhuber{}\%. This translates into a median planet
radius error of \rperrgaia{}\% and \rperrhuber{}\%, respectively. The
disparity is larger for luminosity. Because the \project{K2} stars are brighter
than average for \project{Gaia}, and \project{Gaia} was designed
specifically to measure the distance and brightness of stars,
\project{Gaia} uncertainties on stellar luminosities average just
\lumerrgaia{}\%. However, \cite{hub16} do not estimate luminosity
for the EPIC stars at all, so our luminosity values are derived from
their temperature and radius values and have a median uncertainty of
\lumerrhuber{}\%. This makes our incident flux estimates for M dwarfs
particularly uncertain compared to the larger stars in the
\project{Gaia} sample (median \inserrhuber{}\% and \inserrgaia{}\%,
respectively); however, all incident fluxes are also largely affected
by our median \arerr{}\% uncertainty in the derived $a/R_*$ values.

\begin{figure}[tbp]
\includegraphics[width=\columnwidth]{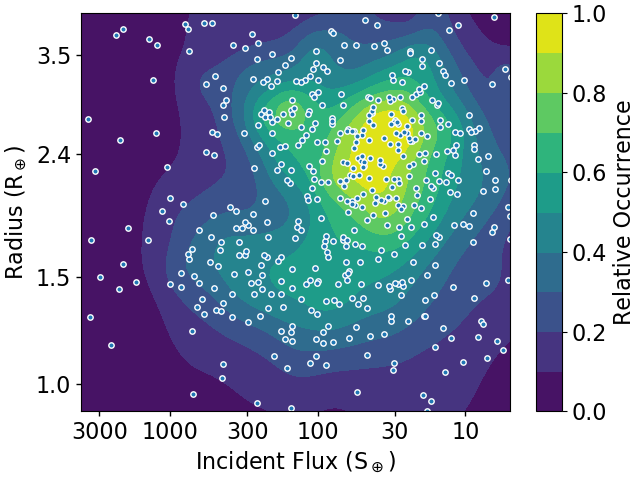}
\caption{ Incident flux vs. planet radius for our \project{K2} planet
  candidates. Each point shows one of our planet candidates, while the
  underlying contours show the distribution accounting for
  uncertainties (typical errors are 50\% in incident flux and 20\% in
  planet radius). The radius gap is identified by the ``pinching'' of
  the contours between the small and large radius groups.  We do not
  correct for pipeline completeness. \label{insolation}}
\end{figure}

Using the stellar luminosity and radius, combined with the derived
$a/R_*$ from the transit fits, we estimate the flux received for each
candidate, assuming a circular orbit; the results are shown in Figure
\ref{insolation}. As with \cite{ful17}, the radius gap in our sample
appears strongest at incident fluxes near 300 S$_\oplus$ and occurs at
a radius of about 2 R$_\oplus$, but our radius and incident flux
errors are slightly too large to make out a significant gap with our
limited planet sample.

With \project{K2}'s larger focus on M dwarfs, our new planet candidates present
an opportunity to explore how the radius gap might change as a
function of host star mass. For example, \cite{wu19} claims a
dependence of the radius gap on the host of the primary star such that
larger stars host larger planets and the gap moves to larger
radii. However, they also note that M dwarfs provide a key anchor to
that relation and there were not enough known planets around M dwarfs
near the radius gap yet.

Our work provides the first step: finding the planets at all. As our
survey of the \project{K2} campaigns doubles the number of planet candidates
around M dwarfs, follow-up characterization with \project{Gaia} and
ground-based spectra can increase the precision of the incident fluxes
and planet radii to measure how the shape and location of the radius
gap might depend on various stellar and environmental properties.

\subsection{The Habitable Zone}
\label{sec:HZ}

Due to the nature of the short \project{K2} campaign durations, most of our
planet candidates are found at very short periods and correspondingly
high instellations. However, a handful of planet candidates around
smaller stars may lie within their star's liquid water ``habitable
zone.'' Unfortunately, because the more precise \project{Gaia}
parameters only consider stars above 0.5 R$_\odot$, most of the
potential habitable zone planets only have EPIC parameters with poorer
luminosity estimates.

\begin{figure}[tbp]
\includegraphics[width=\columnwidth]{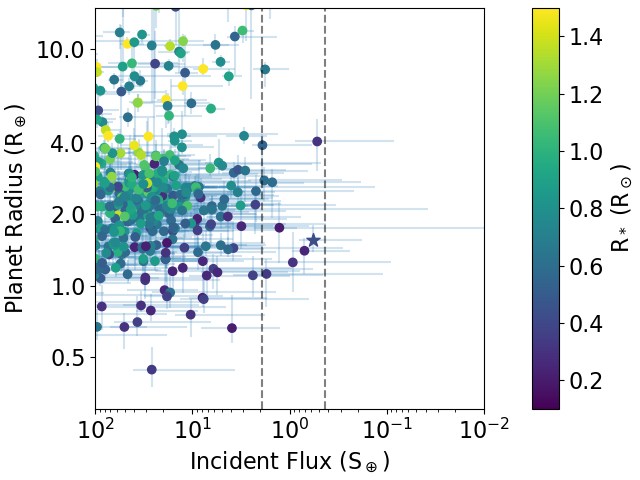}
\caption{View of our \project{K2} candidates near the habitable zone, limited to
  those with an incident flux more than one standard deviation above
  0. The candidates are colored by the host star's radius.  Because of
  the imprecise luminosities available for the M dwarf host stars, the
  incident fluxes near the habitable zone are uncertain and many
  planets span the entire zone and beyond.  Incident fluxes for Venus
  (left dashed) and Mars (right dashed) mark approximate boundaries of
  the habitable zone. The star marks our new planet candidate in the
  habitable zone, the potential K2-83 d: EPIC
  210508766.3. \label{habzone}}
\end{figure}

We plot the incident fluxes near the habitable zone in Figure
\ref{habzone}, limited only to those systems where our measure of the
incident flux is more than 1 standard deviation above 0 (80\% of our
candidates). The majority of planets are well interior to the
habitable zone, although the uncertain incident fluxes allow many to
cover the entire region within one standard deviation.

Many candidates have such uncertain stellar luminosities that their
parameters more than span the entire habitable zone. A better
characterization of our new M dwarf hosts will be necessary to pin
down potential habitable zone planets, as was done by \cite{dre17b}.

Even so, there are five planet candidates in or near the habitable zone
with planet radii small than 2 R$_\oplus$. Of those five, four have
been previously found by other groups (EPIC 201367065.3 [K2-3 d],
EPIC 205489894.1, EPIC 201912552.1 [K2-18 b], EPIC 211579112.1), while one is new:
EPIC 210508766.3.

\begin{figure}[tbp]
\includegraphics[width=\columnwidth]{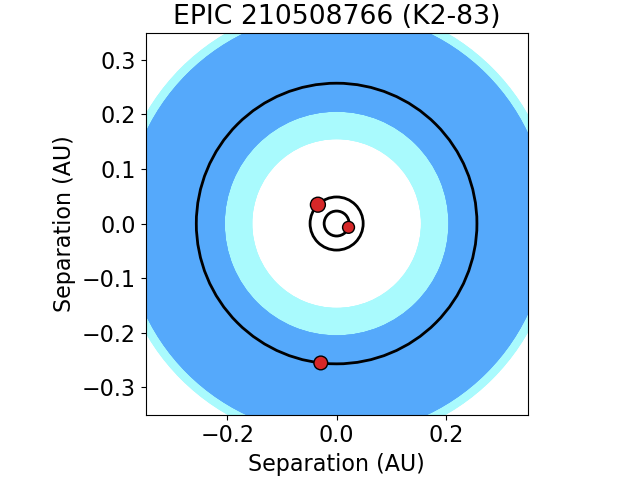}
\caption{ Orbits of the three planets in the K2-83 system relative
  to the habitable zone. The inner two planets have been previously
  confirmed as K2-83 b and K2-83 c while the outer planet is our new
  candidate in the system. The cyan and blue regions represent the
  optimistic and conservative habitable zones from
  \cite{kop13}. \label{fig:hz}}
\end{figure}

Our new small-planet candidate near the habitable zone, EPIC
210508766.3 (if confirmed, it would be K2-83 d), is an outer companion
to the previously confirmed two-planet system K2-83 \citep{cro16,
  dre17}. \cite{dre17b} characterize the star as an M1 dwarf,
slightly larger and brighter than the values we use from the EPIC, but
the planet remains in the habitable zone regardless, with an
instellation between 33\% and 96\% of Earth at 68.3\% confidence.  The
planet has a radius that is about 31\%-73\% larger than that of Earth, which
is in the range where it may be rocky, if planets at shorter period
are any guide \citep{rog15}. We show the architecture of this system
and its habitable zone in Figure \ref{fig:hz}. This planet highlights
the usefulness of \pipeline{EVEREST} and \pipeline{QATS} by finding
outer small, temperate companions that other groups overlooked. We
note that the same three transits used in our discovery of the outer
planet candidate are found in the \pipeline{K2SFF} light curves as
well, but at lower significance and more variable depths. The planet
would not have passed our automated vetting steps if we used the
\pipeline{K2SFF} light curves.  Detection of outer, cooler companions
with a limited number of transits will become especially important
with \project{TESS}, and \pipeline{QATS} could be used to help expand
\project{TESS}'s reach to more habitable zone planets.

\subsection{Reciprocal Transits} \label{sec:reciprocal}

The transit method has become the most successful planet detection
method, especially for rocky, Earth-sized planets. As we move from
planet detection toward detailed characterization and ultimately a
search for life on these transiting planets, a natural question
arises: would theoretical observers in a particular planetary system
be able to detect any planets in our own solar system via transits? We
define any transiting planet geometrically capable from their
perspective of seeing at least one solar system planet transit as a
reciprocally transiting planet.
  
All reciprocally transiting planets will be located near the ecliptic
along the eight planes of our solar system planets' orbits. Because \project{K2}
is the first and only space-based transit mission to observe along the
ecliptic to date, \project{K2} planets will dominate the known population of
reciprocally transiting planets (at least until a potential TESS
extended mission along the ecliptic). This idea was first explored by
\cite{wel18}, who identified 68 known planets or planet candidates (not
all of which transit) around 50 unique stars in zones of the sky that
could detect at least one solar system planet's transit; only seven had
been discovered via \project{K2}.

Among the \nplanet{} planet candidates presented here, 154 of them
(around 138 stars) are reciprocally transiting; 73 of these are new
planet candidates in this work. Just one system, the previously known
K2-101 b (EPIC 211913977, presented in \cite{wel18}), can detect three
transiting solar system planets: it is in the two-in-a-million alignment
capable of observing transits of Jupiter, Saturn, and Uranus. Nineteen other
single-planet systems (nine of them new in this work) are reciprocally
transiting with two solar system planets.

Among our multiplanet systems, nine of our two-planet systems are
capable of seeing one of our solar system planets transit. The
previously known and validated three-planet system K2-58 (EPIC
206026904) would be able to see transits of Venus. The four-planet
system EPIC 211939692 (all new planet candidates in this work) is
capable of seeing Jupiter and Saturn transit from its
perspective. Finally, our new four-planet system EPIC 206135682 (only
one of the planets had been previously found) reciprocally transits
with Earth: planets in this system would see the Sun as a
single-planet transit host with a 365-day habitable zone planet.

\section{Conclusions} \label{conclusions}

We have presented a comprehensive update to the \cite{car13}
\pipeline{QATS} transit detection pipeline with the goal of developing
it into a nearly all-encompassing transit search that can find
periodic planets, those with TTVs, single-transit events, and
ultra-short-period planets --- all without losing sensitivity to the
lowest-S/N planets or multiplanet systems. We applied
this method to \numstars{} stars in \project{K2} campaigns 0--8, using the light
curves from \pipeline{EVEREST 1.0} (\S\ref{methods}). In total, we
found \nplanet{} planet candidates, \nnewplanet{} of which are new and
undetected by all previous searches (\S\ref{sec:general_properties});
along the way we also presented our catalog of \nebs{} EBs (\S\ref{sec:ebs}).

Among our candidates, we did indeed find planets with TTVs
(\S\ref{sec:ttvs}), \singletrantot{} single-transit/eclipse events
(\S\ref{sec:singletran}), and \nnewusp{} new USP candidates
(\S\ref{usps}).  We report two new five-planet systems and a new
six-planet system (\S\ref{sec:multis}). In total, we found \nmultis{}
multiplanet systems, up from the previously known \nothermultis{} in
C0--8. We also identified 154 reciprocally transiting planets (73 new
here): transiting planets that could see at least one solar system
planet transit from its point of view; 12 of our multiplanet systems
are reciprocally transiting, including a four-planet system that could
see transits of Earth (\S\ref{sec:reciprocal}).

As it stands, \project{K2} has doubled the number of transiting planets around M
dwarfs, and we introduce \nmdwarfcands{} new candidates here,
including one potentially rocky and in the habitable zone
(\S\ref{sec:HZ}). This number will double again once campaigns 10--19
are fully searched. Unlike \project{Kepler}, the majority of the \project{K2}
host stars are brighter than 13th magnitude, making follow-up and
characterization more feasible (\S\ref{sec:brighter_smaller}). One of
many potential applications of this smaller, brighter sample of stars
is the future prospect of measuring the planet radius gap as a
function of host star mass (\S\ref{sec:radius_gap}), which will
require improved characterization of the host stars.

Comparing to the previous searches, we were able to recover
\myoverlap{}\% of candidates reported by at least one previous group,
while only \theiroverlap{}\% of our planet candidates had previously
been found by all the other searches combined. As discussed in
\S\ref{previous}, future updates to our pipeline should allow us to
recover more than 90\% of the candidates found by other groups while
continuing to find a comparable number of new candidates missed by the
other searches --- thus approximately doubling \project{K2}'s planet yield.

One of the sources of our missed planets has already been addressed
with the newest \pipeline{EVEREST} pipeline that improves light-curve
precision on saturated, crowded-field, and faint stars
\citep{lug18}. Even with no further updates to \pipeline{QATS}, we can
recover some planets found by other groups just by using their light
curves or the \pipeline{EVEREST 2.0} light curves, even though they
were missed in our \pipeline{EVEREST 1.0} search presented in this
paper.  Future updates to \pipeline{QATS} will address the other two
main difficulties: short-duration transits and properly assessing the
significance of single-transit events. While we successfully detected
some candidates in each category, our efficiency can be significantly
improved. As we look to TESS and its 27-day baseline for most stars,
these single-transit detections will become even more critical. The
more we can improve that particular feature of \pipeline{QATS}, the
further we can expand TESS's sensitivity.

\acknowledgements

The authors thank Jason Curtis, Jeff Coughlin, and the referee for
feedback and helpful discussions.

This research was supported by an appointment to the NASA Postdoctoral
Program at the NASA Goddard Space Flight Center, administered by
Universities Space Research Association under contract with NASA.
This material is based on work supported by the National Science
Foundation Graduate Research Fellowship under Grant DGE 1256082. This
research is part of the Blue Waters sustained-petascale computing
project (Graduate Research Fellowship), which is supported by the
National Science Foundation (awards OCI-0725070 and ACI-1238993) and
the state of Illinois. Blue Waters is a joint effort of the University
of Illinois at Urbana-Champaign and its National Center for
Supercomputing Applications.  E.A. acknowledges support from NSF grant
AST-1615315, NASA grant NNX13AF62G, and from the NASA Astrobiology
Institute's Virtual Planetary Laboratory Lead Team, funded through the
NASA Astrobiology Institute under solicitation NNH12ZDA002C and
Cooperative Agreement No. NNA13AA93A.

\vspace{5mm}
\facility{\project{Kepler}}

\software{\pipeline{EVEREST} \citep{lug16, lug18}, \pipeline{K2SFF}
  \citep{van14, van16b}, \pipeline{QATS} \citep{car13},
  \pipeline{emcee} \citep{for13}, \pipeline{batman} \citep{kre15},
  numpy \citep{numpy}, matplotlib \citep{matplotlib}, scipy
  \citep{scipy}, statsmodels \citep{sea10a}, astropy \citep{astropy1,
    astropy2}.}

%--------------------------BIBLIOGRAPHY---------------------------
\bibliography{ReferenceLibrary}
\clearpage

\begin{landscape}
% planet candidates table is table 7 in reference order in the text
\setcounter{table}{6}
\begin{longrotatetable}
\movetabledown=2.2in
\begin{deluxetable*}{c c c c c c c c c c c }
\tablecaption{Our Sample of Planet Candidates from C0--8 \label{table:candidates}}
\tablewidth{600pt}
\tabletypesize{\scriptsize}
\tablehead{
\colhead{Candidate} & \colhead{C\#} & \colhead{Period} & \colhead{t$_0$} & \colhead{Duration} & \colhead{Depth} & \colhead{$R_p/R_*$} & \colhead{R$_p$} & \colhead{R$_*$} & \colhead{a/R$_*$} &  \colhead{Inc. Flux} \\[-8pt]
 \colhead{}         &    \colhead{}      & \colhead{(days)} & \colhead{(BJD -- 2455000)} & \colhead{(hr)} & \colhead{(ppm)} & \colhead{} & \colhead{(R$_\earth$)} & \colhead{(R$_\sun$)} & \colhead{} & \colhead{(S$_\earth$)}
}
\startdata
201147085.1 & 1 & 1.17589$^{+0.00010}_{-0.00010}$ & 1810.7406$^{+0.0042}_{-0.0040}$ & 1.24$^{+0.14}_{-0.20}$ & 686$^{+80}_{-85}$ & 0.0254$^{+0.0025}_{-0.0052}$ & 0.86$^{+0.18}_{-0.26}$ & 0.309$^{+0.057}_{-0.070}$ & 6.3$^{+2.5}_{-1.4}$ & 190$^{+180}_{-150}$\\
201152065.1 & 1 & 10.6966$^{+0.0020}_{-0.0021}$ & 1817.3147$^{+0.0072}_{-0.0069}$ & 3.25$^{+0.34}_{-0.43}$ & 557$^{+56}_{-59}$ & 0.0226$^{+0.0022}_{-0.0055}$ & 1.45$^{+0.18}_{-0.36}$ & 0.588$^{+0.041}_{-0.015}$ & 21.9$^{+9.6}_{-4.6}$ & 33$^{+29}_{-14}$\\
201155177.1 & 1 & 6.68851$^{+0.00074}_{-0.00075}$ & 1814.6747$^{+0.0047}_{-0.0044}$ & 3.25$^{+0.22}_{-0.31}$ & 1042$^{+69}_{-72}$ & 0.0313$^{+0.0023}_{-0.0047}$ & 2.25$^{+0.27}_{-0.47}$ & 0.658$^{+0.062}_{-0.098}$ & 14.0$^{+5.1}_{-2.5}$ & 108$^{+82}_{-50}$\\
201160662.1 & 1 & 1.5374115$^{+0.0000062}_{-0.0000061}$ & 1810.49714$^{+0.00019}_{-0.00019}$ & 2.380$^{+0.047}_{-0.046}$ & 14564$^{+69}_{-69}$ & 0.259$^{+0.071}_{-0.099}$ & 45$^{+12}_{-17}$ & 1.599$^{+0.047}_{-0.044}$ & 3.59$^{+0.10}_{-0.17}$ & 7170$^{+670}_{-860}$\\
201176672.1 & 1 &  & 1877.80380$^{+0.00050}_{-0.00050}$ & 14.47$^{+0.19}_{-0.21}$ & 30600$^{+110}_{-120}$ & 0.39$^{+0.10}_{-0.14}$ & 30.0$^{+8.3}_{-10.6}$ & 0.707$^{+0.050}_{-0.050}$ &  & \\
201197348.1 & 1 & 14.9139$^{+0.0018}_{-0.0020}$ & 1825.1108$^{+0.0045}_{-0.0042}$ & 3.26$^{+0.23}_{-0.32}$ & 2400$^{+170}_{-180}$ & 0.0460$^{+0.0038}_{-0.0078}$ & 3.24$^{+0.33}_{-0.56}$ & 0.647$^{+0.038}_{-0.024}$ & 32.6$^{+9.7}_{-4.8}$ & 13.7$^{+8.4}_{-4.2}$\\
201205469.1 & 1 & 3.471136$^{+0.000079}_{-0.000079}$ & 1813.3563$^{+0.0011}_{-0.0011}$ & 2.045$^{+0.072}_{-0.105}$ & 6650$^{+160}_{-170}$ & 0.0775$^{+0.0034}_{-0.0063}$ & 5.16$^{+0.49}_{-0.49}$ & 0.610$^{+0.051}_{-0.029}$ & 12.8$^{+2.8}_{-1.3}$ & 46$^{+22}_{-10}$\\
\enddata
\end{deluxetable*}
\end{longrotatetable}
\end{landscape}
\clearpage

% EB table is table 9 in reference order in the text
\setcounter{table}{8}

\startlongtable
\begin{deluxetable*}{c c c c c c }
\tablecaption{Our Sample of Eclipsing Binaries from C0--8 \label{table:ebs}}
\tablewidth{700pt}
\tabletypesize{\scriptsize}
\tablehead{
\colhead{Binary} & \colhead{C\#} & \colhead{Period} & \colhead{t$_0$} & \colhead{Duration} & \colhead{Depth}  \\[-8pt]
   \colhead{}       &    \colhead{}      & \colhead{(days)} & \colhead{(BJD -- 2455000)} & \colhead{(hr)} & \colhead{(ppm)}
}
\startdata
201160323.1 & 1 & 22.27219$^{+0.00020}_{-0.00021}$ & 1811.49231$^{+0.00036}_{-0.00039}$ & 4.270$^{+0.051}_{-0.061}$ & 404200$^{+4100}_{-4300}$\\
201160323.2 & 1 & 22.30082$^{+0.00029}_{-0.00029}$ & 1820.00209$^{+0.00055}_{-0.00055}$ & 4.95$^{+0.13}_{-0.13}$ & 229700$^{+1800}_{-1800}$\\
201161715.1 & 1 & 59.88732$^{+0.00048}_{-0.00048}$ & 1822.51092$^{+0.00033}_{-0.00032}$ & 21.909$^{+0.100}_{-0.117}$ & 89730$^{+140}_{-140}$\\
201161715.2 & 1 &  & 1840.41351$^{+0.00024}_{-0.00025}$ & 19.87$^{+0.17}_{-0.15}$ & 101630$^{+200}_{-190}$\\
201173390.1 & 1 & 16.99562$^{+0.00011}_{-0.00011}$ & 1816.43053$^{+0.00027}_{-0.00027}$ & 12.649$^{+0.088}_{-0.097}$ & 42270$^{+160}_{-150}$\\
201173390.2 & 1 & 16.99471$^{+0.00041}_{-0.00041}$ & 1824.81490$^{+0.00079}_{-0.00080}$ & 9.039$^{+0.057}_{-0.058}$ & 6199$^{+36}_{-37}$\\
201174640.1 & 1 &  & 1812.70325$^{+0.00020}_{-0.00018}$ & 5.696$^{+0.077}_{-0.084}$ & 180830$^{+600}_{-610}$\\
201182911.1 & 1 & 1.9931320$^{+0.0000018}_{-0.0000015}$ & 1811.089992$^{+0.000041}_{-0.000048}$ & 2.298$^{+0.015}_{-0.021}$ & 381690$^{+1420}_{-970}$\\
\enddata
\end{deluxetable*}

\end{document}